\documentclass[12pt,preprint,aps,floatfix,onecolumn]{revtex4}
\usepackage{epsfig,amsmath,bbm,graphicx,color,floatflt}
\usepackage{psfrag}
\usepackage{wrapfig}

\long\def\comment#1{}

\begin{document}
\thispagestyle{empty}
\begin{tabbing}
\\\\
\end{tabbing}
\begin{center}
{\large \bf{
 NUMERICAL SIMULATION OF FOLDING AND UNFOLDING OF PROTEINS}}

\vspace{11 mm}
{\large by

\vspace{11 mm}
Maksim Kouza

\vspace{32mm}
Dissertation directed by: Associate Professor Mai Suan Li
}
\vspace{85 mm}

{\normalsize
Dissertation submitted to the Institute of Physics Polish 

Academy of Sciences in partial fulfillment of 

the requirements for the degree of 

Doctor of Philosophy 

Warsaw

$\, \,$ 2008
}

\end{center}
\newpage
\thispagestyle{empty}


\begin{center} {\large \bf { Acknowledgments}} \end{center}

Probably these first pages of a PhD thesis are the most widely read pages from entire publication. In that place you can find people who means  something in my 5 year life of PhD candidate.

 First and foremost, I  would like to acknowledge my thesis advisor, Assoc. Prof.  Mai Suan Li, for his superb mentorship. His broad knowledge, experience, patience and encouragement helped guide me throughout the duration of this work.  The dedication of his time and energy was one of the main reasons that I was able to finish this challenging  work. He is a excellent advisor who has taught me a lot about things in science to succeed in research. I really enjoyed the time spent with him.

I would like to thank prof. Chin-Kun Hu for providing me with sufficient funds and an opportunity to work in his lab to conduct my research during my visits in Taiwan. 

I also would like to thank P. Bialokozewicz and  P. Janiszewski for the useful discussions and the valuable remarks and tips about linux software.

I am very grateful to the Polish Committee for UNESCO for the financial support.

Lastly, I would like to attribute my largest credit to my family in Poland, Belarus and Russia. Their love and dedication always gave me an enormous amount of power to overcome all the obstacles when I got exhausted.

\thispagestyle{empty}

\newpage
\setcounter{page}{0}
\tableofcontents
\newpage

\begin{center} \section{Introduction} \end{center}

\vspace{1.0cm}
Proteins are biomolecules that
perform and control almost all functions in all living organisms.
Their biological functions include catalysis (enzymes), 
muscle contraction (titin), transport of ions (hemoglobin), transmission of information
 between specific cells and organs (hormones), activities in the
immune system (antibodies), passage of molecules across cell membranes etc.
The long process of life evolution has designed proteins in the natural world
in such a mysterious way that
under normal physiological conditions (pH $\approx$ 7, $T$ = 20-40 C,
atmospheric pressure) they acquire well defined compact three-dimensional 
shapes, known as the native conformations. Only in these conformation proteins
are biologically active. Proteins unfold to more extended conformations, if
the mentioned above conditions are changed or upon application of denaturant
agents like urea or guanidinum chloride.
If the physiological conditions are restored, then most of
proteins refold spontaneously to their native states \cite{Anfinsen_Science73}.
Proteins can also change their shape, if they are subjected
to an 
external mechanical force.
  
The protein folding theory deals with two main problems. One of them is
a prediction of native conformation for a given sequence of amino acids.
This is referred to as the protein folding.
The another one is a design problem (inverse folding), where a target
conformation is known and one has to find what sequence would fold into
this conformation. 
The understanding of folding mechanisms and protein design
have attracted an intensive experimental and theoretical interest over
the past few decades as
they can provide insights into our knowledge
about living bodies. The ability to predict the folded form from its sequence would widen 
the knowledge of genes.
The genetic code is a sequence of nucleotides in DNA that
 determines amino acid sequences for protein synthesis.
Only after synthesis and completion of folding process proteins
 can perform their myriad functions.

In the protein folding problem one achieved two major results. From the kinetics prospect, it is
 widely accepted that folding follows the funnel picture, i.e.
there exist a numerous number of routes to the native state (NS) \cite{Leopold_PNAS92}.
 The corresponding free energy landscape (FEL) looks like a funnel. This new point
of view is in sharp contrast with the picture \cite{Baldwin_Nature94}, which assumes
that the folding progresses along a single pathway. The funnel theory
resolved the so called Lenvithal paradox \cite{Levinthal_JCP68}, according to which folding times
would be astronomically large, while proteins in {\em vivo} fold within
$\mu$s to a few minutes. From the thermodynamics point of view, both
experiment and theory showed that the folding is highly cooperative
\cite{Ptitsyn_book}. The transition from a denaturated state (DS) to the folded
one is first order. However, due to small
free energies of stability of the NS,  relative to the
unfolded states ($5 - 20 k_BT$), the possibility of a marginally second
order transition is not excluded \cite{MSLi_PRL04}.

Recently Fernandez and coworkers \cite{Fernandez_Sci04} have carried out
force clamp experiments in which proteins are forced to refold under the
weak quenched force. Since the force increases the folding time
and initial
conformations can be controlled by the end-to-end distance, this technique
facilitates the study of protein folding mechanisms.
Moreover, by varying the external force
one can estimate the distance between the DS and transition state (TS) \cite{Fernandez_Sci04,MSLi_PNAS06} or,
in other words, the force clamp can serve as a complementary
tool for studying the FEL of biomolecules.

After the pioneering AFM experiment of Gaub {\em et al.}
\cite{Florin_Science94}, the study of mechanical unfolding and
 stability of biomolecules becomes flourish.
Proteins are pulled either by the constant force, or by force ramped
with a constant loading rate. An explanation for this rapidly
developing field is that
single molecules force spectroscopy (SMFS) techniques have a number of advantages compared to conventional folding studies. First,
unlike ensemble measurements, it is possible to observe differences
in nature of individual unfolding events of a single molecule.
Second, the end-to-end distance is a well-defined reaction coordinate
and it makes comparison of
theory with experiments easier. Remember that a choice of a good reaction
coordinate for describing folding remains elusive.
Third, the single molecule technique allows not only for
establishing the mechanical resistance but also for deciphering FEL of biomolecules. Fourth, SMFS is able to reveal the nature of atomic interactions.
It is worthy to note that
studies of protein unfolding are not of academic interest only.
They are very relevant as the unfolding plays
 a critically important role in several processes in cells
\cite{Matoushek_COSP03}. For example, unfolding occurs in process of protein
 translocation across some membranes. There is reversible
 unfolding during action of proteins such a titin. Full or partial
 unfolding is a key step in amyloidosis.

Despite much progress in experiments and theory, many questions remain open. 
What is the molecular mechanism of protein folding of some important proteins?
Can we use approximate theories for them?
Does the size of proteins matter for the cooperativity of the folding-unfolding
transition? 
One of the drawbacks of the force clamp technique \cite{Fernandez_Sci04}
is that it fixes one end of a protein.
While thermodynamic quantities do not depend on what end is anchored,
folding pathways which are kinetic in nature may depend on it.
 Then it is unclear if this technique
probes the same folding pathways as in the case when both termini are free.
Although in single molecule experiments, one does not know what end of
a biomolecule is attached to the surface, it would be interesting to
know the effect of  end
fixation on unfolding pathways. Predictions from
this kind of simulations will be
useful at a later stage of development, when experimentalists can
exactly control
what end is pulled.
Recently, experiments \cite{Brockwell_NSB03,Dietz_PNAS06a} have shown that 
the pulling geometry has a pronounced effect on the unfolding free 
energy landscape. The question is can one describe this phenomenon
theoretically. The role of non-native interactions in mechanical unfolding of proteins remains largely unknown. It is well known that an external
force increases folding barriers making the configuration sampling difficult.
A natural question arises is if one can can develop a efficient method
to overcome this problem. Such a method would be highly useful
for calculating thermodynamic quantities of a biomolecule subjected to an 
mechanical external force.

In this thesis we address the following questions.
\begin{enumerate}

\item
We have
studied the folding mechanism of the protein domain hbSBD
(PDB ID: 1ZWV)  of the mammalian
mitochondrial branched-chain $\alpha$-ketoacid dehydrogenase
(BCKD) complex in detail, using Langevin simulation and CD experiments.
Our results support its two-state behavior.

\item
The cooperativity of the denaturation transition of proteins was investigated
with the help of lattice as well as off-lattice models. Our studies
reveal that the sharpness of this transition enhances as the number of amino acids grows. The corresponding scaling behavior is governed by an universal
critical exponent.

\item
It was shown that refolding pathways of single
$\alpha\beta$-protein ubiquitin (Ub) depend on what end is anchored to
the surface. Namely, the fixation of the N-terminal changes refolding
pathways but anchoring the C-terminal leaves them unchanged.
Interestingly, the end fixation has no effect on multi-domain Ub.

\item
The FEL of Ub and fourth domain of {\em Dictyostelium discoideum}
filamin (DDFLN4) was deciphered.
We have studied the effect of pulling direction on the FEL of Ub.
In agreement with the experiments, pulling at Lys48 and C-terminal
increases the distance between the NS and TS about 
two-fold compared to the case when the force is applied to two termini.

\item
A new force replica exchange (RE) method was developed for efficient configuration
sampling of biomolecules pulled by an external mechanical force. Contrary to
the standard temperature RE, the exchange is carried out between different
forces (replicas). Our method was successfully applied to study thermodynamics of
a three-domain Ub.

\item
Using the Go modeling and all-atom models with explicit water, we have studied
the mechanical unfolding mechanism of DDFLN4 in detail. We predict that, contrary to the experiments
of Rief group \cite{Schwaiger_NSB04}, an additional unfolding peak
would occur at the end-to-end $\Delta R \approx 1.5 $nm in the
force-extension curve. Our study reveals the important role of non-native interactions which are responsible for a peak located at $\Delta R \approx 22 $nm. This peak can not be encountered by the Go models in which the non-native
interactions are neglected. Our finding may stimulate further experimental and theoretical studies on this protein.

\end{enumerate}

My thesis is organized as follows:

Chapter 2 presents basic concepts about proteins.
Experimental and theoretical tools for studying  protein folding and unfolding are discussed in Chapter 3.
Our theoretical results on the size dependence of the cooperativity
index which characterizes the sharpness of the melting transition are
provided in Chapter 4.
Chapter 5 is devoted to the simulation of the hbSBD domain using the
Go-modeling. Our new force RE and its application to a three-domain Ub
are presented in Chapter 6. In Chapter 7 and 8 I presented results concerning refolding under quench force and unfolding of ubiquitin and its trimer. Both, mechanical and thermal unfolding pathways will be presented.
The last Chapters 9 and 10  discuss the results of all-atom molecular dynamics
and Go simulations for mechanical unfolding of the protein DDFLN4.
The results presented in this thesis are based on the following works:
\begin{enumerate}

\item
M. Kouza, C.-F. Chang, S. Hayryan, T.-H. Yu, M. S. Li, T.-H. Huang,
 and C.-K. Hu,
Biophysical Journal {\bf 89}, 3353 (2005).

\item 
M. Kouza, M. S. Li, E. P. O'Brien Jr., C.-K. Hu, and D. Thirumalai,
Journal of Physical Chemistry A \textbf{110}, 671 (2006)

\item 
M. S. Li, M. Kouza, and C.-K. Hu,
Biophysical Journal {\bf 92}, 547 (2007)

\item
M. Kouza, C.-K. Hu and M. S. Li, 
Journal of Chemical Physics {\bf 128}, 045103 (2008).

\item
M. S. Li and M. Kouza,
Dependence of protein mechanical unfolding pathways on pulling speeds,
submitted for publication.

\item
M. Kouza, and M. S. Li, Protein mechanical unfolding: importance of non-native interactions,
submitted for publication.
\end{enumerate}

\newpage

\begin{center} \section{Basic concepts} \end{center}

\subsection {What is protein?}

\begin{figure}[!hbtp]
\epsfxsize=4.5in
\vspace{5 mm}
\centerline{\epsffile{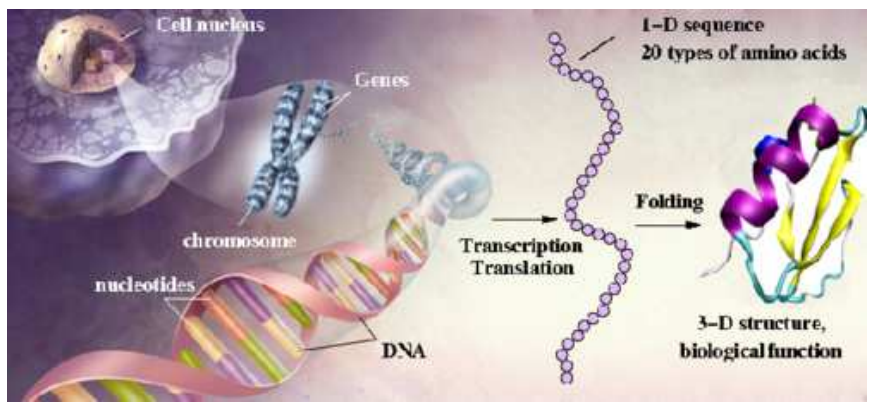}}
\linespread{0.8}
\caption{The connection between genetic information, DNA and protein.
This image and the rest of molecular graphics in this dissertation were made using
 VMD \cite{VMD}, xmgrace, xfig and gimp software.}
\label{mediator}
\vspace{5 mm}
\end{figure}

The word "protein" which comes from Greek means "the primary importance". As mentioned above,
they play a crucial role in living organisms.
Our muscles, organs, hormones, antibodies and enzymes are
made up of proteins. They are about 50\% of the dry weight of cells. 
Proteins are used as a mediator in the process
of how the genetic information moves around the cell or in another words transmits
from parents to children (Fig. \ref{mediator}). Composed of DNA, genes keep the genetic
 code as it is a basic unit of heredity. 
Our various characteristics such as color of hair, 
eyes and skin are determined after very complicated processes.
In brief, at first linear strand of DNA
in gene is transcribed to mRNA and this information is then "translated" into a 
protein sequence. Afterwards proteins start to fold up to get biologically functional
three-dimensional structures, such as various pigments, enzymes and hormones. One protein is responsible for skin color, another
one - for hair color. Hemoglobin gives the color of our blood and carry out the transport
functions, etc. Therefore, proteins perform a lot of diverse 
functions and understanding of mechanisms of their
folding/unfolding is essential to know how a living body works.

\begin{figure}[!hbtp]
\epsfxsize=4.2in
\vspace{5 mm}
\centerline{\epsffile{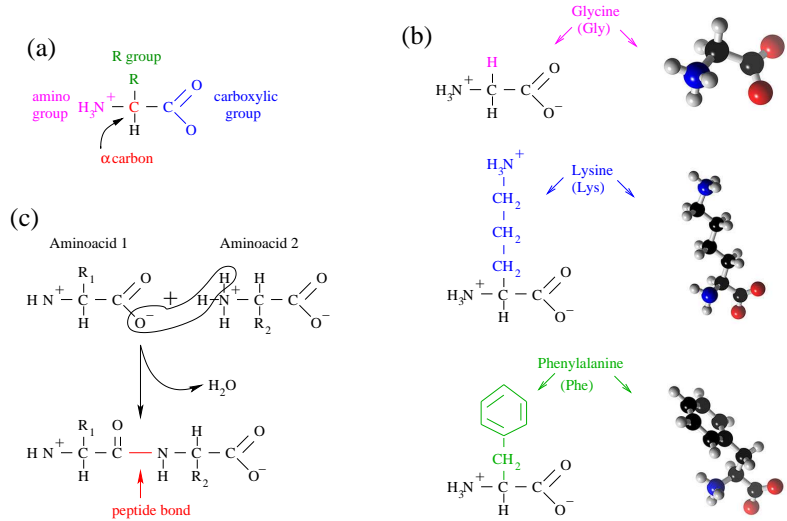}}
\linespread{0.8}
\caption{ (a) Components of an amino acid: C - central carbon
atom, H - hydrogen atom, $H_{3}N$ - amino group, $COO^{-}$ -
carboxyl group, R - radical group.
(b) Examples of three amino acids, which shows the differences in radical groups.
(c) Formation of a peptide bond. The carboxyl group of amino acid
 1 is linked to the adjacent amino group of amino acid 2.} \label{peptbond}
\vspace{5 mm}
 \end{figure}

The number of proteins is huge. The protein data bank (http://www.rcsb.org) contains
about 54500 protein entries (as of November 2008) and this number keeps growing
rapidly.
Proteins are complex compounds that are typically constructed from one 
set of 20 amino acids. Each amino
 acid has an amino end ( $-NH_2$) and an acid end 
(carboxylic group -COOH). In the middle of amino acid there is an alpha carbon
to which hydrogen and one of 20 different side groups are attached
(Fig. \ref{peptbond}a).
The structure of side group determines which of 20 amino acids we have.
 The simplest amino acid is Glysine, which  has only a single
 hydrogen atom in its side group. Other aminoacids have more 
complicated construction, that can contain carbon, hydrogen, oxygen, nitrogen 
or sulfur (e.g., Fig. \ref{peptbond}b).

Amino acids are denoted either by one letter or by three letters.
Phenylalanine, for example, is labeled as Phe or F.
There are several ways for classification of amino acids. Here we divide them
into four groups basing on their interactions
with water, their natural solvent. 
These groups are:
\begin{enumerate}
\item \label{en1} Alanine (Ala/A), Isoleucine (Ile/I), Leucine (Leu/L), Methionine (Met/M),
 Phenylalanine (Phe/F), Proline (Pro/P),
Tryptophan (Trp/W), Valine (Val/V).
\item \label{en2} Asparagine (Asn/N), Cysteine (Cys/C), Glutamine (Gln/Q), Glycine (Gly/G),
 Serine (Ser/S), Threonine (Thr/T),
Tyrosine (Tyr/Y).
\item \label{en3} Arginine (Arg/R), Histidine (His/H), Lysine (Lys/K).
\item \label{en4} Aspartic acid (Asp/D), Glutamic acid (Glu/E).
\end{enumerate}

Here one and three-letter notations of amino acids are given in brackets.
Group \ref{en1} is made of non
polar hydrophobic residues. The three other groups are made of
hydrophilic residues. From an electrostatic point of view, groups
\ref{en2}, \ref{en3} and \ref{en4} contain polar neutral, positively charged and negatively charged residues,
respectively.

\begin{figure}
\epsfxsize=5.5in
\vspace{5 mm}
\centerline{\epsffile{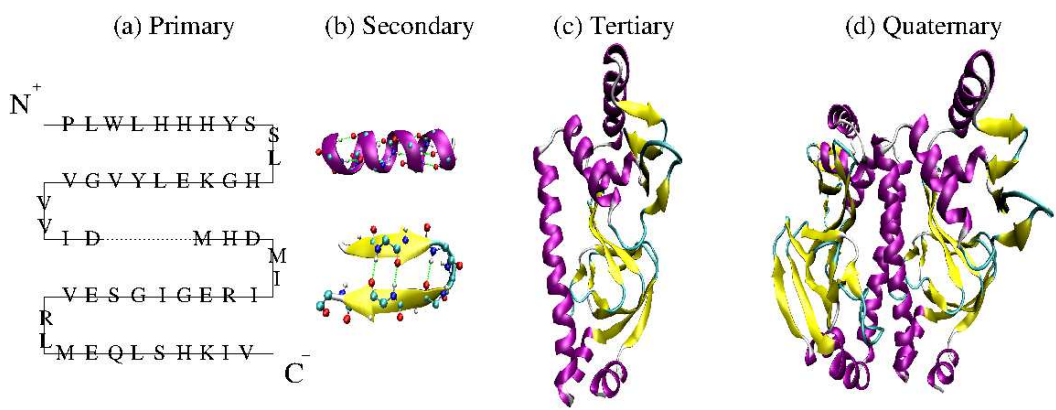}}
\linespread{0.8}
\caption{Levels of protein structures. (a) An example of primary structures
or sequences. (b) Alpha helix and beta strand are main secondary structures. The green dashed lines shows HBs.
(c) Tertiary structure of protein (PDB ID: 2CGP).
(d) Quaternary structure from two domains (PDB ID: 1CGP).
}
\label{str}
\vspace{5 mm}
\end{figure}

 In order to make proteins, amino acids link together in long chains 
by a chemical reaction in which a water molecule is released and thus 
peptide bond is created (Fig. \ref{peptbond}c).
Hence, protein is a chain of amino acids connected via peptide bonds having free amino group at
one end and carboxylic group at the other one.
 The sequence of linked amino acids is known
as a {\bf primary structure} of a protein (Fig. \ref{str}a). The structure is
 stabilized by hydrogen 
bonding between the amine and carboxylic groups.
Pauling  and Corey\cite{Pauling_PNAS51a,Pauling_PNAS51b} theoretically predicted that proteins should
exhibit some local ordering, now known as {\bf secondary structures}.
Based on energy considerations, they showed that
there are certain
regular structures which maximize the number of hydrogen bonds (HBs)
between the C-O  and the H-N groups of the backbone.
 Depending on angles 
between the carbon and the nitrogen, and the carbon and carboxylic group, the 
secondary structures may be either alpha-helices or beta-sheets
(Fig. \ref{str}b). Helices are one-dimensional structures, where the HBs are aligned
 with its axis. There are 3.6 amino acids per helix
turn, and the typical size of a helix is 5 turns. $\beta$-strands
are quasi two-dimensional structures. The H-bonds are
perpendicular to the strands. A typical $\beta$-sheet has a length of
8 amino acids, and consists of approximately 3 strands.
In addition to helices and beta strands, secondary structures may be
turns or loops.
The third type
of protein structure is called  {\bf tertiary structure} (Fig. \ref{str}c). It is an overall 
topology of the folded polypeptide chain. A variety of bonding interactions 
between the side chains of the amino acids determines this structure.
Finally, the {\bf quaternary structure} (Fig. \ref{str}d) involves 
multiple folded protein molecules as a multi-subunit complex.

\subsection{The possible states of proteins}

Although it was long believed that proteins are either denaturated or native,
it seems now well established that they may exist in at least three different phases.
The following classification is widely accepted:
\begin{enumerate}
\item {\bf Native state}\\ 
In this state, the protein  is said to be folded and has its full biological activity. Three dimensional
native structure is well-defined and unique, having a compact and globular shape. Basically, the conformational entropy of the NS is zero.

\item {\bf Denaturated states}\\ 
These states of proteins lack their biological activity.
Depending on external conditions, there
exists at least two denaturated phases:
\begin{enumerate}
\item {\it Coil state} \\
In this state, a denaturated protein has no definite shape.
Although there might be local aggregation phenomena, it is fairly well
described as the swollen phase of a homopolymer in a good solvent.
Coil state has large conformational entropy.

\item {\it Molten globule} \\
At low pH (acidic conditions),  some proteins may exist in a compact
state, named ``molten globule'' \cite{Ptitsyn_book}. This state is compact
having a 
globular shape, but it does not have a well defined structure and bears
strong resemblance to the collapsed phase of a homopolymer in a
bad solvent. It is slightly less compact
than the NS, and has finite conformational
entropy.
\end{enumerate}
\end{enumerate}

In vitro, the transition between the various phases is controlled by
temperature, pH, denaturant agent such as urea or guanidinum chloride.

\subsection{Protein folding}

{\em Protein folding is a process in which a protein reaches the NS
starting from denaturated ones.}
Understanding this complicated process has attracted attention of researchers
for over forty years.
Although a number of issues remain unsolved, several universal features have been obtained.
Here we briefly discuss the state of art of this field.

\subsubsection{Experimental techniques}

To determine protein structures one mainly uses 
the X-ray crystallography \cite{Kendrew_Nature60} and NMR \cite{Bax_JBNMR97}.
 About 85\% of structures that have been deposited in Protein Data Bank was determined
by X-ray diffraction method.
NMR generally gives a
worse resolution compared to X-ray crystallography and it is limited to relatively
small biomolecules. However, this method has the advantage that it does not require crystallization
and permits to study proteins in their natural environments.

Since proteins fold within a few microseconds to seconds, the folding process
can be studied using the fluorescence, circular dichroism (CD) {\em etc}
\cite{Nolting_book}. CD, which is directly related to this thesis,
is based on the differential absorption of left- and
right-handed circularly polarized light. 
It allows for determination of secondary structures and also for changes
in protein structure, 
providing possibility to observe folding/unfolding transition
experimentally.
As the fraction of the folded conformation $f_N$ depends on the ellipticity
$\theta$ linearly (see Eq. \ref{theta_fN_eq} below),
one can obtain it as a function of
$T$ or chemical denaturant by measuring $\theta$.

\subsubsection{ Thermodynamics of folding}

The protein folding is a spontaneous process which obeys the main 
thermodynamical principles. Considering a protein and solvent as a isolated
system, in accord with the second thermodynamic law,
their total entropy has the tendency to increase,
 $\Delta S_{prot}+\Delta S_{sol} \ge 0$. Here $\Delta S_{prot}$ and
$\Delta S_{sol}$ are the protein and solvent entropy.
If a protein absorpts from the environment  heat $Q$, then 
$\Delta S_{prot}=-\frac{Q}{T}$ ($-Q$ is the heat obtained by the solvent
from the protein). Therefore, we have $Q - T\Delta S_{prot} \le 0$.
In the isobaric process, $\Delta H=Q$ as the system does not perform work,
where $H$ is the enthalpy. Assuming $\Delta G = \Delta H - T\Delta S_{prot}$,
we obtain
\begin{equation}
\Delta G = \Delta H - T\Delta S_{prot} \le 0.
\label{folding_thermo_eq}
\end{equation}
In the isothermic process ($T$=const), $G$ in Eq. (\ref{folding_thermo_eq})
 is the
Gibbs free energy of protein ($G=H-TS_{prot}$). Thus the folding
proceeds in such a way that the Gibbs free energy decreases.
This is reasonable because the system always tries to get a state with
minimal free energy.
As the system progresses to the NS, $\Delta S_{prot}$ should decrease
disfavoring the condition (\ref{folding_thermo_eq}).
However, this condition can be fulfilled, provided $\Delta H$ decreases. 
One can show that this is the case
taking into account the hydrophobic effect which increases the solvent
entropy (or decrease of $H$) by  burying hydrophobic residues
in the core region \cite{Fersht_book}.
Thus, from the thermodynamics point of view the protein folding process
is governed by the interplay of two conflicting factors:
(a) the decrease of configurational entropy humps the folding
and (b) the increase of the solvent entropy speeds it up. 

\subsubsection{Levinthal's paradox and funnel picture of folding}

Let us consider a  protein which has only 100 amino acids.
Using a trivial model where there are just two possible orientations per
 residue, we obtain $2^{100}$
possible conformational states. If one assumes that an jump from
one  conformation to the another one requires
100 picoseconds, then  it would take about $5\times10^8$ years
to check up all the conformations before acquiring
the NS. However, in reality, typical folding times range from
microseconds to seconds. It is quite surprising that
proteins are designed in such a way, that they can  find  correct NS in very
short time. This puzzle is known as Levinthal's paradox\cite{Levinthal_JCP68}.

\begin{wrapfigure}{r}{0.42\textwidth}
\includegraphics[width=0.40\textwidth]{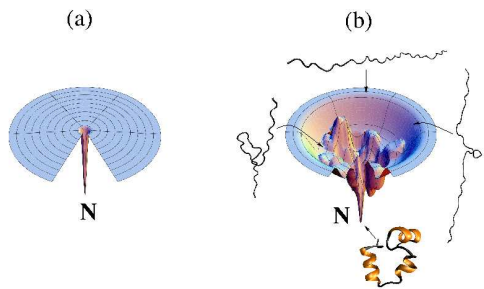}
\hfill\begin{minipage}{6.8 cm}
\linespread{0.8}
\caption{(a) Flat energy landscape, which
corresponds to blind search for the NS.
(b) Funnel-like FEL proposed by Wolynes and co-workers. \label{free_energy}}
\end{minipage}
\end{wrapfigure}

To resolve this paradox, Wolynes
and coworkers \cite{Leopold_PNAS92,Onuchic_COSB04} propose
the theory based on the folding FEL.
According to their theory, the Levinthal's scenario or the {\em old view}
corresponds to random search for the NS on a flat FEL (Fig. \ref{free_energy}a)
traveling along
a single deterministic pathway.
Such a blind search would lead to astronomically large folding times.
Instead of the old view, the {\em new view} states that the FEL
has a "funnel"-like shape (Fig. \ref{free_energy}b) and folding pathways are multiple. If some pathways get stuck
somewhere, then other pathways would lead to the NS. In the funnel one can
observe a bottleneck region which corresponds to an ensemble of conformations
of TS. By what ever pathway a protein folds, it has to overcome
the TS (rate-limiting step). The folding on a rugged FEL is slower
than on the smooth one due to kinetic traps. 

It should be noted that very likely that the funnel FEL occurs only in
systems which satisfy the principle of {\em minimal frustration} 
\cite{Bryngelson_PNAS87}. Presumably, Mother Nature selects only those sequences that
fulfill this principle.
Nowadays, the funnel theory was confirmed both theoretically 
\cite{Clementi_JMB00,Koga_JMB01}
and experimentally \cite{Jin_Structure03} and it is widely accepted
in the scientific community.

\subsubsection{Folding mechanisms}

The funnel theory gives a global picture about folding. In this section we 
are interested in pathways navigated by an ensemble of denaturated
states of a polypeptide chain en route to the native conformation. The quest to
answer this question has led to discovering diverse mechanisms by which 
proteins fold.

\paragraph{Diffusion-collision mechanism.}

This is one of the earliest mechanisms, in which folding
pathway is not unbiased
\cite{Kim_ARB90}.
Local secondary structures are assumed to
form independently, then they diffuse until a collision in which a
native tertiary structure is formed.\\

\paragraph{Hydrophobic-collapse mechanism.}

Here one assumes that a proteins collapses
quickly around hydrophobic residues forming an intermediate state (IS)
\cite{Ptitsyn_TBCS95}.
 After
that, it rearranges in such a way that secondary structures gradually
appear.\\

\paragraph{Nucleation-collapse mechanism.}

This was suggested by Wetlaufer long time ago \cite{Wetlaufer_PNAS73}
to explain the efficient folding of proteins.
In this mechanism several 
neighboring residues are suggested to
form a secondary structure as a folding nucleus.
Starting from this nucleus, occurrence of secondary structures propagates
to remaining amino acids leading to formation of the native conformation.
In the other words, after formation of a well defined nucleus, a protein 
collapses quickly to the NS. Thus, this mechanism with
a single nucleus is probably
applied to those proteins which fold fast and without intermediates.

Contrary to the old picture of single nucleus 
\cite{Wetlaufer_PNAS73,Shakhnovich_Nature96}, simulations \cite{Guo_FD97}
and experiments \cite{Viguera_NSB96}
showed that there are several nucleation regions.
The contacts between the residues in these regions occur with
varying probability in the TS. This observation allows one
to propose the multiple folding nuclei mechanism, which asserts that, in the
folding nuclei, there is a distribution of contacts , with some occurring
with higher probability than others \cite{Klimov_JMB98}.
The rationale for this mechanism is that sizes of nuclei are small
(typically of 10-15 residues
\cite{Guo_Biopolymer95,Wolynes_PNAS97}) and the linear density of hydrophobic
amino acids along a chain is roughly constant.
The nucleation-collapse mechanism with multiple nuclei is also called
as {\em nucleation-condensation} one.\\

\paragraph{Kinetic partitioning mechanism.}

It should be noted that topological frustration is an inherent property
of all polypeptide chains. It is a direct consequence of the polymeric nature
of proteins, as well as of the competing interactions (hydrophobic residues,
which prefer the formation of compact structures, and hydrophilic residues,
which are better accommodated by extended conformations. It is for this reason
that an ideal protein, which has complete compatibility between local and
nonlocal interactions, does not exists, as was first recognized by Go
\cite{Go_ARBB83}.
The basic consequences of the complex free energy surface arising from 
topological frustration leads naturally to the kinetic partitioning mechanism
\cite{Thirumalai_TCA97}. The main idea of this mechanism is as follows.
Imagine en ensemble of denaturated molecules in search of the native conformation.
It is clear that the partition factor $\Phi$ would reach the NS 
rapidly without being trapped in the low energy minima. 
The remaining fraction (1-$\Phi$) would be trapped in one or more minima
and reach the native basin by activated transitions on longer times scales
\cite{Thirumalai_JPI95}. Structures of trap-minima are intermediates
that slow the folding process. So, the fraction of molecules $\Phi$ that reaches
the native basin rapidly follows a two-state scenario without population
of any intermediates. A detailed kinetic analysis of the remaining
fraction of molecules (1-$\Phi$) showed that they reach the NS
 through a three-stage multipathway mechanism \cite{Veitshans_FD97}.
Experiments on hen-egg lysozyme \cite{Thirumalai_TCA97}
, e.g.,
seem to support 
the kinetic partitioning mechanism, which is valid for
folding via intermediates.

\subsubsection{Two- and multi-state folding}

Folding pathways and rates are defined by functions of proteins.
They could not fold too fast, as this may hump cells which continuously
synthesize chains. Presumably, by evolution sequences were selected
in such a way that there is neither universal
nor the most efficient mechanism for
all of them. Instead, the folding process may share features of
different mechanisms mentioned above. For example, the pool of molecules on
the fast track in the kinetic partitioning mechanism, reaches the
native basin through the nucleation collapse mechanism.

Regardless of the folding mechanism is universal or not, it is useful to divide
proteins into two groups. One of them includes two-state
 molecules that fold
without intermediates, i.e.
they get folded after crossing a single
TS. Proteins which fold via intermediates belong to the
another group. These multi-state proteins have more than one TS.
The list of two- and three-state folders 
is available in Ref. \cite{Jackson_FD98}.
Recently, it was suggested that the folding may proceed in down-hill manner
without any TS \cite{Munoz_Science02}. This problem is under debate.

\subsection{Mechanical unfolding of protein}

The last ten years have witnessed an intense activity SMFS experiments in detecting
inter and
intramolecular forces of biological systems to understand
their functions and structures.  Much of the research has been focused
on the elastic properties of proteins, DNA, and RNA, i.e, their response to an
 external force, following the seminal papers by Rief {\em et al.} \cite{Rief_Science97},
 and Tskhovrebova {\em et al.} \cite{Tskhovrebova_Nature97}.
The main advantage of the SMFS is its ability
to separate out the fluctuations of individual molecules from the
ensemble average behavior observed in  traditional bulk biochemical
experiments. Thus, using the SMFS one can
measure detailed distributions, describing certain molecular properties
(for example, the distribution of unfolding forces of biomolecules
\cite{Rief_Science97}) and observe possible intermediates in
chemical reactions. This technique can be used to decipher the unfolding
FEL of biomolecules \cite{Bustamante_ARBiochem_04}.
 The SMFS studies
provided unexpected insights into the strength of forces driving biological
processes as well as determined various biological interactions which leads
to the mechanical stability of biological structures.

\subsubsection{Atomic force microscopy}
There are a number of techniques for manipulating single molecules:
\begin{wrapfigure}{r}{0.48\textwidth}\centering
\includegraphics[width=0.42\textwidth]{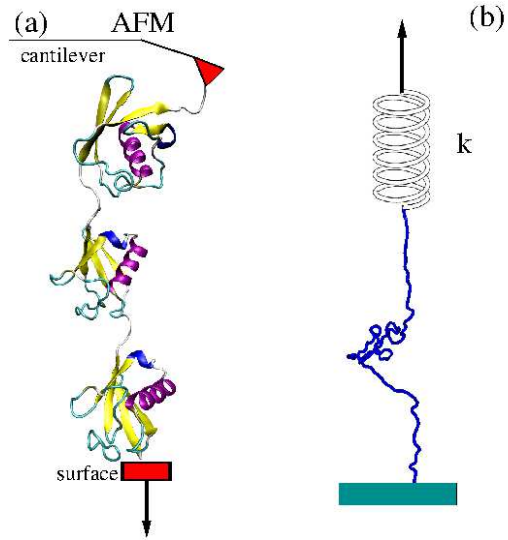}
\hfill \begin{minipage}{7.4 cm}
\linespread{0.8}
\caption{ (a) Schematic representation of AFM technique. (b) Cartoon for the spring constant of the cantilever.\label{AFM_fig}}
\end{minipage}
\end{wrapfigure}
the atomic force microscopy (AFM) \cite{Binnig_PRL86},
the laser optical tweezer (LOT), magnetic tweezers
, bio-membrane force probe, {\em etc}.
In this section we briefly discuss the AFM 
which is used to probe the mechanical response of proteins
under external force.

In AFM, one terminal
 of a biomolecules is anchored to a surface and the another one
 to a force sensor (Fig. \ref{AFM_fig}a).
The molecule is stretched by increasing the distance between the surface
and the force sensor, which is a micron-sized cantilever.
The force measured on experiments
is proportional to the
displacement of the cantilever.

If the stiffness of the cantilever $k$ is known,
then a biomolecule experiences the force $f = k\delta x$,
where $\delta x$ is a cantilever bending which is detected by the laser.
In general, the resulting force versus extension curve is used in combination
with theories for obtaining
mechanical properties of biomolecules.
The spring constant of AFM cantilever tip is typically
$k = 10 - 1000$ pN/nm. The value of $k$ and thermal
fluctuations  define spatial and force resolution in AFM experiments
because when the cantilever is kept at a fixed position the force
acting on the tip and the distance between the substrate and the tip fluctuate. The respective fluctuations are
\begin{equation}
 <\delta x^2> = k_BT/k,
\label{fluc_dx_eq}
\end{equation}
 and
\begin{equation}
<\delta f^2> = kk_BT.
\label{fluc_f_eq}
\end{equation}
Here $k_B$ is the Boltzmann constant.
 For $k=10$ pN/nm and the room temperature
$k_BT \approx 4$ pN nm we have $\sqrt{<\delta x^2>} \approx 0.6$ nm
and $\sqrt{<\delta f^2>} \approx 6$ pN. Thus, AFM can probe
unfolding of proteins which have unfolding force
of $\sim 100$ pN, but it is not precise enough for studying, nucleic acids and molecular motors as these biomolecules have lower mechanical resistance.
For these biomolecules, one can use, e.g.  
LOT which has the resolution $\sqrt{<\delta f^2} \sim 0.1$ pN.

\subsubsection{Mechanical resistance of proteins}

\begin{figure}[!htbp]
\vspace{7 mm}
\epsfxsize=6.3in
\centerline{\epsffile{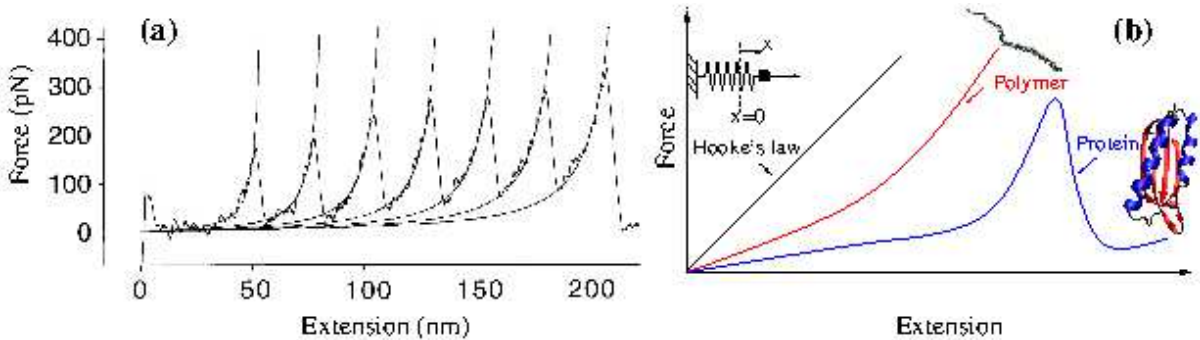}}
\linespread{0.8}
\caption{(a) Force-extension curve obtained by stretching
of a Ig8 titin fragment. Each peak corresponds to unfolding of a
single domain. Smooth curves are fits to the worm-like chain model. Taken from
Ref. \cite{Rief_Science97}. (b) Sketch of dependence of the force on
the extension for a spring, polymer and proteins.
}
\label{force_ext_Sci97_fig}
\vspace{7 mm}
\end{figure}

Proteins are pulled either by a constant force, $f$=const, or by
a force ramped linearly with time, $f=kvt$, where $k$ is the cantilever
stiffness, and $v$ is a pulling speed. In AFM experiments typical
$v \sim 100$ nm/s is used \cite{Rief_Science97}.
Remarkably, the force-extension curve obtained in the constant
rate pulling experiments has the saw-tooth shape due to domain by domain unfolding
(Fig. \ref{force_ext_Sci97_fig}a).
 Here each peak corresponds to unfolding
of one domain.
Grubmuller {\em et al} \cite{Grubmuller_Science96} and Schulten {\em et al}
\cite{Izrailev_BJ97} were first to reproduce this remarkable
result by steered MD (SMD) simulations. The saw-tooth shape is not trivial
if we recall that
 a simple spring displays
the linear dependence of $f$ on extension obeying the Hooke law, while for polymers one has
a monotonic dependence which may be fitted to the worm-like chain (WLC)
model \cite{Marko_Macromolecules95} (Fig. \ref{force_ext_Sci97_fig}b).
A non-monotonic behavior is clearly caused by complexity of the
native topology of proteins.

To characterize protein mechanical stability,
one use the unfolding force $f_u$, which
is identified as the maximum force, $f_{max}$, in
the force-extension profile, $f_u \equiv f_{max}$. If this profile has
several local maxima, then we choose the largest one. Note that
$f_u$ depends on pulling speed logarithmically,
$f_u \sim \ln v$ \cite{Evans_BJ97}.
Most of the proteins studied so far display varying
degree of mechanical resistance.
Accumulated experimental and 
theoretical results \cite{Sulkowska_BJ08,MSLi_BJ07a}
have revealed a number of factors that govern mechanical resistance.
 As a consequence of the local nature of applied force,
the type of secondary structural motif is thought to be important, with
$\beta$-sheet structures being more mechanically resistant than all
$\alpha$-helix ones \cite{MSLi_BJ07a}.
 For example, $\beta$-protein I27 and
$\alpha/\beta$-protein Ub have $f_u \approx 200$ pN which is
considerably higher than $f_u \approx 30$ pN for purely $\alpha$-spectrin
\cite{Rief_JMB99}.
Since the secondary
structure content is closely related to the contact order \cite{Plaxco_JMB98},
$f_u$ was shown to depend on the later linearly \cite{MSLi_BJ07a}.
In addition to secondary structure, tertiary structure may influence
the mechanical resistance. The 24-domain ankyrin, e.g., is mechanically more
stable than single- or six-domain one \cite{Lee_Nature06}.
The mechanical stability depends on pulling geometry
\cite{Dietz_PNAS06}.
The points of application of the force to a protein
and
the pulling direction do matter. If a force is applied parallel to HBs (unzipping), then
$\beta$-proteins are less stable than the case where the force direction
is orthogonal to them (shearing).
The mechanical stability
can be affected by ligand binding
\cite{Cao_PNAS07}
and disulphide bond formation
\cite{Wiita_Nature07}. Finally, note that the mechanical resistance of proteins
can be captured not only by all-atom SMD \cite{Sotomayor_Science07}, but
also by simple Go models \cite{Sulkowska_BJ08,MSLi_BJ07a}.
This is because the mechanical unfolding is mainly governed by the native
topology and native topology-based Go models suffice. However, in this thesis, we will show that in some cases non-native interactions can not be neglected.

\subsubsection{Construction of unfolding free energy landscape by SMFS}

Deciphering FEL is a difficult task as it is a function of many variables.
Usually, one projects it into one- or two-dimensional space. The validity of
such approximate mapping is not {\em a priory}
clear and experiments should be used
to justify this. In the mechanical unfolding case, however, the end-to-end
extension $\Delta R$ can serve as a good reaction coordinate and FEL can be mapped 
into this dimension. Thus, considering FEL as a function of $\Delta R$,
one can estimate the distance between the NS and TS,
$x_u$, using either the dependencies of unfolding rates on the
external force \cite{MSLi_BJ07}
or the dependencies of $f$ on pulling speed $v$ 
\cite{Carrion-Vasquez_PNAS99}. Unfolding barriers may be also extracted
with the help of the non-linear kinetic theory
\cite{Dudko_PRL06} (see below).

Experiments and simulations \cite{MSLi_BJ07a} showed that $x_u$ varies
between 2 - 15 \AA , depending on the secondary structure content or
the contact order. The smaller CO , the larger is $x_u$. It is remarkable
that $x_u$ and unfolding force $f_u$ are mutually related. Namely,
using a simple network model, Dietz and Rief \cite{Dietz_PRL08} argued that
$x_uf_u$ $\approx$ 50 pN nm for many proteins.

\newpage

\begin{center} \section{Modeling, Computational tools and theoretical background} \end{center}

\subsection{Modeling of Proteins}

In this section we briefly discuss main models used to study
protein dynamics.

\subsubsection{Lattice models}

In last about fifteen years, considerable insight into
thermodynamics and kinetics of protein
folding has been gained due to simple lattice models
\cite{Dill_ProteinSci95, Kolinski_book96}.
Here
amino acids are
represented by single beads which are located at vertices
of a cubic lattice. The most important difference
from homopolymer models is that  amino acid sequences and the role of
contacts should be taken into account. Due to the constraint that a contact
is formed if two residues are nearest neighbors, but not
successive in sequence, a contacts between residues $i$ and $j$ is
allowed provided $|i-j| \ge 3$. In the simple Go modeling \cite{Go_ARBB83},
the interaction between two beads which form a native contact is
assumed to be attractive, while the non-native interaction is repulsive.
This energy choice guarantees that the native conformation has the lowest
energy. 
In more realistic models specific interactions between amino acids are taken into account.
Several kinds of potentials 
\cite{Miyazawa_Macromolecules85,Kolinski_JCP93,Betancourt_ProSci99} are used to
describe these interactions.

\begin{figure}
\epsfxsize=5.5in
\vspace{5 mm}
\centerline{\epsffile{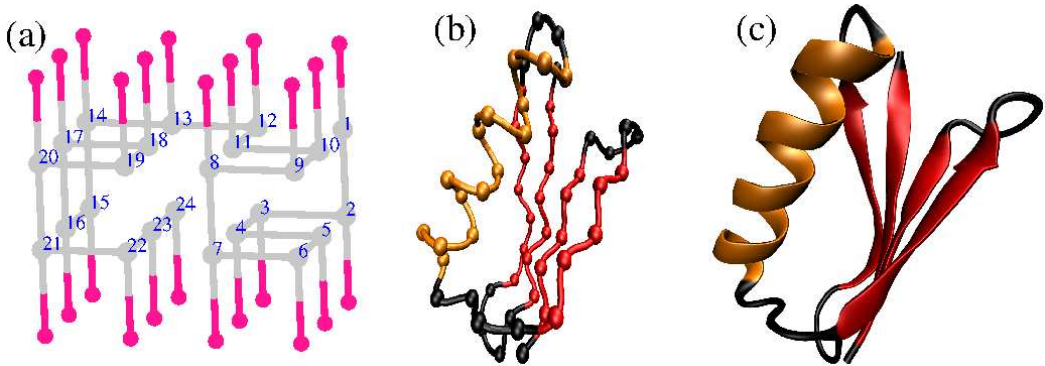}}
\linespread{0.8}
\caption{Representation of protein conformation by lattice model with side chain (a), off-lattice C$_\alpha$-Go model (b) and all-atom model (c). }
\label{models}
\vspace{5 mm}
\end{figure}

A next natural step to mimic more realistic features of
proteins  such as a dense core packing
is to include the rotamer degrees of freedom \cite{Kolinski_Proteins96}. One of
the simplest models is a cubic lattice of a backbone sequence
of $N$ beads,
to which a side bead representing a side chain is attached
\cite{Bromberg_ProteinSci94} (Fig. \ref{models}).
 The system has in total 2$N$ beads. Here we consider a Go model, where the energy of a conformation is \cite{Kouza_JPCA06}
\begin{eqnarray}
E \; = \;  \epsilon _{bb} \sum_{i=1,j>i+1}^{N} \,
 \delta _{r_{ij}^{bb},a}
+ \epsilon _{bs} \sum_{i=1,j\neq i}^{N} \, \delta _{r_{ij}^{bs},a}
+ \epsilon _{ss} \sum_{i=1,j>i}^{N} \, \delta _{r_{ij}^{ss},a} \; ,
\label{energy_eq_lattice}
\end{eqnarray}
where $\epsilon _{bb}, \epsilon _{bs}$ and $\epsilon _{ss}$ are
backbone-backbone(BB-BB), backbone-side chain
(BB-SC) and side chain-side chain (SC-SC) contact energies, respectively.
The distances $r_{ij}^{bb}, r_{ij}^{bs}$ and $r_{ij}^{ss}$  are between
BB, BS and SS beads, respectively.
The contact energies $\epsilon _{bb}, \epsilon _{bs}$
and $\epsilon _{ss}$ are taken to be -1 (in units of k$_{b}$T) for native
and 0 for non-native interactions. The neglect of interactions between residues
not present in the NS is the approximation used in the Go model.

In order to monitor protein dynamics usually one use the standard move set
which includes the tail flip,
corner flip, 
and crankshaft
for backbone beads. 
The Metropolis
criterion is applied to accept or reject moves \cite{Kolinski_book96}.
While lattice models have been widely used in the protein folding
problem \cite{Kolinski_book96}, they attract little attention in the mechanical unfolding
simulation \cite{Socci_PNAS99}.
In present thesis, we employed this model to study the cooperativity of
the folding-unfolding transition.

\subsubsection{Off-lattice coarse-grained Go modeling}

The major shortcoming of lattice models is that beads are confined to
lattice vertices and it does not allow for describing the protein shape
accurately. This can be remedied with the help of off-lattice models in which
beads representing amino acids can occupy any positions (Fig. \ref{models}b).
A number of off-lattice coarse-grained models with realistic
interactions (not Go) between amino acids 
 have been developed to study the
mechanical resistance of proteins
\cite{Klimov_PNAS00,Kirmizialtin_JCP05}.
However, it is not an easy task to construct such models for
long proteins.

In the pioneer paper \cite{Go_ARBB83} Go introduced a very simple model
in which non-native interactions are ignored. This native topology-based
model turns out to be highly useful in predicting the folding
mechanisms and deciphering the free energy landscapes of two-state proteins
\cite{Takaga_PNAS99,Clementi_JMB00,Koga_JMB01}.
On the other hand, in mechanically unfolding one stretches
a protein from its native conformation, unfolding properties are mainly
governed by its native topology.
 Therefore, the native-topology-based or Go modeling is suitable
for studying the mechanical unfolding. Various versions of Go models
\cite{Clementi_JMB00,Cieplak_Proteins02,Karanicolas_ProSci02,West_BJ06,Hyeon_Structure06,MSLi_BJ07} have been applied to this problem.
In this thesis we will focus on
the variant of Clementi {\em et al.}
\cite{Clementi_JMB00}.
Here one uses coarse-grained continuum representation for a protein
in which only the positions of C$_{\alpha}$-carbons are retained.
The interactions between residues are assumed to be Go-like and
the energy of such a model is as follows \cite{Clementi_JMB00}
\begin{eqnarray}
E \; &=& \; \sum_{bonds} K_r (r_i - r_{0i})^2 + \sum_{angles}
K_{\theta} (\theta_i - \theta_{0i})^2 \nonumber \\
&+& \sum_{dihedral} \{ K_{\phi}^{(1)} [1 - \cos (\phi_i -
\phi_{0i})] +  K_{\phi}^{(3)} [1 - \cos 3(\phi_i - \phi_{0i})] \}
\nonumber \\
& + &\sum_{i>j-3}^{NC}  \epsilon_H \left[ 5\left(
\frac{r_{0ij}}{r_{ij}} \right)^{12} - 6 \left(
\frac{r_{0ij}}{r_{ij}}\right)^{10}\right] + \sum_{i>j-3}^{NNC}
\epsilon_H \left(\frac{C}{r_{ij}}\right)^{12} + E_f
. \label{Hamiltonian}
\end{eqnarray} 
Here $\Delta \phi_i=\phi_i - \phi_{0i}$,
$r_{i,i+1}$ is the distance between beads $i$ and $i+1$, $\theta_i$
is the bond angle
 between bonds $(i-1)$ and $i$,
and $\phi_i$ is the dihedral angle around the $i$th bond and
$r_{ij}$ is the distance between the $i$th and $j$th residues.
Subscripts ``0'', ``NC'' and ``NNC'' refer to the native
conformation, native contacts and non-native contacts,
respectively. Residues $i$ and $j$ are in native contact if
$r_{0ij}$ is less than a cutoff distance $d_c$ taken to be $d_c =
6.5$ \AA, where $r_{0ij}$ is the distance between the residues in
the native conformation.

\begin{figure}[!hbtp]
\epsfxsize=6.5in
\vspace{0.2in}
\centerline{\epsffile{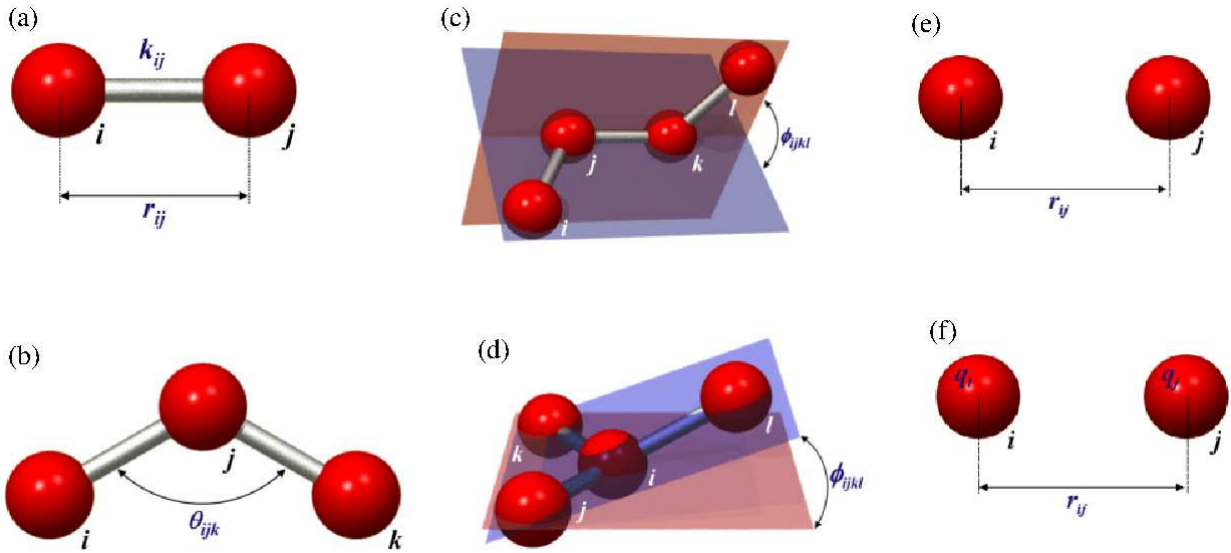}}
\caption{Schematic representation for covalent bonding (a), bond angle interactions (b), proper torsion potential (c), improper dihedral angles (d), long range Van der Waals (e) and electrostatic interactions (f).}
\label{interactions}
\end{figure}

The local interaction in Eq. (\ref{Hamiltonian}) involves three first terms.
The harmonic term 
accounts for chain
connectivity (Fig. \ref{interactions}a), while the second term represents the bond angle potential (Fig. \ref{interactions}b).
The potential for the
dihedral angle degrees of freedom (Fig. \ref{interactions}c) is given by the third term in
Eq. (\ref{Hamiltonian}). The non-local interaction energy between residues that are
separated by at least 3 beads is given by 10-12 Lennard-Jones potential (Fig. \ref{interactions}e).
A soft sphere repulsive potential
(the fifth term in Eq. \ref{Hamiltonian})
disfavors the formation of non-native contacts.
The last term accounts for the force applied to C and N termini
along the end-to-end
vector $\vec{R}$.
We choose $K_r =
100 \epsilon _H/\AA^2$, $K_{\theta} = 20 \epsilon _H/rad^2,
 K_{\phi}^{(1)} = \epsilon _H$, and
$K_{\phi}^{(3)} = 0.5 \epsilon _H$, where $\epsilon_H$ is the
characteristic hydrogen bond energy and $C = 4$ \AA.

In the constant force simulations the last term in Eq. (\ref{Hamiltonian})
is
\begin{equation}
E_f \; = \; -\vec{f}.\vec{r},
\label{E_f_constant_eq}
\end{equation} 
where $\vec{r}$ is the end-to-end vector and $\vec{f}$ is the force applied
either to both termini or to one of them.
In the constant velocity force simulation we fix the N-terminal and pull the
C-terminal by force
\begin{equation}
 f \; =  \; k(vt -x),
\label{E_f_velocity_eq}
\end{equation}
where $x$ is the displacement of the pulled atom from its original position
\cite{Lu_BJ98}, and
the pulling direction was chosen along the vector from fixed atom to
pulled atom.
In order to mimic AFM experiments (see
section {\em Experimental technique}), throughout this thesis we used the $k= K_r = 100 \epsilon _H/\AA^2 \; \approx 100$ pN/nm, which has the same order of magnitude as those for cantilever stiffness.

\subsubsection{All-atom models}

The intensive theoretical study of
protein folding has been performed with the help of all-atom simulations
\cite{Isralewitz_COSB01,Gao_PCCP06,Sotomayor_Science07}.
All-atom models include the local interaction
and the non-bonded terms. The later include
the (6-12) Lenard-Jones potential,
the electro-static interaction,
and the interaction with environment. The all-atom model with the CHARMM
force field \cite{Brooks_JCC83} and explicit TIP3 water \cite{Jorgenson_JCP83}
has been employed first by Grubmuller {\em et al.}
\cite{Grubmuller_Science96} to compute the rupture force of the
streptavidin-biovitin complex. Two years later a similar model was
successfully applied by
Schulten and coworkers \cite{Lu_BJ98} to the titin domain I27.
The NAMD software \cite{Phillips_JCC05} developed by this group is now widely
used for stretching biomolecules by the constant mechanical force and by the
force with constant loading rate (see
recent review \cite{Isralewitz_COSB01}
 for more references).
NAMD works with not only CHARMM but also with AMBER potential parameters
\cite{Weiner_JCC81},
and file formats.
Recently, it becomes possible to use the GROMACS software \cite{Gunstren_96}
for all-atom simulations of mechanical unfolding of proteins in explicit water.
 As we will present results obtained for mechanical unfolding of DDFLN4
using the Gromacs software, we discuss it in more detail.

Gromacs force field we use provides parameters for all atoms in a system, including water molecules and hydrogen atoms.
The general functional form of a force filed consists of two terms:
\begin{equation}
E_{total}=E_{bonded} + E_{nonbonded}
\end{equation}
where $E_{bonded}$ is the bonded term which is  related to atoms that are linked by covalent bonds and $E_{nonbonded}$ is
the nonbonded one which is described the long-range electrostatic and van der Waals forces.

{\bf Bonded interactions}.
The potential function for bonded interactions can be subdivided into four parts: covalent bond-stretching, angle-bending, improper dihedrals and proper dihedrals. The bond stretching between two covalently bonded atoms $i$ and $j$ is represented by a harmonic potential
\begin{equation}
V_b(r_{ij})=\frac{1}{2}k_{ij}^b(r_{ij}-b_{ij})^2
\end{equation}
where $r_{ij}$ is the actual bond length, $b_{ij}$ the reference bond lengh, $k_{ij}$ the bond stretching force constant. Both reference bond lengths and force constants are specific for each pair of bound atoms and they are usually extracted from experimental data or from quantum mechanical calculations.

The bond angle bending interactions between a triplet of atoms {\it i-j-k} are also represented by a harmonic potential on the angle $\theta_{ijk}$

\begin{equation}
V_a(\theta_{ijk})=\frac{1}{2}k_{ijk}^{\theta}(\theta_{ijk}-\theta_{ijk}^0)^2
\end{equation}
where $k_{ijk}^{\theta}$ is the angle bending force constant, $\theta_{ijk}$ and $\theta_{ijk}^0$ are the actual and reference angles, respectively. Values of $k_{ijk}^{\theta}$ and $\theta_{ijk}^0$ depend on chemical type of atoms.

Proper dihedral angles are defined according to the IUPAC/IUB convention (Fig. \ref{interactions}c), where $\phi$ is the angle between the {\it ijk} and the {\it ikl} planes, with zero corresponding to the {\it cis} configuration ({\it i} and {\it l} on the same side). To mimic rotation barriers around the bond the periodic cosine form of potential is used.
\begin{equation}
V_d(\phi_{ijkl})=k_{\phi}(1+cos(n\phi-\phi_s))
\end{equation}
where $k_{\phi}$ is dihedral angle force constant, $\phi_s$ is the dihedral angle (Fig. \ref{interactions}c), and $n$=1,2,3 is a coefficient of symmetry.

Improper potential is used to maintain planarity in a molecular structure. The torsional angle definition is shown in the figure \ref{interactions}d. The angle $\xi_{ijkl}$ still depends on the same two planes ijk and jkl, as can be seen in the figure with the atom i in the center instead on one of the ends of the dihedral chain. Since this potential used to maintain planarity, it only has one minimum and a harmonic potential can be used:

\begin{equation}
V_{id}(\xi_{ijkl})=\frac{1}{2}k_{\xi}(\xi_{ijkl}-\xi_0)^2
\end{equation}
where $k_{\xi}$ is improper dihedral angle bending force constant, $\xi_{ijkl}$ - improper dihedral angle.

{\bf Nonbonded interactions}. They act between atoms within the same protein as well as between different molecules in large protein complexes. Non bonded interactions are divided into two parts: electrostatic (Fig. \ref{interactions}f) and Van der Waals (Fig. \ref{interactions}e) interactions.
The electrostatic interactions are modeled by Coulomb potential:
\begin{equation}
V_c(r_{ij})=\frac{q_iq_j}{4\pi\epsilon_0r_{ij}}
\end{equation}
where $q_i$ and $q_j$ are atomic charges, $r_{ij}$ distance between atoms i and j, $\epsilon_0$ the electrical permittivity of space.
The interactions between two uncharged atoms are described by the Lennard-Jones potential
\begin{equation}
V_{LJ}(r_{ij})=\frac{C_{ij}^{12}}{r_{ij}^{12}} - \frac{C_{ij}^6}{r_{ij}^6}
\end{equation}
where $C_{ij}^{12}$ and $C_{ij}^6$ are specific Lennard-Jones parameters which depend on pairs of atom types.

{\bf SPC water model.} To calculate the interactions between molecules in solvent, we use a model of the individual water molecules what tell us where the charges reside. Gromacs software uses SPC or Simple Charge Model to represent water molecules. The water molecule has three centers of concentrated charge: the partial positive charges on the hydrogen atoms are balanced by an appropriately negative charge located on the oxygen atom. An oxygen atom also gets the Lennard-Jones parameters for computing intermolecular interactions between different molecules. Van der Waals interactions involving hydrogen atoms are not calculated.

\subsection{Molecular Dynamics}

One of the important tools that have been employed to study the 
biomolecules are the molecular dynamics (MD) simulations. 
It was first introduced by Alder and Wainwright in 1957 to study 
the interaction of hard spheres. In 1977, the first biomolecules, 
the bovine pancreatic trypsin inhibitor (BPTI) protein, was simulated 
using this technique.
Nowadays, the MD technique is quite common in the 
study of biomolecules such as solvated proteins, protein-DNA 
complexes as well as lipid systems addressing a variety of issues 
including the thermodynamics of ligand-binding, the folding and unfolding 
of proteins.

It is important to note that biomolecules exhibit a wide range 
of time scales over which specific processes take place. For 
example, local motion which involves atomic fluctuation, side 
chain motion, and loop motion occurs in the length scale of 0.01 
to 5 {\AA} and the time involved in such process is of the order 
of 10$^{-15}$ to 10$^{-12}$ s. The motion of a helix, protein domain 
or subunit falls under the rigid body motion whose typical length 
scales are in between 1 -- 10 {\AA} and time involved in such motion 
is in between 10$^{-9}$ to $10^{-6}$ s.
 Large-scale motion consists of helix-coil 
transitions or folding unfolding transition, which is more than 
5 {\AA} and time involved is about 10$^{-7}$ to 10$^{1}$ s. 
Typical time scales for protein folding are 10$^{-6}$ to 10$^{1}$ s
\cite{Kubelka_COSB04}. In unfolding experiments, to stretch out
a protein of length $10^2$ nm, one needs time $\sim$ 1 s using
a pulling speed
$v \sim 10^2$ nm/s \cite{Rief_Science97}.

The steered MD (SMD) that combines the stretching condition with the
standard MD was initiated by Schulten and coworkers \cite{Isralewitz_COSB01}.
They simulated the force-unfolding of a number of proteins
showing atomic details of the molecular motion under force. The focus was on the
rupture events of HBs that stabilized the structures. The structural and energetic analysis enabled them to identify the origin of free energy
barrier and intermediates during mechanical unfolding.
However, one has to notice that there is enormous difference
between the simulation condition used in SMD and real experiment.
In order to stretch out proteins within a reasonable amount of CPU time,
SMD simulations at constant pulling speed use eight to ten orders of higher
pulling speed, and one to two orders of larger spring constant than
those of AFM experiments. Therefore, effective force acting on the molecule
is about three-four orders higher. It is unlikely, that the dynamics under 
such an extreme condition can mimic real experiments, and one has to be very
careful about comparison of simulation results with experimental ones.
In literature the word "steered" also means MD at extreme conditions,
where constant force and constant pulling speed are chosen very high.

Excellent reviews on MD and its 
use in biochemistry and biophysics are numerous (see, e.g.,
\cite{Adcock_ChemRev06} and references therein).
Below, we only focus on the Brownian dynamics as
well as on the second-order Verlet   
method for the Langevin dynamics simulation , which have been intensively used
to obtain main results presented in this thesis.

\subsubsection{Langevin dynamics simulation}

The Langevin equation is a stochastic differential equation which introduces
friction and noise
terms into Newton's second law to approximate effects of
temperature and environment:

\begin{equation}
m\frac{d^2 \vec{r}}{d t^2} = \vec{F}_c - \gamma \frac{d\vec{r}}{dt} + \vec{\Gamma}  \equiv \vec{F}.
\label{Langevin_eq}
\end{equation}
where $\Gamma$ is a random force, $m$  the mass of a bead, $\gamma$ the friction
coefficient, and
$\vec{F}_c = -d\vec{E}/d\vec{r}$. Here the configuration energy $E$ for the Go model,
for example, is given by Eq. (\ref{Hamiltonian}).
 The random force $\Gamma$ is taken to be a Gaussian random
variable with white noise spectrum and is related to the friction coefficient by the
fluctuation-dissipation relation:

\begin{equation}
<\Gamma (t) \Gamma (t')> = 2\gamma k_BT\delta(t-t')
\label{Noise}
\end{equation}
where $k_B$ is a Boltzmann's constant, $\gamma$ friction coefficient, $T$ temperature and $\delta(t-t')$ the Dirac delta function.
The friction term only influences kinetic but not thermodynamic properties.

In the low friction regime, where $\gamma < 25\frac{m}{\tau_L}$
(the time unit $\tau_L = (ma^2/\epsilon_H)^{1/2} \approx 3$ ps),
Eq. (\ref{Langevin_eq})
can be solved using the second-order
Velocity Verlet algorithm \cite{Swope_JCP82}:
\begin{eqnarray}
x(t+\Delta t) \; = \; x(t) + \dot{x}(t)\Delta t + \frac {1}{2m}F(t)(\Delta t)^2,
\end{eqnarray}
\begin{eqnarray}
\dot{x}(t+\Delta t) \; = \; \left(1-\frac{\gamma\Delta t}{2m}\right)\left[1 - \frac{\gamma\Delta t}{2m} +
\left(\frac{\gamma\Delta t}{2m}\right)^2\right]\dot{x}(t) + \qquad \nonumber\\
\left(1 - \frac{\gamma\Delta t}{2m} + \left(\frac{\gamma\Delta t}{2m}\right)^2\right)(F_c(t) + \Gamma (t) +
F_c(t+\Delta t) + \Gamma (t+\Delta t))\frac{\Delta t}{2m} + o(\Delta t^2),
\end{eqnarray}
with the time step $\Delta t = 0.005 \tau_L$.

\subsubsection{Brownian dynamics}

In the overdamped limit ($\gamma > 25\frac{m}{\tau_L}$) the inertia term can be neglected, and we obtain a much simpler equation:
\begin{equation}
\frac{dr}{dt} = \frac{1}{\gamma}(F_c + \Gamma).
\label{overdamped_eq}
\end{equation}
This equation may be solved using the simple Euler method which gives
the position of a biomolecule at the time $t + \Delta t$ as follows:

\begin{equation}
x(\Delta t+t) = x(t) + \frac{\Delta t}{\gamma} (F_c + \Gamma).
\label{Euler}
\end{equation}
Due to the large value of $\gamma$ we can choose the time step
$\Delta t = 0.1 \tau_L$ which is 20-fold larger than the low viscosity
case. Since the water has $\gamma \approx 50\frac{m}{\tau_L}$
\cite{Veitshans_FD97},
the Euler method is
valid for studying protein dynamics.

\subsection{Theoretical background}

In this section we present basic formulas used throughout my thesis.

\subsubsection{Cooperativity of folding-unfolding transition}

The sharpness of the fold-unfolded transition might be characterized
quantitatively
via the cooperativity index $\Omega _c$ which is defined as follows
\cite{MSLi_Polymer04}
\begin{equation}
 \Omega_c=\frac{T_F^2}{\Delta T}
\biggl(\frac{df_N}{dT}\biggr)_{T=T_F},
\label{cooper_index_eq}
\end{equation}
where $\Delta T$ is the transition width and $f_N$ the probability of being
in the NS. The larger $\Omega _c$, the sharper is the transition.
$f_N$ is defined as the thermodynamic average of the
fraction of native contacts $\chi$, $f_N = <\chi >$. For off-lattice models, $\chi$ is \cite{Camacho_PNAS93}:
\begin{equation}
\chi \; = \frac{1}{Q_{total}} \sum_{i<j+1}^N \,\;
\theta (1.2r_{0ij} - r_{ij}) \Delta_{ij}
\label{chi_eq_Go}
\end{equation}
where
$\Delta_{ij}$ is equal to 1 if residues $i$ and $j$ form a native
contact and 0 otherwise and $\theta (x)$ is the Heaviside
function. The argument of this function guarantees that
a native contact between $i$ and $j$ is classified as formed
when $r_{ij}$ is shorter than 1.2$r_{0ij}$ \cite{Clementi_JMB00}.
In the lattice model with side chain (LMSC) case, we have
\begin{eqnarray}
\chi \; = \;  \frac{1}{2N^{2} - 3N + 1} \left[ \sum_{i<j} \,
 \delta (r_{ij}^{ss} - r_{ij}^{ss,N})
+ \sum_{i<j+1} \, \delta (r_{ij}^{bb} - r_{ij}^{bb,N})
+ \sum_{i \neq j} \, \delta (r_{ij}^{bs} - r_{ij}^{bs,N}) \; \right].
\label{chi_eq_lattice}
\end{eqnarray}
Here $bb$, $bs$ and $ss$ refer to backbone-backbone, backbone-side chain
and side chain-side chain pairs, respectively.

\subsubsection{Kinetic theory for mechanical unfolding of biomolecules}

One of the notable aspects in force experiments on single biomolecules
is that the end-to-end extension $\Delta R$ is directly measurable or controlled by
instrumentation. $\Delta R$ becomes a natural reaction coordinate for describing
mechanical processes. 

The theoretical framework for understanding the effect of external constant
force on
rupture rates was first discussed in the context
of cell-cell adhesion by Bell in 1978 \cite{Bell_Sci78}.
Evans and Rirchie have extended his theory to the case when the loading force
increases linearly with time
\cite{Evans_BJ97}. The phenomenological Bell theory is based on the assumption that the
TS does not move under stretching. Since this assumption is not true,
Dudko {\em et al} \cite{Dudko_PRL06}
have developed the microscopic theory which is free from this shortcoming.
In this section we discuss the phenomenological as well as microscopic
kinetics theory.

\paragraph{Bell theory for constant force case.}

\begin{figure}
\vspace{5 mm}
\epsfxsize=4.2in
\centerline{\epsffile{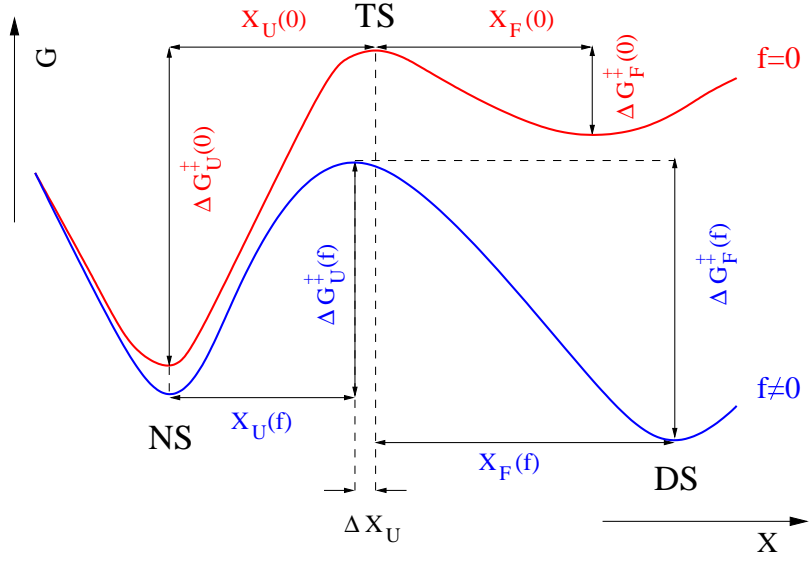}}
\linespread{0.8}
\caption{Conceptual plot for the FEL without (blue) and
under (red) the external force. $x_u$ is the shift of $x_u$
in the presence of force.}
\label{conceptual_TS_fig}
\vspace{5 mm}
\end{figure}

Suppose the external constant force, $f$, is applied to the termini of a biomolecule.
The deformation of the FEL under force is schematically
shown in Fig. \ref{conceptual_TS_fig}.
Assuming that the force does not change
the distance between the NS and TS ($x_u(f)=x_u(0)$),
Bell \cite{Bell_Sci78} stated that
the activation energy is changed to
$\Delta G^{\ddag}_u(f) = \Delta G^{\ddag}_u(0)
- fx_u$, where $x_u  =x_u(0)$. In general, the proportionality factor $x_u$
has the dimension of length
and may be viewed as the width of the potential.
Using the Arrhenius law, Bell obtained the following
formula for the unfolding/unbinding rate constant \cite{Kramers_Physica40}:
\begin{equation}
 k_{u} (f) \; = \; k_{u}(0) \exp (fx_{u}/k_BT),
\label{Bell_Ku_eq}
\end{equation}
where $k_{u}(0)$ is the rate constant is the unfolding rate constant
in the absence of a force. If a reaction takes place in condensed phase,
then according to the Kramers theory the prefactor $k_{u}(0)$ is equal
\begin{equation}
k_{u}(0) \; = \; \frac{\omega_0\omega_{ts}}{2\pi \gamma}\exp(-\Delta G^{\ddag}_u(0)/k_BT).
\label{Omega}
\end{equation}
Here $\gamma$ is a solvent viscosity, $\omega_0$ the angular frequency (curvature) at the reactant bottom, and $\omega_{ts}$ the curvature at barrier top of
the effective reaction coordinate \cite{Kramers_Physica40}.
For biological reactions, which belong to the Kramers category,
$\frac{\omega_0\omega_{ts}}{2\pi \gamma} \sim 1 \mu$s \cite{Levinthal_JCP68}.
It is important to note that the unfolding rate grows exponentially
with the force. This is the hallmark of the Bell model.
Even Eq. (\ref{Bell_Ku_eq}) is very simple, as we will
see below, it fits most of experimental data very well.
Using Eq. (\ref{Bell_Ku_eq}),
one can extract the distance $x_u$, or the location of the TS.

\paragraph{Bell theory for force ramp case.}

Assuming that the force increases linearly
with a rate $v$,
Evans and Rirchie
in their seminal paper \cite{Evans_BJ97}, have shown that the
distribution of unfolding force $P(f)$ obeys the following equation:
\begin{equation}
P(f) \; = \; \frac{k_{u}(f)}{v} \exp \{ \frac{k_BT}{x_uv}
\left[ k_u(0)-k_u(f) \right]\},
\end{equation}
where $k_u(f)$ is given by Eq. (\ref{Bell_Ku_eq}).
Then, the most probable unbinding force or the maximum of force distribution
 $f_{max}$, obtained from the condition  $dP(f)/df|_{f=f_{max}}=0$,
is
\begin{equation}
f_{max}=\frac{k_BT}{x_u}ln\frac{kvx_u}{k_u(0)k_BT}.
\label{f_logV_eq}
\end{equation}
The logarithmic dependence of $f_{max}$ on the pulling speed $v$ was
confirmed by enumerous experiments and simulations
 \cite{Klimov_PNAS99,Kouza_JCP08}.

\paragraph{Beyond Bell approximation.}

The major shortcoming of the the Bell approximation is
the assumption that $x_u$ does not depend on
the external force. Upon force application the location of TS
should move
closer to
the NS reducing $x_u$ (Fig \ref{conceptual_TS_fig}),
as postulated by Hammond in the context
of chemical reactions of small organic molecules
\cite{Hammond_JACS53}.
The Hammond behavior has been
observed in protein folding experiments
 \cite{Matouschek_Biochemistry95}
and simulations \cite{Lacks_BJ05}.

Recently, assuming that $x_u$ depends on the external force
and using the Kramers theory,
several groups \cite{Schlierf_BJ06,Dudko_PRL06}
have tried to go beyond the Bell approximation. We follow
Dudko {\em et al.} who proposed
the following force dependence for the unfolding time \cite{Dudko_PRL06}:
\begin{eqnarray}
\tau _u \; = \; \tau _u^0
\left(1 - \frac{\nu x_u}{\Delta G^{\ddagger}}\right)^{1-1/\nu}
\exp\lbrace -\frac{\Delta G^{\ddagger}}{k_BT}[1-(1-\nu x_uf/\Delta G^{\ddagger})^{1/\nu}]\rbrace.
\label{Dudko_eq}
\end{eqnarray}
Here, $\Delta G^{\ddagger}$ is the unfolding barrier, and $\nu = 1/2$ and 2/3
for the cusp \cite{Hummer_BJ03} and the
linear-cubic free energy surface \cite{Dudko_PNAS03}, respectively.
Note that
$\nu =1$ corresponds to the phenomenological
Bell theory (Eq. \ref{Bell_Ku_eq}), where $\tau_u=1/k_u$.
An important consequence following from
Eq. (\ref{Dudko_eq}), is that one can apply it to estimate not only
$x_u$, but also $\Delta G^{\ddagger}$, if $\nu \ne 1 $. Expressions
 for the distribution
of unfolding forces and the $f_{max}$ for arbitrary $\nu$ may be found in
\cite{Dudko_PRL06}.

\subsubsection{Kinetic theory for refolding of biomolecules.}

In force-clamp experiments \cite{Fernandez_Sci04}, a protein refolds under
the quenched force. Then,
in the Bell approximation,
the external force increases the folding barrier
(see Fig. \ref{conceptual_TS_fig}) by amount
$\Delta G^{\ddag}_f = fx_f$, where $x_f = x_f(0)$ is
a distance between the DS 
and the TS. Therefore, the refolding time
reads as
\begin{equation}
\tau_{f} (f) \; = \; \tau_{f}(0) \exp (fx_{f}/k_BT).
\label{Bell_Kf_eq}
\end{equation}
Using this equation and the force dependence of $\tau_{f} (f)$,
one can extract $x_f$ 
\cite{Fernandez_Sci04,MSLi_PNAS06,MSLi_BJ07}.
One can extend the nonlinear theory of Dudko {\em et al} \cite{Dudko_PRL06}
to the refolding case by replacing $x_u \rightarrow -x_f$
in, e.g.,
Eq. (\ref{Dudko_eq}). Then the folding barriers can be estimated using the
microscopy theory with $\nu \neq 1$.

\subsection{Progressive variable}

In order to probe folding/refolding pathways, for $i$-th trajectory
we introduce the progressive variable
\begin{equation}
 \delta _i =
t/\tau^i_{f}.
\label{progress_fold_eq}
\end{equation}
Here $\tau^i_{f}$ is the folding time, which is is defined
as a time to get the NS starting from the denaturated one
for the $i$-th trajectory.
Then one
can average the fraction of native contacts over many trajectories
in a unique time window
$0 \le \delta _i \le 1$ and monitor the folding sequencing with
the  help of the progressive variable $\delta$.

In the case of unfolding, the progressive variable is defined in a similar
way:
\begin{equation}
 \delta _i =
t/\tau^i_{u}.
\label{progress_unfold_eq}
\end{equation}
Here $\tau^i_{u}$ is the folding time, which is is defined
as a time to get a rod conformation starting from the NS for the $i$-th
trajectory.
The unfolding time, $\tau_u$,
is the average of first passage times to reach a rod conformation.
Different trajectories start from the same native
conformation but, with different random number seeds.
In order
to get the reasonable estimate for $\tau_u$,
for each case we have generated 30 - 50 trajectories.
Unfolding pathways were probed by monitoring
the fraction of native contacts of secondary structures as a
function of  progressive variable $\delta$.

\clearpage

\begin{center}\section{ Effect of finite size on cooperativity and rates of protein folding}\end{center}

\subsection{Introduction}

Single domain globular proteins are mesoscopic systems that self-assemble,
under folding conditions, to a compact state with definite topology.  Given
that the folded states of proteins are only on the order of tens of
Angstroms
(the radius of gyration $R_g \approx 3 N^{\frac {1}{3}}$ \AA $~$
\cite{Dima_JPCB04} where $N$ is
the number of amino acids) it is surprising that they undergo highly
cooperative transitions from an ensemble of unfolded states to the NS 
\cite{Privalov_APC79}.
Similarly, there is a wide spread in the folding
times as well \cite{Galzitskaya_Proteins03}.
The rates of folding vary by nearly nine orders of
magnitude.  Sometime ago it was shown theoretically that the folding time
,$\tau_F$, should depend on $N$
\cite{Finkelstein_FoldDes97}
 but only recently has experimental
data confirmed this prediction
\cite{Galzitskaya_Proteins03,MSLi_Polymer04,Ivankov_PNAS04}.
It has been shown that $\tau_F$
can be approximately evaluated using $\tau_F \approx \tau_F^0
\exp(N^{\beta})$ where $1/2 \le \beta < 2/3$ with
the prefactor $\tau_F^0$ being on the order
of a $\mu s$.

Much less attention has been paid to finite size effects on the
cooperativity of transition from unfolded states to the
native basin of attraction (NBA). Because 
$N$ is finite, large conformational fluctuations are possible but
require careful examination \cite{Klimov_JCC02,MSLi_Polymer04}.
  For large enough $N$ it is likely that the
folding or melting temperature itself may not be unique
\cite{Holtzer_BJ97}.
 Although substantial variations in $T_m$ are unlikely it has already been shown that the there
is a range of temperatures over which individual residues in a protein achieve their
NS ordering \cite{Holtzer_BJ97}.
  On the other hand, the global cooperativity, as measured by the
dimensionless parameter $\Omega_c$ (Eq. \ref{cooper_index_eq}) has been
shown to scale as \cite{MSLi_PRL04}
\begin{equation}
\Omega_c \approx N^{\zeta}
\label{omega}
\end{equation}
 Having used the scaling arguments and analogy with a magnetic system, it was shown that \cite{MSLi_PRL04}
\begin{equation}
\zeta= 1+ \gamma \approx 2.2
\label{2dot2}
\end{equation}
where the magnetic susceptibility exponent $\gamma \approx 1.2$. This result in not trivial because the protein melting transition is first order \cite{Privalov_APC79}, for which $\zeta=2$ \cite{Naganathan_JACS05}. Let us mention the main steps leading to Eq. (\ref{2dot2}).
The folding temperature can be identified with the peak in $d<\chi>/dT$
or in the fluctuations in $\chi$, namely, $\Delta \chi = <\chi^2> - <\chi>^2$. Using an analogy to magnetic systems, we identify $T(\partial <\chi>/\partial h)=\Delta \chi$ where $h$ is an "ordering field" that is conjugate to $\chi$. Since $\Delta \chi$ is dimensionless, we expect $h \approx T$ for proteins, and hence, $T(\partial <\chi>/\partial T)$ is like susceptibility. Hence, the scaling of $\Omega_c$ on $N$ should follow the way $(T_F/\Delta T) \Delta \chi$ changes with $N$ \cite{Kohn_PNAS04}.

For efficient folding in proteins $T_F \approx T_\Theta$ \cite{Klimov_PRL97}, where $T_\Theta$ is the temperature at which the coil-globule transition occurs. It has been argued that $T_F$ for proteins may well be a tricritical point, because the transition at $T_F$ is first-order while the collapse transition is (typically) second-order.
Then, as temperature approaches from above, we expect that the characteristics of polypeptide chain at $T_\Theta$ should manifest themselves in the folding cooperativity.
At or above $T_F$, the susceptibility $\Delta \chi$ should scales with $\Delta T$ as $\Delta \chi \sim \Delta T^{- \gamma}$ as predicted by the scaling theory for second order transitions \cite{Fisher_PRB82}. Therefore, $\Omega_c \sim \Delta T^{-(1+\gamma)}$. taking into account that $\Delta T \sim N^{-1}$ \cite{Grosberg_Book94} we come to Eqs. \ref{omega} and \ref{2dot2}.

In this chapter we use LMSC,
off-lattice Go models for 23 proteins and
experimental results for a number of proteins to further confirm the
theoretical predictions (Eq. \ref{omega} and \ref{2dot2}).  Our results show that $\zeta \approx 2.22$ which
is \textit{distinct from the expected
result} ($\zeta = 2.0$) \textit{for a strong first order transition} \cite{Fisher_PRB82}. 
Our another goal is to study the dependence of the folding time on the number of amino acids.
The larger data set of proteins for which folding rates are available
shows that the folding time scales as
\begin{equation}
\tau_F = \tau_0 \exp(cN^{\beta})
\end{equation}
with $c \approx 1.1$, $\beta = 1/2$ and $\tau_0 \approx 0.2 \mu s$.

The results presented in this chapter are taken from Ref. \cite{Kouza_JPCA06}.

\subsection{Models and methods}

The LMSC (Eq. \ref {energy_eq_lattice}) and coarse-grained off-lattice model (Eq. \ref{Hamiltonian}) \cite{Clementi_JMB00} were used.
For the LMSC we performed Monte Carlo simulations using
the previously well-tested move set MS3 \cite{Li_JPCB02}. This move set ensures that
ergodicity is obtained efficiently even for $N=50$, it uses
single, double and triple bead moves \cite{Betancourt_JCP98}.
Following standard practice the thermodynamic properties are computed
using the multiple histogram method \cite{Ferrenberg_PRL89}. The kinetic simulations are carried out
by a quench from high temperature to a temperature at which the NBA
is preferentially populated. The folding times are calculated
from the distribution of first passage times.

For off-lattice models, we assume the dynamics of the polypeptide chain obeys the Langevin
equation. The equations of motion were integrated using the velocity form 
of the Verlet algorithm with the time step $\Delta t = 0.005 \tau_L$,
where $\tau_L = (ma^2/\epsilon_H)^{1/2} \approx 3$ ps.
In order to calculate the thermodynamic quantities we collected
histograms for the energy and native contacts
at five or six different temperatures
(at each temperature 20 - 50 trajectories were generated depending on proteins).
As with the LMSC we used the multiple histogram method \cite{Ferrenberg_PRL89} 
to obtain the thermodynamic parameters at all temperatures.
For off-lattice and LMSC models the probability of being in the NS is computed
using Eq. (\ref{chi_eq_Go}) and Eq. (\ref{chi_eq_lattice}), respectively.

The extent of cooperativity of the transition to the NBA from the ensemble of
unfolded states is measured using the dimensionless parameter $\Omega_c$ (Eq. \ref{cooper_index_eq}). Two points about $\Omega_c$ are noteworthy. (1) For
proteins that melt by a two-state transition it is trivial  to show that
$\Delta H_{vH} = 4k_B\Delta T\Omega _c$, where $\Delta H_{vH}$ is the
van't Hoff enthalpy at $T_F$. For an infinitely sharp two-state transition
there is a latent heat release at $T_F$, at which $C_p$ can be approximated
by a delta-function. In this case $\Omega_c \rightarrow \infty$ which implies
that $\Delta H_{vH}$ and the calorimetric enthalpy $\Delta H_{cal}$
(obtained by integrating the temperature dependence of the specific heat
$C_p$ ) would coincide. It is logical to infer
that as $\Omega_c$ increases the ratio $\kappa = \Delta H_{vH}/\Delta H_{cal}$
should approach unity. 
(2)  Even for moderate sized proteins that undergo a two-state transition
$\kappa \approx 1$ \cite{Privalov_APC79}.
It is known that the extent of cooperativity depends on external
conditions as has been demonstrated for thermal denaturation of CI2 at
several values of pH \cite{Jackson_Biochemistry91}. The values of $\kappa$ for all
pH values are $\approx 1$.
However, the variation in cooperativity of CI2 as pH varies are
reflected in the changes in  $\Omega _c$ \cite{Klimov_FD98}.
Therefore, we believe that $\Omega _c$, that varies in the
range $0 < \Omega _c < \infty$, is a better descriptor of the extent of
cooperativity than $\kappa$. The latter merely tests the applicability
of the two-state approximation.
\vspace{5 mm}
\begin{figure}[!htbp]
\includegraphics[width=0.47\textwidth]{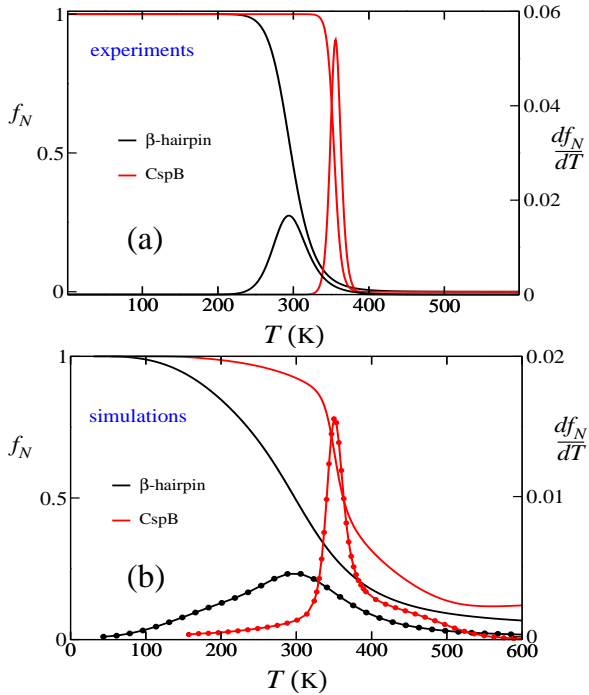}
\hfill
\linespread{0.8}
\parbox[b]{0.47\textwidth}{\caption{The temperature dependence of $f_N$ and $df_N/dT$ for $\beta$-hairpin
($N=16$) and CpsB ($N=67$). The scale for $df_N/dT$ is given on the right.
(a): the experimental curves were
obtained using
$\Delta H = 11.6$ kcal/mol,
$T_m=297$ K and $\Delta H = 54.4$ kcal/mol and $T_m= 354.5$ K for
$\beta$-hairpin and CpsB, respectively.
(b): the simulation results were calculated from $f_N = <\chi (T)>$.
The Go model gives only a qualitatively reliable estimates of $f_N(T)$.\\\\\\\\\\\\}\label{hairpin_CspB_fig}}
\\
\end{figure}

\subsection{Results}

\subsubsection{Dependence of cooperativity $\Omega_c$ on number of aminoacids $N$}

\begin{wrapfigure}{r}{0.47\textwidth}
\includegraphics[width=0.40\textwidth]{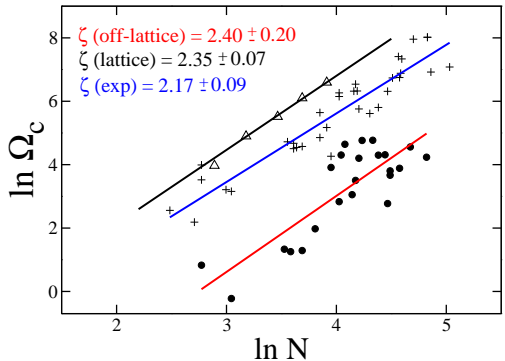}
\hfill\begin{minipage}{7.7 cm}
\linespread{0.8}
\caption{Plot of ln$\Omega_c$ as a function of ln$N$.
The red line is a fit to the simulation data for the 23
off-lattice Go proteins from  which we estimate
$\zeta =2.40 \pm 0.20$. The black line is a fit to the lattice models
with side chains ($N = 18, 24, 32, 40$ and 50) with
$\zeta = 2.35 \pm 0.07$.
The blue line is a fit to the experimental values of
$\Omega_c$ for 34 proteins (Table \ref{scalling_table2})
with $\zeta = 2.17 \pm 0.09$. The larger deviation in $\zeta$ for the minimal
models is due to lack of all the interactions that stabilize the NS. \label{Scal_Omega_fig}}
\end{minipage}
\\
\end{wrapfigure}

For the 23 Go proteins listed in Table \ref{scalling_table1}, we calculated $\Omega_c$ from
the temperature dependence of $f_N$.
In Fig. \ref{hairpin_CspB_fig}
we compare the temperature dependence of $f_N(T)$ and $df_N(T)/dT$ for
$\beta$-hairpin ($N=16$) and {\it Bacillus subtilis} (CpsB, $N=67$).
It is clear that the transition width and the amplitudes of $df_N/dT$
obtained using Go models, compare only qualitatively well with experiments.
As pointed out by Kaya and Chan \cite{Kaya_SFG00,Kaya_JMB03,Chan_ME04,Kaya_PRL00},
the simple Go-like models consistently
underestimate the extent of cooperativity. Nevertheless, both the models and
experiments show that $\Omega_c$ increases dramatically as $N$ increases
(Fig. \ref{hairpin_CspB_fig}).
The variation of $\Omega_c$ with $N$ for the 23 proteins obtained from
the simulations of Go models is given in Fig. \ref{Scal_Omega_fig}.
From the ln$\Omega_c$-ln$N$ plot we obtain $\zeta = 2.40 \pm 0.20$
and $\zeta = 2.35 \pm 0.07$ for off-lattice models and LMSC, respectively. These
values of $\zeta$ deviate  from the theoretical prediction
$\zeta \approx 2.22$.
We suspect that this is due to large fluctuations in the NS of
polypeptide chains that are represented using minimal models.
Nevertheless, the results for the minimal models rule out
the value of $\zeta = 2$ that is predicted for systems that undergo first
order transition. The near coincidence of $\zeta$ for both models show that
the details of interactions are not relevant.
For the thirty four proteins (Table \ref{scalling_table2}) for which we could find thermal
denaturation data, we calculated $\Omega_c$ using the $\Delta H$,
and $T_F$ (referred to as the melting temperature $T_m$ in the experimental
literature).

 From the plot of ln$\Omega_c$ versus ln$N$ we find that
$\zeta = 2.17 \pm 0.09$. The experimental value of $\zeta$, which also
deviates from $\zeta = 2$, is in much better agreement with the theoretical
prediction. The analysis of experimental data requires care because the
compiled results were obtained from a number of different laboratories around
the world. Each laboratory uses different methods to analyze the raw
experimental data which invariably lead to varying methods to
estimate errors in
$\Delta H$ and $T_m$. To estimate the error bar for $\zeta$ it is important
to consider the errors in the computation of $\Omega_c$.
Using the reported experimental errors in $T_m$ and
$\Delta H$ we calculated the variance $\delta^2\Omega_c$ using the standard
expression for the error propagation \cite{MSLi_PRL04}.

\subsubsection{Dependence of folding free energy barrier on number of amino acids $N$}

The simultaneous presence of stabilizing (between hydrophobic residues) and
destabilizing interactions involving polar and charged residues in 
polypeptide chain renders the NS only marginally stable
\cite{Poland_book}.
The hydrophobic residues enable the formation of compact structures while
polar and charged residues, for whom water is a good solvent, are better
accommodated by extended conformations. Thus, in the folded state the
average energy gain per residue (compared to expanded states) is 
$-\epsilon _H (\approx (1 - 2)$ kcal/mol) whereas due to chain connectivity
and surface area burial the loss in free energy of exposed residues is
$\epsilon _P \approx \epsilon _H$. Because there is a large number of
solvent-mediated interactions that stabilize the NS,
 even when $N$ is small, it follows from the
central limit theorem that the barrier height $\beta \Delta G^{\ddagger}$,
whose lower bound is the stabilizing free energy should scale as
$\Delta G^{\ddagger} \sim k_BT\sqrt{N}$ \cite{Thirumalai_JPI95}.
\begin{wrapfigure}{r}{0.47\textwidth}
\includegraphics[width=0.40\textwidth]{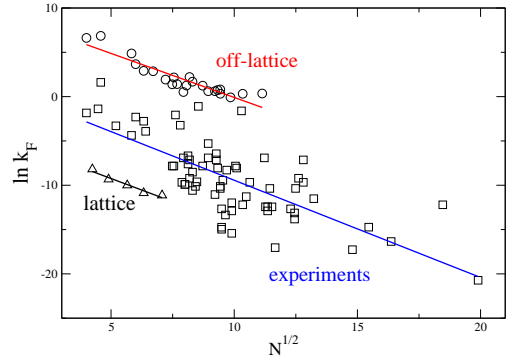}
\hfill\begin{minipage}{7.5 cm}
\linespread{0.8}
\caption{Folding rate of 69 real proteins (squares) is plotted as
a function of $N^{1/2}$ (the straight line represent the fit
$y = 1.54 -1.10x$ with the correlation coefficient $R=0.74$).
The open circles represent the data obtained for 23
off-lattice Go proteins (see Table \ref{scalling_table1})
(the linear fit $y = 9.84 - x$ and $R=0.92$).
The triangles denote the data obtained for lattice models with side
chains ($N = 18, 24, 32, 40$ and 50, the linear fit
$y = -4.01 - 1.1x$ and $R=0.98$). For real proteins
and off-lattice Go proteins $k_F$ is measured in $\mu s^{-1}$, whereas
for the lattice models it is measured in MCS$^{-1}$
where MCS is Monte Carlo steps. \label{real_pro_fig}}
\end{minipage}
\end{wrapfigure}
A different physical picture has been used to argue that 
$\Delta G^{\ddagger} \sim k_BTN^{2/3}$ \cite{Finkelstein_FoldDes97,Wolynes_PNAS97}.
Both the scenarios show that the barrier to folding rates 
scales sublinearly with $N$. 

The dependence of ln$k_F$ ($k_F = \tau_F^{-1}$) on $N$ using experimental
data for 69 proteins \cite{Naganathan_JACS05}
and the simulation results for the 23 proteins is
consistent with the predicted behavior that
$\Delta G^{\ddagger} = ck_BT\sqrt{N}$ with $c \approx 1$ (Fig. \ref{real_pro_fig}). The correlation
between the experimental results and the theoretical fit is 0.74
which is similar to the previous analysis using a set of 57
proteins \cite{MSLi_Polymer04}. It should be noted that the data can also be fit using
$\Delta G^{\ddagger} \sim k_BTN^{2/3}$. 
The prefactor $\tau_F^0$ using the $N^{2/3}$ fit is over
an order of magnitude larger than
for the $N^{1/2}$ behavior. In the absence of accurate
measurements for a larger data set of proteins it is difficult to
distinguish between the two power laws for $\Delta G^{\ddagger}$.
\begin{table}[h]
\scriptsize{
\begin{tabular}{|c|c|c|c|c|}
\hline
Protein&$N$&PDB code$^{\rm a}$&$\Omega_c^{\rm b}$&$\delta \Omega_c^{\rm c}$ \\
\hline
$\beta$-hairpin&$16$&1PGB&2.29&0.02\\
\hline
$\alpha$-helix&$21$&no code&0.803&0.002\\
\hline
WW domain&$34$&1PIN&3.79&0.02\\
\hline
Villin headpiece&$36$&1VII&3.51&0.01\\
\hline
YAP65&$40$&1K5R&3.63&0.05\\
\hline
E3BD&$45$& &7.21&0.05 \\
\hline
hbSBD&$52$&1ZWV&51.4&0.2\\
\hline
Protein G&$56$&1PGB&16.98&0.89\\
\hline
SH3 domain ($\alpha$-spectrum)&$57$&1SHG&74.03&1.35\\
\hline
SH3 domain (fyn)&$59$&1SHF&103.95&5.06\\
\hline
IgG-binding domain of streptococcal protein L&$63$&1HZ6&21.18&0.39\\
\hline
Chymotrypsin Inhibitor 2 (CI-2)&$65$&2CI2&33.23&1.66\\
\hline
CspB (Bacillus subtilis)&$67$&1CSP&66.87&2.18\\
\hline
CspA&$69$&1MJC&117.23&13.33\\
\hline
Ubiquitin&$76$&1UBQ&117.8&11.1\\
\hline
Activation domain procarboxypeptidase A2&$80$&1AYE&73.7&3.1\\
\hline
His-containing phosphocarrier protein&$85$&1POH&74.52&4.2\\
\hline
hbLBD&$87$&1K8M&15.8&0.2\\
\hline
Tenascin (short form)&$89$&1TEN&39.11&1.14\\
\hline
Twitchin Ig repeat 27&$89$&1TIT&44.85&0.66\\
\hline
S6&$97$&1RIS&48.69&1.31\\
\hline
FKBP12&$107$&1FKB&95.52&3.85\\
\hline
Ribonuclease A&$124$&1A5P&69.05&2.84\\
\hline
\end{tabular}
\linespread{0.8}
\caption{List of 23 proteins used in the simulations.
(a) The NS for use in the Go model is obtained from the structures deposited in the Protein Data Bank. (b) $\Omega _c$ is calculated
using equation (\ref{cooper_index_eq}).
(c) 2 $\delta \Omega _c = |\Omega _c - \Omega _{c_1}| + |\Omega _c - \Omega _{c_2}|$, where $\Omega _{c_1}$ and $\Omega _{c_2}$ are
values of the cooperativity measure obtained by retaining only one-half the conformations used to compute $\Omega _c$.}
\label{scalling_table1}
\vspace{3 mm}
}
\end{table}
Previous studies \cite{Klimov_JCP98}
 have shown that there is a correlation between folding
rates and $Z$-score which can be defined as
\begin{equation}
Z_G \; = \; \frac{G_N - <G_U>}{\sigma} ,
\label{Zscore_eq}
\end{equation}
where $G_N$ is the free energy of the NS, $<G_U>$ is the average free
energy of the unfolded states and $\sigma$ is the dispersion in the free
energy of the unfolded states. From the fluctuation formula it follows that
$\sigma = \sqrt{k_BT^2C_p}$ so that
\begin{equation}
Z_G \; = \; \frac{\Delta G}{\sqrt{k_BT^2C_p}} .
\label{Zscore1_eq}
\end{equation}
Since $\Delta G$ and $C_p$ are extensive it follows that $Z_G \sim N^{1/2}$.
This observation establishes an intrinsic connection between the
thermodynamics and kinetics of protein folding that involves formation and
rearrangement of non-covalent interactions. In an interesting
 recent note \cite{Naganathan_JACS05}
it has been argued that the finding
 $\Delta G^{\ddagger} \sim k_BT\sqrt{N}$ can be
interpreted in terms of $n_{\sigma}$ in which $\Delta G$ in
 Eq. (\ref{Zscore1_eq}) is replaced by $\Delta H$. In either case, there
appears to be a thermodynamic rationale for the sublinear scaling
of the folding free energy barrier.

\begin{table}[!htbp]
{\scriptsize
\begin{tabular}{|c|c|c|c|c|c|c|c|c|}
\hline
Protein&$N$&$\Omega_c^a$&$\delta \Omega_c^b$& &Protein&$N$&$\Omega_c^a$&$\delta \Omega_c^b$ \\
\cline{1-4} \cline{6-9}
BH8 $\beta$-hairpin \cite{Dyer}&12&12.9&0.5& &SS07d \cite{Knapp_JMB96}&64&555.2&56.2\\
\cline{1-4} \cline{6-9}
HP1 $\beta$-hairpin \cite{Xu_JACS03}&15&8.9&0.1& &CI2 \cite{Jackson_Biochemistry91}&65&691.2&17.0  \\
\cline{1-4} \cline{6-9}
MrH3a $\beta$-hairpin \cite{Dyer}&16&54.1&6.2& &CspTm \cite{Wassenberg_JMB99} &66&558.2&56.3 \\
\cline{1-4} \cline{6-9}
$\beta$-hairpin \cite{Honda_JMB00}&16&33.8&7.4& &Btk SH3 \cite{Knapp_Proteins98} &67&316.4&25.9\\
\cline{1-4} \cline{6-9}
Trp-cage protein \cite{Qui_JACS02}&20&24.8&5.1& &binary pattern protein \cite{Roy_Biochemistry00} &74&273.9&30.5 \\
\cline{1-4} \cline{6-9}
$\alpha$-helix \cite{Williams_Biochemistry96}&21&23.5&7.9& &ADA2h \cite{Villegas_Biochemistry95}  &80&332.0&35.2\\
\cline{1-4} \cline{6-9}
villin headpeace \cite{Kubelka_JMB03}&35&112.2&9.6& &hbLBD \cite{Naik_ProtSc04} &87&903.1&11.1 \\
\cline{1-4} \cline{6-9}
FBP28 WW domain$^c$ \cite{Ferguson_PNAS01}&37&107.1&8.9& &tenascin Fn3 domain \cite{Clarke_JMB97} &91&842.4&56.6\\
\cline{1-4} \cline{6-9}
FBP28 W30A WW domain$^c$ \cite{Ferguson_PNAS01} &37&90.4&8.8& &Sa RNase \cite{Pace_JMB98} &96&1651.1&166.6 \\
\cline{1-4} \cline{6-9}
WW prototype$^c$ \cite{Ferguson_PNAS01}&38&93.8&8.4& &Sa3 RNase \cite{Pace_JMB98}&97&852.7&86.0\\
\cline{1-4} \cline{6-9}
YAP WW$^c$ \cite{Ferguson_PNAS01}&40&96.9&18.5& &HPr \cite{VanNuland_Biochemistry98}&98&975.6&61.9 \\
\cline{1-4} \cline{6-9}
BBL \cite{Ferguson_p}&47&128.2&18.0& &Sa2 RNase \cite{Pace_JMB98} &99&1535.0&156.9 \\
\cline{1-4} \cline{6-9}
PSBD domain \cite{Ferguson_p}&47&282.8&24.0& &barnase \cite{Martinez_Biochemistry_94}&110&2860.1&286.0 \\
\cline{1-4} \cline{6-9}
PSBD domain \cite{Ferguson_p}&50&176.2&13.0& &RNase A \cite{Arnold_Biochemistry97}&125&3038.5&42.6 \\
\cline{1-4} \cline{6-9}
hbSBD \cite{Kouza_BJ05} &52&71.8&6.3& &RNase B \cite{Arnold_Biochemistry97}&125&3038.4&87.5\\
\cline{1-4} \cline{6-9}
B1 domain of protein G \cite{Alexander_Biochemistry92} &56&525.7&12.5& &lysozyme \cite{Hirai_JPC99} &129&1014.1&187.3 \\
\cline{1-4} \cline{6-9}
B2 domain of protein G \cite{Alexander_Biochemistry92} &56&468.4&20.0& &interleukin-1$\beta$ \cite{Makhatadze_Biochemistry94} &153&1189.6&128.6\\\hline
\end{tabular}
}
\linespread{0.8}
\caption{List of 34 proteins for which $\Omega _c$ is calculated
using experimental data. The calculated $\Omega _c$ values from experiments
are significantly larger than those obtained using the Go models (see Table \ref{scalling_table1}).
a) $\Omega _c$ is computed at $T = T_F = T_m$ using the experimental values
of $\Delta H$ and $T_m$.
b) The error in $\delta \Omega_c$ is computed using the proceedure given in \cite{MSLi_PRL04,Gutin_PRL96}.
c) Data are averaged over two salt conditions at pH 7.0.
\vspace{5 mm}}
\label{scalling_table2}
\end{table}

\subsection{Conclusions}

We have reexamined the dependence of the extent of cooperativity as a function
of $N$ using lattice models with side chains, off-lattice models and experimental data on thermal denaturation.
The finding that $\Omega _c \sim N^{\zeta}$ at $T \approx T_F$ with $\zeta > 2$
provides additional support for the earlier theoretical predictions \cite{MSLi_PRL04}. More
importantly, the present work also shows that the theoretical value for
$\zeta$ is independent of the precise model used which implies that $\zeta$
is universal. It is surprising to find such general characteristics for
proteins for which specificity is often an important property. We should note
that accurate value of $\zeta$ and $\Omega _c$ can only be obtained using
more refined models that perhaps include desolvation 
penalty \cite{Kaya_JMB03,Cheung_PNAS02}

In accord with a number of theoretical predictions
\cite{Thirumalai_JPI95,Finkelstein_FD97,Wolynes_PNAS97,Gutin_PRL96,Li_JPCB02,Koga_JMB01}
we found that the folding free energy barrier scales only sublinearly
with $N$. The relatively small barrier is in accord with the marginal stability
of proteins. Since the barriers to global unfolding is relatively small it
follows that there must be large conformational fluctuations even when the
protein is in the NBA. Indeed, recent experiments show that such dynamical
fluctuations that are localized in various regions of a monomeric protein might
play an important functional role. These observations suggest that small barriers in proteins and RNA \cite{Hyeon_Biochemistry05}
might be an evolved characteristics of all natural sequences.

\newpage
\begin{center} \section{Folding of the protein hbSBD} \end{center}

\subsection{Introduction}

\begin{figure}
\epsfxsize=5.2in
\vspace{5 mm}
\centerline{\epsffile{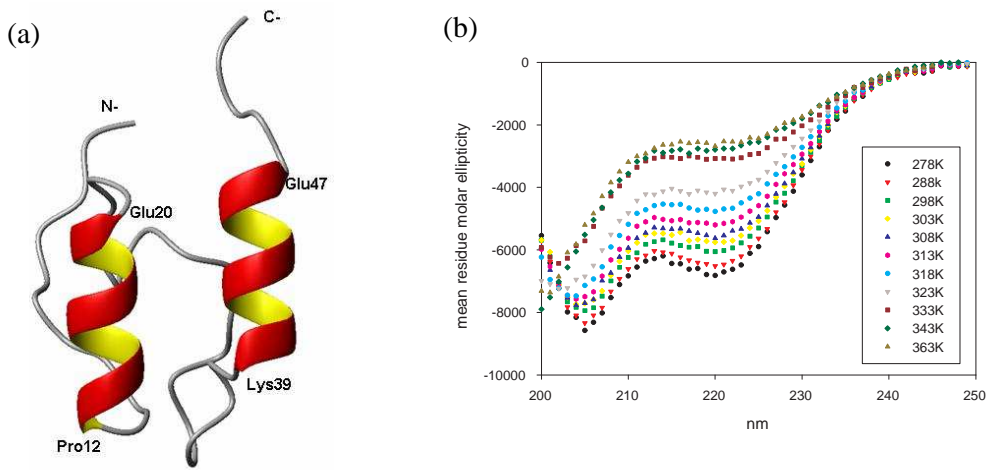}}
\linespread{0.8}
\caption{ (a) Ribbon representation of the structure
of hbSBD domain. The helix region H$_{1}$ and H$_2$ include residues
Pro12 - Glu20 and Lys39 - Glu47, respectively. (b) Dependence of the mean residue molar ellipticity on the wave length for 18 values of temperatures between 278 and 363 K.}
\label{hbSBD_12}
\vspace{5 mm}
\end{figure}

Understanding the dynamics and mechanism of protein
folding remains one of the  most challenging problems in molecular
biology \cite{Dagget_Trends03}.  Single domain $\alpha$ proteins attract
much attention of researchers because most of them fold faster
than $\beta$ and $\alpha\beta$ proteins \cite{Jackson_FD98,Kubelka_COSB04}
due to relatively simple energy landscapes and one can, therefore,
use them to probe main aspects of the funnel theory
\cite{Bryngelson_Proteins1995}. Recently, the study of this class of proteins
becomes even more attractive because the one-state or downhill
folding may occur in some small $\alpha$-proteins
\cite{Munoz_Science02}.
The mammalian mitochondrial branched-chain $\alpha$-ketoacid
dehydrogenase (BCKD) complex catalyzes the oxidative decarboxylation
of branched-chain $\alpha$-ketoacids derived from leucine, isoleucine
and valine to give rise to branched-chain acyl-CoAs.
In patients with inherited maple syrup urine disease,
the activity of the BCKD complex is deficient, which
is manifested by often fatal acidosis and mental retardation \cite{ccf1}.
 The BCKD multi-enzyme complex (4,000 KDa in size) is organized about
a cubic 24-mer core of dihydrolipoyl transacylase (E2), with multiple
 copies of hetero-tetrameric decarboxylase (E1), a homodimeric
dihydrogenase (E3), a kinase (BCK) and a phosphatase attached
through ionic interactions.  The E2 chain of the human BCKD
complex, similar to other related multi-functional enzymes
\cite{ccf2}, consists of three domains:  The amino-terminal
lipoyl-bearing domain (hbLBD, 1-84), the interim E1/E3
subunit-binding domain (hbSBD, 104-152) and the carboxy-terminal
inner-core domain. The structures of these domains serve as bases
for modeling interactions of the E2 component with other
components of $\alpha$-ketoacid dehydrogenase complexes. The
structure of hbSBD (Fig. \ref{hbSBD_12}a) has been determined by NMR
spectroscopy, and the main function of the hbSBD is to attach both
E1 and E3 to the E2 core \cite{ccf3}. The two-helix structure of
this domain is reminiscent of the small protein BBL
\cite{Ferguson_JMB04} which may be a good candidate for observation of
downhill folding \cite{Munoz_Science02,Munoz_JACS04}. So the study of hbSBD is
interesting not only because of the important biological role of
the BCKD complex in human metabolism but also for illuminating
folding mechanisms.

From the biological point of view, hbSBD could be less stable than
hbLBD and  one of our goals is, therefore, to check this by the
CD experiments. In this paper we study the
thermal folding-unfolding transition in the hbSBD by the CD
technique in the absence of urea and pH=7.5. Our thermodynamic
data do not show evidence for the downhill folding and they are
well fitted by the two-state model. We obtained folding
temperature $ T_F = 317.8 \pm 1.95$ K and the transition enthalpy
$\Delta H_G = 19.67 \pm 2.67$ kcal/mol. Comparison of such
thermodynamic parameters of hbSBD with those for hbLBD shows that
hbSBD is indeed less stable as required by its biological
function. However, the value of $\Delta H_G$ for hbSBD is still
higher than those of two-state $\alpha$-proteins reported in
\cite{Eaton_COSB04}, which indicates that the folding process in the
hbSBD domain is highly cooperative.

From the theoretical point of view it is very interesting to
establish if the two-state foldability of hbSBD can be captured by
some model. The all-atom model would be the best choice for a
detailed description of the system but the study of hbSBD requires
very expensive CPU simulations. Therefore we employed the
off-lattice coarse-grained Go-like model \cite{Go_ARBB83,Clementi_JMB00}
which is simple and allows for a thorough characterization of
folding properties. In this model amino acids are represented by
point particles or beads located at positions of $C_{\alpha}$
atoms. The Go model is defined through the experimentally
determined native structure \cite{ccf3}, and it captures essential
aspects of the important role played by the native structure
\cite{Clementi_JMB00,Takada_PNAS1999}.

It should be noted that
the Go model by itself can not be employed to ascertain the two-state behavior
of proteins.
However, one can use it in conjunction with experiments providing
the two-state folding because this model does not {\it always} provide
the two-state behavior as have been clearly shown in the seminal work
of Clementi {\it et al.} \cite{Clementi_JMB00}. In fact,
the Go model correctly captures not only the two-state folding of
proteins CI2 and SH3 (more two-state Go folders may be found
in Ref. \cite{Koga_JMB01})
but also intermediates of the three-state folder
barnase, RNAse H and CheY \cite{Clementi_JMB00}.
The reason for this is that
the simple Go model ignores the energetic frustration but it still takes
the topological frustration into account.
Therefore, it can capture intermediates 
that occur due to topological constraints but not those
emerging from the frustration of the contact interactions. 
 With the help of Langevin dynamics
simulations and the histogram method \cite{Ferrenberg_PRL89} we have
shown that, in agreement with our CD data, hbSBD is a two-state
folder with a well-defined TS in the free
energy landscape. The two helix regions were found to be
highly structured in the TS. The two-state behavior of hbSBD  
is also supported by our kinetics study
showing that the folding kinetics follows the single exponential scenario.
The two-state folding obtained in our simulations suggests that for hbSBD
the topological frustration is more important than the energetic factor.

 The dimensionless quantity, $\Omega _c$
\cite{Klimov_FD98}, which characterizes the structural
cooperativity of the thermal denaturation transition was computed
and the reasonable agreement between the CD experiments and Go
simulations was obtained. Incorporation of side chains may give a
better agreement \cite{Klimov_FD98,Li_Physica05} but this
problem is beyond the scope of the thesis.

The material presented in this chapter is based on our work \cite{Kouza_BJ05}.

\subsection{Materials and Methods}

\subsubsection{Sample Preparation}

 hbSBD protein was purified from the BL21(DE3) strain of
\textit{E. coli }containing a plasmid that carried the gene of
hbLBD(1-84), a TEV cleavage site in the linker region, and hbSBD
(104-152), generously provided to us by Dr. D.T. Chuang of the
Southwestern Medical Center, University of Texas.  There is an
extra glycine in front of Glu104 which is left over after TEV
cleavage, and extra leucine,
 glutamic acid at the C-terminus before six histidine residues.
 The protein was purified by Ni-NTA affinity chromatography, and the
 purity of the protein was found to be better than 95\%,
based on the Coomassie blue-stained gel. The complete sequence of
$N=52$ residues for hbSBD is\\
(G)EIKGRKTLATPAVRRLAMENNIKLSEVVGSGKDGRILKEDILNYLEKQT(L)(E).

\subsubsection{Circular Dichroism}

 CD measurements were carried out in Aviv CD spectrometer model 202
with temperature and stir control units at different temperature
taken from
 260nm to 195nm. All experiments were carried at 1 nm bandwidth
 in 1.0 cm quartz square cuvette thermostated to  $\pm 0.1^o$C.
Protein concentration ($\sim$ 50 uM) was determined by UV absorbance at 280nm
using $\epsilon _{280nm}$=1280 M$^{-1}$cm$^{-1}$ with 50mM phosphate buffer at pH7.5.
 Temperature control was achieved using a circulating water bath system,
 and the equilibrium time was three minutes for each temperature point.
The data
was collected at each 2K increment in temperature. The study was
 performed at heating rate
of 10$^{o}$C/min and equilibration time of 3 minutes.
 The volume changes as a result of thermal expansion as well
 as evaporation of water were neglected.

\subsubsection{Fitting Procedure}

Suppose the thermal denaturation is a two-state
transition, we can write the ellipticity as
\begin{equation}
\theta \; \, = \; \, \theta _D + (\theta _N - \theta _D)f_N \, ,
\label{theta_fN_eq}
\end{equation}
where $\theta _D$ and $\theta _N$ are values for the denaturated and folded states. The fraction of the folded conformation
$f_N$ is expressed as  \cite{Privalov_APC79}
\begin{eqnarray}
f_N \; \, &=& \; \, \frac{1}{1 + \exp (-\Delta G_T/T)} \, ,\nonumber \\
\Delta G_T \; \, &=& \; \, \Delta H_T - T\Delta S_T \; = \;
\Delta H_G\left(1 - \frac{T}{T_G}\right) \nonumber \\
&+&\Delta C_p \left[(T-T_G) - T \ln \frac{T}{T_G}\right] \, .
\label{fN_twostate_eq}
\end{eqnarray}
Here $\Delta H_G$ and $\Delta C_p$ are jumps of the enthalpy
and heat capacity at the mid-point temperature $T_G$ (also known
as melting or
folding temperature) of thermal transition, respectively.
Some other thermodynamic characterization of stability
such as the temperature of maximum stability ($T_S$), the
temperature with zero enthalpy ($T_H$), and the conformational
stability ($\Delta G_S$) at $T_S$ can be computed
from results of regression analysis \cite{Becktel_Biopolymers87}
\begin{eqnarray}
\ln \frac{T_G}{T_S} \; &=& \; \frac{\Delta H_G}{T_G\Delta C_p}, \\
T_H \; &=& \; T_G - \frac{\Delta H_G}{\Delta C_p}, \\
\Delta G_S \; &=& \; \Delta C_p (T_S - T_H).
\label{parameters_eq}
\end{eqnarray}

Using Eqs. (\ref{theta_fN_eq}) - (\ref{parameters_eq})
we can obtain all thermodynamic parameters from CD data.

It should be noted that the fitting of Eq. (\ref{fN_twostate_eq})
with $\Delta C_p > 0$ allows for an additional cold denaturation
\cite{Privalov_CRBMB90} at temperatures much lower than the room
temperature . The temperature of such a transition, $T_G'$, may be
obtained by the same fitting procedure with an additional
constraint of $\Delta H_G <0$. Since the cold denaturation
transition is not seen in Go models, to compare the simulation
results to the experimental ones we also use the approximation in
which $\Delta C_p=0$.

\subsubsection{Simulation}

We use coarse-grained continuum representation for hbSBD
protein, in which only the positions of 52 C$_{\alpha}$-carbons
are retained. We adopt the off-lattice version of the Go model
\cite{Go_ARBB83} where the interaction between residues forming native
contacts is assumed to be attractive and the non-native
interactions - repulsive (Eq. \ref{Hamiltonian}).

The nativeness of any configuration is measured by the number of
native contacts $Q$. We define that the $i$th and $j$th residues
are in the native contact if $r_{0ij}$ is smaller than a cutoff
distance $d_c$ taken to be $d_c = 7.5$ \AA,
where $r_{0ij}$ is the distance between the $i$th and $j$th residues in
the native conformation. Using this definition and the native
conformation of Ref. \onlinecite{ccf3}, we found that the total
number of native contacts $Q_{total} = 62$. To study the
probability of being in the NS we use the following
overlap function as in Eq. (\ref{chi_eq_Go}).

The overlap function $\chi$, which is one if the
conformation of the polypeptide chain coincides with the native
structure and zero for unfolded conformations, can serve as an
order parameter for the folding-unfolding transition. The
probability of being in the NS, $f_N$, which can be
measured by the CD and other experimental techniques, is defined
as $f_N = <\chi>$, where $<...>$ stands for a thermal average.

The dynamics of the system is obtained by integrating the following Langevin
equation \cite{Allen_book} (Eq. \ref{Langevin_eq}). The Verlet algorithm \cite{Swope_JCP82} was employed.
It should be noted that the folding thermodynamics
does not depend on the environment viscosity (or on $\zeta$)
but the folding kinetics depends
on it \cite{Klimov_PRL97}.  We chose the dimensionless parameter
$\tilde{\zeta} = (\frac{a^2}{m\epsilon_H})^{1/2}\zeta = 8$, where
$m$ is the mass of a bead and $a$ is the bond length between successive beads.
One can show that this value of $\tilde{\zeta}$ belongs to the
interval of the viscosity where the folding
kinetics is fast. We have tried other values of $\tilde{\zeta}$
but the results
remain unchanged qualitatively.
All thermodynamic quantities are obtained by the histogram method
\cite{Ferrenberg_PRL89}.

\subsection{Results}

\subsubsection{CD  Experiments}

\noindent
 The structure of hbSBD is shown in Figure \ref{hbSBD_12}a.
 Its conformational stability is investigated
in present study by analyzing the unfolding transition induced by
temperature as monitored
 by CD, similar to that described previously \cite{Naik_FEBS02,Naik_ProtSc04}.
The reversibility of thermal denaturation was ascertained by monitoring
the return of the CD signal upon cooling from 95$^{o}$C to 22 $^{o}$C;
immediately after the conclusion of the thermal transition.
The transition was found to be more than 80\% reversible.
 Loss in reversibility to greater extent was observed on prolonged
 exposure of the sample to higher temperatures.
 This loss of reversibility is presumably due to irreversible
 aggregation or decomposition. Figure \ref{hbSBD_12}b
 shows the wavelength dependence  of mean residue molar
ellipticity  of hbSBD at various temperatures between 278K and 363K.
 In a separate study, the thermal unfolding transition as monitored
 by ellipticity at 228 nm was found to be independent of hbSBD
 concentration in the range of 2 uM to 10 uM. It was also found to be
 unaffected by change in heating rate between 2$^{o}$C/min to 20$^{o}$C/min.
These observations suggest absence of stable intermediates in heat
 induced denaturation of hbSBD. A valley at around 220 nm,
 characteristics of the helical secondary structure is evident for
 hbSBD.

\begin{figure}
\epsfxsize=5.2in
\vspace{5 mm}
\centerline{\epsffile{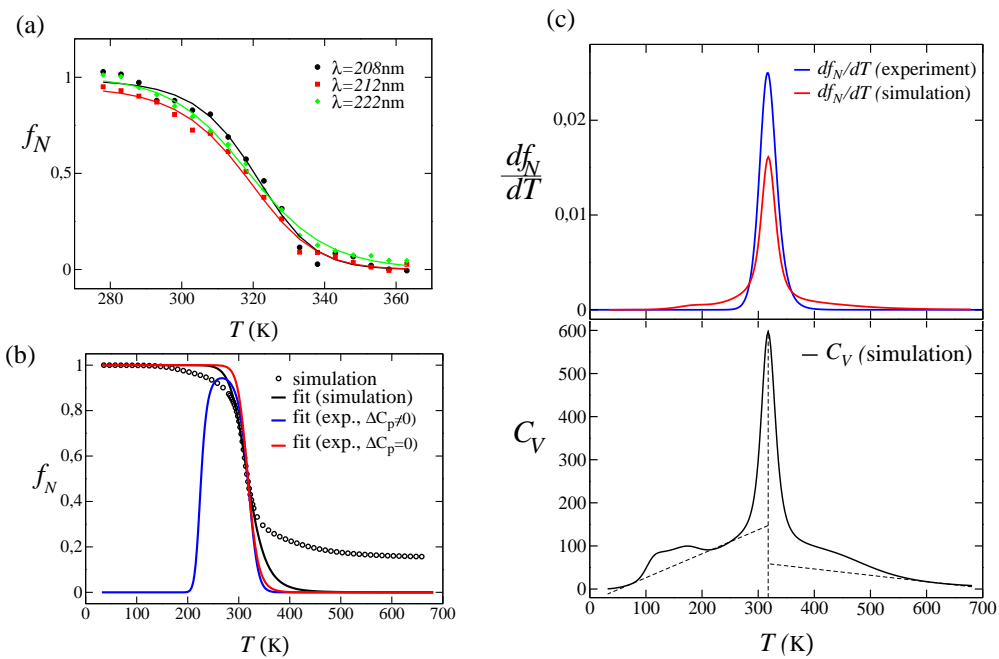}}
\linespread{0.8}
\caption{ (a) Temperature dependence of the fraction
of folded conformations $f_N$, obtained from the ellipticity
$\theta$ by Eq. (\ref{fN_twostate_eq}), for wave lengths $\lambda$
= 208 (blue circles), 212 (red squares)  and 222 nm (green
diamonds). The solid lines corresponds to the two state fit given
by Eq. (\ref{fN_twostate_eq}) with $\Delta C_p \ne 0$. We obtained
$T_G=T_F= 317.8 \pm 1.9$ K, $\Delta H_G = 19.67 \pm 2.67$ kcal/mol
and $\Delta C_p = 0.387\pm0.054$. (b) The dependence of $f_N$ for various sets
of parameters. The blue and red curves correspond to the
thermodynamic parameters presented on the first and the second
rows of Table \ref{hbSBD_table1}, respectively. Open circles refer to simulation
results for the Go model. The solid black curve is the two-state
fit ($\Delta C_p=0$) which gives $\Delta H_G= 11.46$ kcal/mol and
$T_F=317.9$. (c) The upper part refers to the temperature
dependence of $df_N/dT$ obtained by the simulations (red) and the
CD experiments (blue). The experimental curve is plotted using
two-state parameters with $\Delta C_p = 0$ (see, the second row on
Table \ref{hbSBD_table1}). The temperature dependence of the heat capacity $C_V(T)$
is presented in the lower part. The dotted lines illustrate the
base line substraction. The results are averaged over 20 samples. }
\label{hbSBD_345}
\vspace{5 mm}
\end{figure}

Figure \ref{hbSBD_345}a shows the temperature dependence of the
population of the native conformation, $f_N$, for wave lengths
$\lambda = 208, 212$ and 222 nm. We first try to fit these data to
Eq. (\ref{fN_twostate_eq}) with $\Delta C_p \ne 0$. The fitting
procedure gives slightly different values for the folding (or
melting) temperature and the enthalpy jump for three sets of
parameters. Averaging over three values, we obtain $ T_G = 317.8
\pm 1.95$ K and $\Delta H_G = 19.67 \pm 2.67$ kcal/mol. Other
thermodynamic quantities are shown on the first row of Table \ref{hbSBD_table1}.
The similar fit but with $\Delta C_p=0$ gives the
thermodynamic parameters shown on the second row of this table.
Since the experimental data are nicely fitted to the two-state
model we expect that the downhill scenario does not applied to the
hbSBD domain.

For the experimentally studied temperature interval two types of
the two-state fit (\ref{fN_twostate_eq}) with $\Delta C_p=0$ and
$\Delta C_p \ne 0$ give almost the same values for $T_G$, $\Delta
H_G$ and $\Delta S_G$. However, pronounced different behaviors of
the population of the native basin, $f_N$, occur when we
interpolate results to the low temperature region (Fig. \ref{hbSBD_345}b).
For the $\Delta C_p=0$ case, $f_N$ approaches
the unity as $T \rightarrow 0$ but it goes down for $\Delta C_p
\ne 0$. This means that the $\Delta C_p \ne 0$ fit is valid if the
second cold denaturation transition may occur at $T_G$'. This
phenomenon was observed in single domains as well as in
multi-domain globular proteins \cite{Privalov_CRBMB90}. We predict that
the cold denaturation of hbSBD may take place at $T_G' \approx
212$ K which is lower than $T_G' \approx 249.8$ K for hbLBD
shown on the 4th row of Table \ref{hbSBD_table1}.
It would be of great interest to carry out the cold denaturation
experiments in cryo-solvent to elucidate this issue.

To compare the stability of the hbSBD domain with the hbLBD domain
which has been studied in detail previously \cite{Naik_ProtSc04} we also
present the thermodynamic data of the latter on Table \ref{hbSBD_table1}. Clearly,
hbSBD is less stable than hbLBD by its smaller $\Delta G_S$ and
lower $T_G$ values. This is consistent with their respective
backbone dynamics as revealed by $^{15}$N-T$_1$, $^{15}$N-T$_2$,
and  $^{15}$N-$^1$H NOE studies of these two domains using
uniformly  $^{15}$N-labeled protein samples (Chang and Huang,
unpublished results). Biologically, hbSBD must bind to either E1
or E3 at different stages of the catalytic cycle, thus it needs to
be flexible to adapt to local environments of the active sites of
E1 and E3. On the other hand, the function of hbLBD is to permit
its Lys44 residue to channel acetyl group between donor and
acceptor molecules and only the Lys44 residue needs to be flexible
\cite{Chang_JBC02}. In addition, the NMR observation for the
longer fragment (comprising residues 1-168 of the E2 component)
also showed that the hbLBD region would remain structured after
several months while the hbSBD domain could de-grate in a shorter
time.
\begin{center}\begin{table}[htbp] 
\begin{tabular}{|c|c|c|c|c|c|c|c|c|}
\hline
&&\scriptsize$\Delta{H}_G$&\scriptsize$\Delta{C_p}$&\scriptsize$\Delta{S_G}$&&&\scriptsize$\Delta{G_S}$&\\
\scriptsize Domain&\scriptsize $T_G(K)$&\scriptsize kcal/mol/K)&\scriptsize (kcal/mol/K)&\scriptsize(cal/mol/K)&\scriptsize$T_S(K)$&\scriptsize$T_H(K)$&\scriptsize(kcal/mol)&\scriptsize$T'_G(K)$\\
\hline
\scriptsize SBD(exp)&\scriptsize$317.8\pm1.9$&\scriptsize$19.67\pm2.67$&\scriptsize$0.387\pm0.054$&\scriptsize$61.64\pm7.36$&\scriptsize$270.9\pm2.0$&\scriptsize$267.0\pm2.1$&\scriptsize$1.4\pm0.1$&\scriptsize$212\pm2.5$\\
\hline
\scriptsize SBD(exp)&\scriptsize$317.9\pm2.2$&\scriptsize$20.02\pm3.11$&\scriptsize$0.0$&\scriptsize$62.96\pm9.92$&\scriptsize$-$&\scriptsize$-$&\scriptsize$-$&\scriptsize-\\
\hline
\scriptsize SBD(sim)&\scriptsize$317.9\pm7.95$&\scriptsize$11.46\pm0.29$&\scriptsize$0.0$&\scriptsize$36.05\pm1.85$&\scriptsize$-$&\scriptsize$-$&\scriptsize$-$&\scriptsize-\\
\hline
\scriptsize LBD(exp)&\scriptsize$344.0\pm0.2$&\scriptsize$78.96\pm1.28$&\scriptsize$1.51\pm0.04$&\scriptsize$229.5\pm3.7$&\scriptsize$295.7\pm3.7$&\scriptsize$291.9\pm1.3$&\scriptsize$5.7\pm0.2$&\scriptsize$249.8\pm1.1$\\
\hline
\end{tabular}
\linespread{0.8}
\caption{Thermodynamic parameters obtained from
the CD experiments and simulations for hbSBD domain. The results
shown on the first and fourth rows were obtained by fitting
experimental data to the two-state equation (\ref{fN_twostate_eq})
with $\Delta C_p \ne 0$. The second and third rows corresponding
to the fit with $\Delta C_p = 0$. The results for hbLBD are taken
from Ref. \cite{Naik_ProtSc04} for comparison.}
\label{hbSBD_table1}
\end{table}
\end{center}

\subsubsection{Folding Thermodynamics from simulations}

In order to calculate the thermodynamics quantities we have collected
histograms for the energy and native contacts
at six values of temperature: $T = 0.4, 0.5, 0.6, 0.7,0.8$
and 1.0 $\epsilon_H/k_B$. For sampling,
at each temperature 30 trajectories
of $16\times 10^7$ time steps have been generated with initial
$4\times 10^7$ steps discarded for thermalization. 
The reweighting histogram method \cite{Ferrenberg_PRL89} was used 
to obtain the thermodynamics parameters at all temperatures.

 Figure \ref{hbSBD_345}b (open circles) shows the temperature
dependence of population of the NS, defined as the
renormalized number of native contacts 
for the Go model. Since there is no cold denaturation for this model,
to obtain the thermodynamic parameters we fit $f_N$ to the
two-state model (Eq. \ref{fN_twostate_eq}) with $\Delta C_p=0$.

The fit (black curve) works pretty well around the transition
temperature but it gets worse at high $T$ due to slow decay of
$f_N$ which is characteristic for almost all of theoretical
models. In fitting we have chosen the hydrogen bond energy
$\epsilon_H = 0.91$ kcal/mol in Hamiltonian (\ref{Hamiltonian}) so
that $T_G = 0.7 \epsilon_H/k_B$ coincides with
 the experimental value 317.8 K. From the
fit we obtain $\Delta H_G = 11.46$ kcal/mol which is smaller than
the experimental value indicating that the Go model is less
stable compared to the real hbSBD.

Figure \ref{hbSBD_345}c shows the temperature dependence of derivative of the
fraction of native contacts with respect to temperature  $df_N/dT$
and the
specific heat $C_v$ obtained from the Go simulations. The collapse
temperature $T_{\theta}$, defined as the temperature at which
$C_v$ is maximal, almost coincides with the folding temperature
$T_F$ (at $T_F$ the structural susceptibility has maximum).
According to Klimov and Thirumalai \cite{Klimov_PRL96},
the dimensionless parameter $\sigma = \frac{|T_{\theta}-T_F|}{T_F}$
may serve as an indicator for foldablity of proteins. Namely,
sequences with $\sigma \leq 0.1$ fold much faster that  
those which have the same number of residues but with $\sigma$
exceeding 0.5. From this perspective, having $\sigma \approx 0$ hbSBD
is supposed to be a
good folder {\it in silico}. However, one has to be cautious about
this conclusion because 
the pronounced correlation between folding times $\tau _F$
and the equilibrium parameter
$\sigma$, observed for simple on- and off-lattice models 
\cite{Klimov_PRL96,Veitshans_FD97} may be not valid for proteins in laboratory
\cite{Gillepse_ARB04}. In our opinion, since the data collected from
theoretical and
experimental studies are limited, further studies are required to
clarify the relationship between $\tau _F$ and $\sigma$.    

 Using experimental values for
$T_G$ (as $T_F$) and $\Delta H_G$ and the two-state model with $\Delta C_p
=0$ (see Table \ref{hbSBD_table1}) we can obtain the temperature dependence of the
population of NS $f_N$ and, therefore, $df_N/dT$ for
hbSBD (Fig. \ref{hbSBD_345}c). Clearly, the folding-unfolding transition
{\it in vitro} is sharper than
in the Go modeling. One of possible reasons is that our Go
model ignores the side chain which can enhance the cooperativity of
the denaturation transition \cite{Klimov_FD98}.

The sharpness of the fold-unfolded transition might be characterized
quantitatively via the cooperativity index $\Omega _c$ (Eq. \ref{cooper_index_eq}).
From Fig. \ref{hbSBD_345}c, we obtain
$\Omega_c = 51.6$ and 71.3 for the Go model and CD experiments,
respectively. Given the simplicity of the Go model used here the
agreement in $\Omega_c$ should be considered reasonable. We can
also estimate $\Omega _c$ from the scaling law suggested in Ref.
\onlinecite{MSLi_PRL04}, $\Omega _c = 0.0057 \times N^{\mu}$, where
exponent $\mu$ is universal and expressed via the random walk
susceptibility exponent $\gamma$ as $\mu = 1+\gamma
\approx 2.22 (\gamma
\approx 1.22$). Then we get $\Omega_c \approx 36.7$ which is lower
than the experimental as well as simulation result. This means
that hbSBD {\em in vitro} is, on average, more cooperative than
other two-state folders.

Another measure for the cooperativity is
$\kappa _2$ which is defined as  \cite{Kaya_PRL00}
$\kappa _2 = \Delta H_{vh}/\Delta H_{cal}$, where
$\Delta H_{vh} \;  = \; 2T_{max}\sqrt{k_B C_V(T_{max})}$
and $\Delta H_{cal} \; = \;  \int_0^{\infty} C_V(T)dT$,
are the van't Hoff and the calorimetric enthalpy, respectively,
$C_V(T)$ is the specific heat. Without the baseline substraction
in $C_V(T)$ \cite{Chan_ME04}, for the Go model of hbSBD we
obtained $\kappa _2 \approx 0.25$. Applying the baseline
substraction
as shown in the lower part of Fig. \ref{hbSBD_345}c
we got $\kappa _2 \approx 0.5$ which is still much lower than
$\kappa _2 \approx 1$ for a truly all-or-none transition.
Since $\kappa _2$ is an extensive parameter, its low value
is due to the shortcomings of the off-lattice
Go models but not due to the finite size effects.
More rigid lattice models give better results for
the calorimetric cooperativity \cite{Li_Physica05}.
Thus, for the hbSBD domain the Go model gives
the better agreement with our CD experiments for the
structural cooperativity
$\Omega_c$ than for the calorimetric measure $\kappa _2$.

\subsubsection{Free Energy Profile}

\begin{figure}
\epsfxsize=4.2in
\centerline{\epsffile{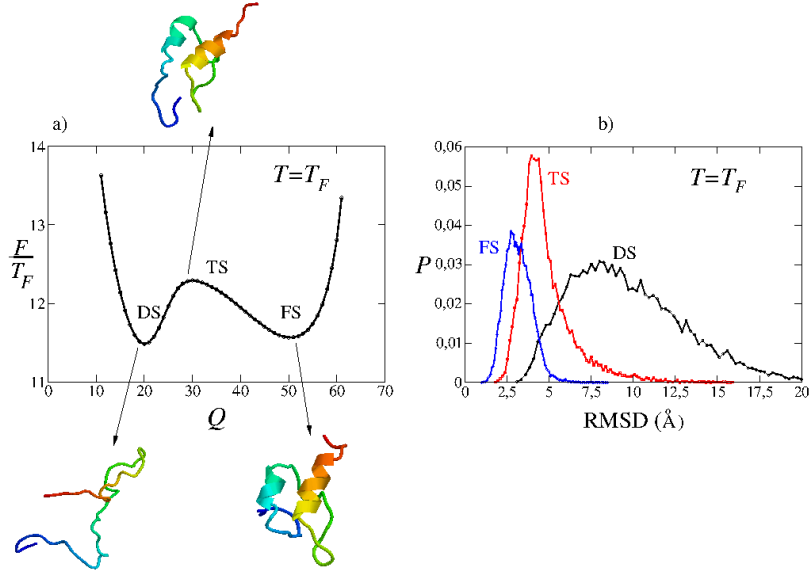}}
\linespread{0.8}
\caption{ (a) The dependence of free energy on the
number of native contacts $Q$ at $T=T_F$. The typical structures
of the DS , TS and folded
state are also drawn. The helix regions
H$_1$ (green) and H$_2$ (orange) of the TS structure involve residues 13 - 19
and 39 - 48, respectively. For the folded state structure H$_2$ is the same as for
the TS structure but H$_1$ has two residues more (13 - 21).
(b) Distributions of RMSD  for three
ensembles shown in (a). The average values of RMSD are equal to 9.8,
4.9 and 3.2 \AA$~$ for the DS, TS and folded state, respectively.}
\label{hbSBDfig6}
\vspace{3 mm}
\end{figure}

To get more evidence that hbSBD is a two-state folder we
study the free energy profile using some quantity as a reaction
coordinate. The precise reaction coordinate for a
multi-dimensional process such as protein folding is difficult to
ascertain. However, Onuchic and coworkers \cite{Nymeyer_PNAS98}
have argued that, for minimally frustrated systems such as Go
models, the number of native contact $Q$ may be appropriate. Fig.
\ref{hbSBDfig6}a shows the dependence of free energy on $Q$ for $T=T_F$. Since
there is only one local maximum corresponding to the transition
state (TS), hbSBD is a two-state folder. This is not unexpectable
for hbSBD which contains only helices. The fact that the simple Go model
correctly captures the two-state behavior as was observed in the CD
experiments, suggests that the energetic frustration ignored in this model
plays a minor role compared to the topological frustration
\cite{Clementi_JMB00}.

We have sorted out structures of the DS , TS
and the folded state at $T=T_F$ generating $10^4$
conformations in equilibrium. The distributions of the RMSD,
$P_{\rm RMSD}$,
 of these states are
plotted in  Fig. \ref{hbSBDfig6}b. As expected, $P_{\rm RMSD}$ for
the DS spreads out more than that for the TS and folded state. According to
the free energy profile in Fig. \ref{hbSBDfig6}a, the TS conformations
have 26 - 40 native contacts. We have found that the size (number
of folded residues) \cite{Bai_ProtSc04} of the TS is equal to 32. Comparing this size
with the total number of residues ($N=52$) we see that the fraction of
folded residues in the TS is higher than the typical value
for real two-state proteins
\cite{Bai_ProtSc04}. This is probably an artifact of Go models \cite{Kouza_BJ05}.
The TS conformations are relatively compact having
the ratio $<R_g^{TS}>/R_g^{NS} \approx 1.14$, where $<R_g^{TS}>$
is the average radius of gyration of the TS ensemble and $R_g^{NS}$
is the radius of gyration of the native conformation shown in Fig. \ref{hbSBD_12}a.
Since the RMSD, calculated only for two helices, is about 0.8 \AA
the structures of two helices in the TS
are not distorted much. It is also evident from
the typical structure of the TS shown in Fig.  \ref{hbSBDfig6}b where
the helix regions H$_1$ and H$_2$ involve residues 13 - 19 and 39 - 48,
respectively (a residue is considered to be in the helix state if its
dihedral angle is about 60$^o$).
Note that H$_1$ has two residues less compared to
H$_1$ in the native conformation (see the caption to Fig. \ref{hbSBD_12}a)
but H$_2$ has even one bead more than its NS counterpart.
Overall, the averaged RMSD of the TS conformations from the
native conformation (Fig. \ref{hbSBD_12}a) is about
4.9 \AA$~$ indicating that the TS is not close to the native one. As seen
from Figs. \ref{hbSBDfig6}a and \ref{hbSBD_12}a, the main difference comes
from the tail parts. The most probable conformations
(corresponding to maximum of $P_{\rm RMSD}$ in Fig. \ref{hbSBDfig6}b of the folded state have RMSD about 2.5 \AA. This value is reasonable from the point
of view of the experimental structure resolution.

\subsubsection{Folding Kinetics}

The two-state foldability, obtained from the thermodynamics simulations
may be also  probed by studying
the folding kinetics. For this purpose we monitored the
time dependence of the fraction of unfolded trajectories $P_u(t)$ defined
as follows \cite{Klimov_COSB99}
\begin{equation}
P_u(t) \, = \, 1 - \int_0^t P^{\textstyle{N}}_{fp}(s)ds,
\label{Pu_eq}
\end{equation}
where $P^{\textstyle{N}}_{fp}$ is the distribution of first passage folding
times
\begin{equation}
P^{\textstyle{N}}_{fp} \, = \, \frac{1}{M} \sum_{i=1}^{M}
\delta (s - \tau _{f,1i}).
\label{Pn_eq}
\end{equation}
Here $\tau _{f,1i}$ is time for the $i$th trajectory
to reach the NS for the first time,
$M$ is the total number of trajectories used in simulations.
A trajectory is said to be folded if all of native contacts form.
 As seen from
Eqs. \ref{Pu_eq} and \ref{Pn_eq},
$P_u(t)$ is the fraction of trajectories which do not reach
the NS at time $t$.
In the two-state scenario the folding becomes triggered after 
overcoming only one free energy barrier between the TS
and the denaturated one.
 Therefore, $P_u(t)$ should be a single exponential,
i.e. $P_u(t) \sim \exp(-t/\tau_F)$ (a multi-exponential
behavior occurs in the case when the folding proceeds via intermediates)
\cite{Klimov_COSB99}.
\begin{wrapfigure}{r}{0.48\textwidth}\centering
\includegraphics[width=0.45\textwidth]{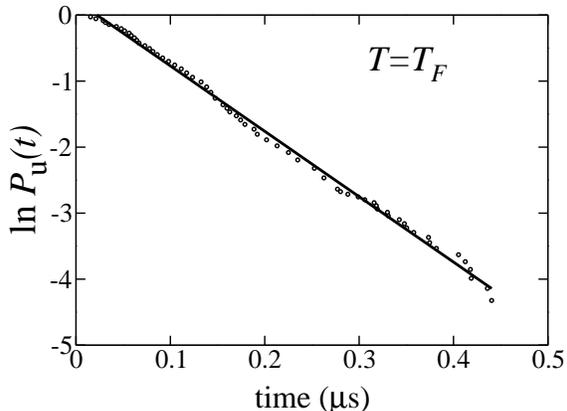}
\hfill \begin{minipage}{7 cm}
\linespread{0.8}
\caption{The semi-logarithmic plot of the time
dependence of the fraction of unfolded trajectories at $T=T_F$.
The distribution $P_u(t)$
was obtained from first passage times of 400 trajectories, which
start from random conformations.
The straight line corresponds to the fit ln $P_{\rm u}(t) =
-t/\tau_F$, where $\tau _F = 0.1 \mu$s. \label{hbSBDfig7}}
\end{minipage}
\end{wrapfigure}
 Since the function $P_u(t)$ can be measured directly by a number of experimental techniques \cite{Greene_Methods04,Dyson_ME2005},
the single exponential kinetics of two-state folders
is supported by a large body of experimental work (see, i.e. Ref.
\cite{Naik_FEBS02} and references there).
Fig. \ref{hbSBDfig7} shows the semi-logarithmic plot for $P_u(t)$ at $T=T_F$ for
the Go model. 
Since the single exponential fit works pretty well, one can expect
that intermediates do not occur on the folding pathways.
Thus, together with the thermodynamics data our kinetic study supports 
the two-state behavior of the hbSBD domain as observed
on the CD experiments.

From the linear fit in Fig. \ref{hbSBDfig7} we obtain the folding time
$\tau_F \approx 0.1 \mu$s.
This value is consistent with the
estimate of the folding time defined as the average value of the
first passage times.
If we use the empirical formula for the folding time
$\tau_F = \tau_F^0 \exp(1.1N^{1/2})$, where prefactor
$\tau_F^0=0.4 \mu$s and $N$ is a number of amino acids \cite{MSLi_Polymer04}
then $\tau_F = 1.1\times 10^3 \mu$s for $N=52$. This value is
about four orders of magnitude larger than that obtained from the
Go model.
 Thus the Go model can capture the two-state feature of
the denaturation transition for hbSBD domain but not folding
times.

\subsection{Discussion}

We have used CD technique and the Langevin dynamics to study the
mechanism of folding of hbSBD. Our results suggest that this
domain is a two-state folder. The CD experiments reveal that the
hbSBD domain is less  stable than the hbLBD domain in the same
BCKD complex, but it is more stable and cooperative compared to
other fast folding $\alpha$ proteins. 

Both the thermodynamics and
kinetics results, obtained from the Langevin dynamics
simulations, show that the simple Go model correctly captures
the two-state feature of folding.
It should be noted that the two-state behavior is not the natural
consequence of the Go modeling because it allows for fishing folding
intermediates caused by the topological frustration. From this standpoint
it may be used to decipher the foldability of model proteins
for which the topological frustration dominates.
The reasonable agreement between
the results obtained by the Go modeling 
and our CD experiments,
suggests that the NS topology of hbSBD is more important
than the energetic factor. 

The theoretical model gives the reasonable
agreement with the CD experimental data for the structural
cooperativity $\Omega _c$. However, the calorimetric cooperativity
criterion $\kappa _2 \approx 1$ for two-state folders is hard to
fulfill within the Go model. From the $\Delta C_p \ne 0$ fitting
procedure we predict that the cold denaturation of hbSBD may occur
at $T \approx 212$ K and it would be very interesting to verify
this prediction experimentally. We are using the package SMMP
\cite{Eisenmenger_CPC01} and a parallel algorithm \cite{Hayryan_JCC01} to
perform all-atom simulation of hbSBD to check the relevant
results. \vskip 2 mm

\newpage

\begin{center}
\section{Force-Temperature phase diagram of single and three domain ubiquitin. New force replica exchange method} \end{center}
\subsection{Introduction}

Protein Ub continues to attract the attention of researchers because
there exist many processes in living systems where it plays
the vital role. Usually, Ub presents  in the form of a polyubiquitin chain
 that is conjugated to other proteins.
Different Ub linkages lead to  different biological functions.
In  case of Lys48-C and N-C linkages  polyubiquitin chain serves as a signal for
degradation  proteins \cite{Thrower_EMBO2000, Kirisako_EMBO2006},  whereas
in the Lys63-C case it plays completely different functions, including
DNA repair,
polysome stability and endocytosis
 \cite{Hofmann_Cell1999, Spence_Cell2000, Galan_EMBO1997}.

When one studies thermodynamics of a large system like multi-domain
Ub the problem of slow dynamics occurs,
due to the rough FEL.
This
problem might be remedied using
the standard RE method in
the temperature space in the absence of external force
 \cite{Hukushima_JPSC96,Sugita_ChemPhysLett99,Phuong_Proteins05}
as well as in the presence of it \cite{Li_JPCB06}.
However,
if one wants to construct the force-temperature phase diagram,
then this approach becomes inconvenient because one has to collect
data at different values of forces.
Moreover, the external force increases unfolding barriers and a system may
get trapped in some local minima. In order to have better sampling for a system
subject to external force we propose a new RE method \cite{Kouza_JCP08} in which
the exchange is carried not in the temperature space but in the force space,
 i.e.
the exchange between different force values. This procedure would help
the system to escape from local minima efficiently.

In this chapter we address two topics. First, we develop
a new version of the RE method to study thermodynamics of a
large system under the force. The basic idea is that for
a given temperature we perform simulation at
different values of force and the exchange between them is carried out according
to the Metropolis rule.
This new approach has been employed to obtain the force-temperature phase
diagram of the three-domain Ub, which will be referred to as trimer . Within our choice
of force replicas it speeds up computation about four times
compared to the conventional simulation.
Second, we construct the temperature-force $T-f$ phase diagram of Ub and its trimer which
 allows us to to determine the equilibrium critical force $f_c$ separating
 the folded and unfolded regions.

 This chapter is based on Ref. \cite{Kouza_JCP08}.

\subsection{Model}
\begin{figure}
\epsfxsize=4.5in
\vspace{0.2in}
\centerline{\epsffile{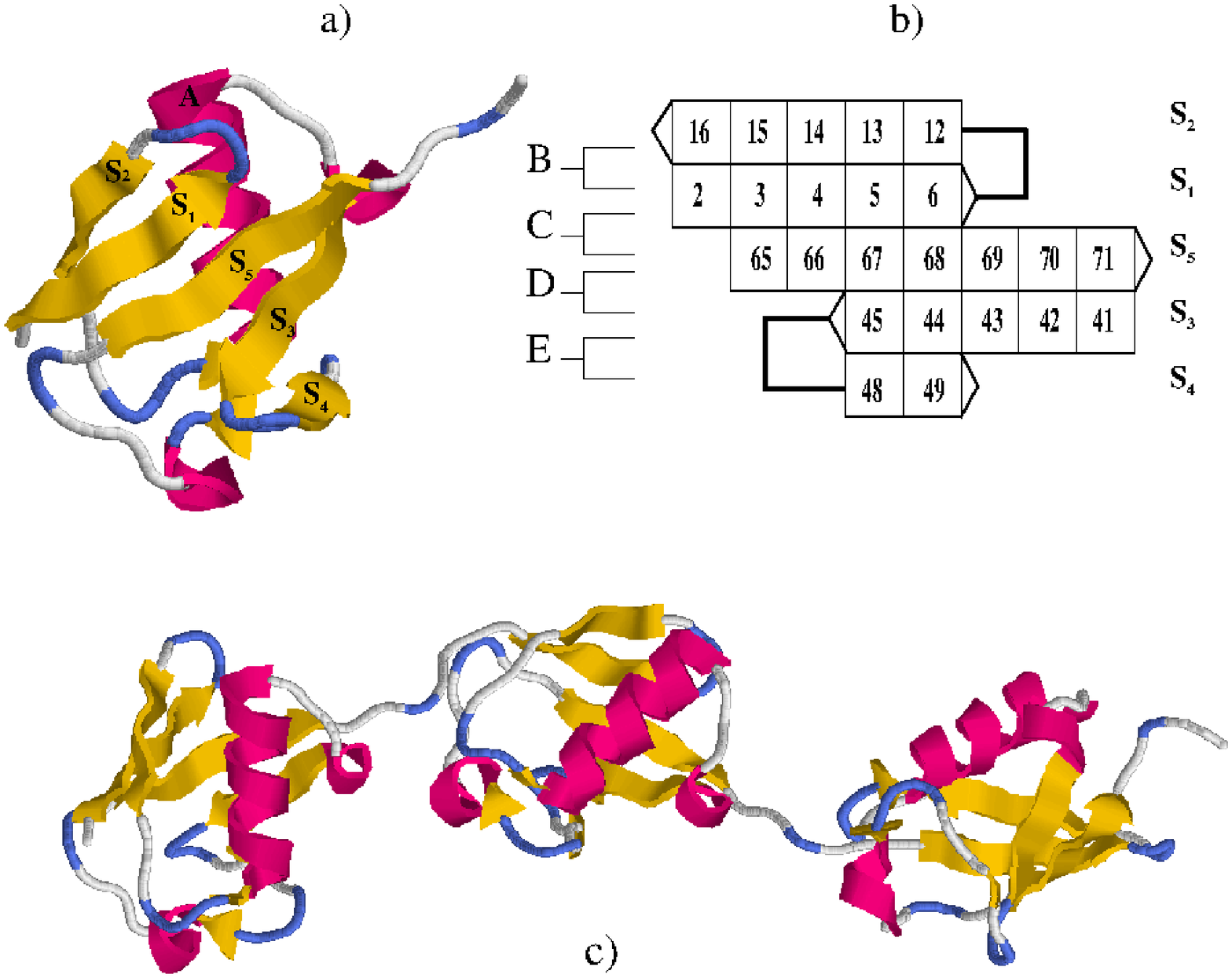}}
\linespread{0.8}
\caption{(a) NS conformation of Ub taken from the PDB
(PDB ID: 1ubq). There are five $\beta$-strands: S1 (2-6), S2 (12-16),
S3 (41-45), S4 (48-49) and S5 (65-71), and one helix A (23-34).
(b) Structures B, C, D and E consist of pairs of strands (S1,S2),
(S1,S5), (S3,S5) and (S3,S4), respectively. In the text we also refer to helix A as the structure A.  (c) The native conformation of trimer was designed as described in section 6.2. There are 18 inter-  and 297 intra-domain native contacts}
\label{ubiquitin_struture_fig}
\end{figure}

Figure \ref{ubiquitin_struture_fig} shows native conformations for single Ub and trimer. Native conformation of Ub is taken form the PDB (1UBQ) and with the choice of cutoff distance $d_c=6.5\AA$ it has 99 native contacts. NS of three-domain Ub. is not available yet and we have to construct it for Go modeling. To make it we translate one unit by the distance $a=3.82\AA$ and slightly rotate it, then translate and rotate one more to have nine interdomain contacts (about 10\% of the intra-domain contacts). There are 18 inter- and 297 intradomain native contacts.

We use coarse-grained continuum representation for Ub and trimer in which only the positions of $C_\alpha$-carbons are retained. The energy of Go-type model \cite{Clementi_JMB00} is described by Eq. (\ref{Hamiltonian}).
In order to obtain the $T-f$ phase diagram, we use the fraction of native contacts or
the overlap function as in Eq. (\ref{chi_eq_Go}).
 The $T-f$ phase diagram ( a plot of $1-f_N$ as
a function of $f$ and $T$) and thermodynamic quantities were
obtained by the multiple histogram method \cite{Ferrenberg_PRL89}
extended to the case when the external force is applied to the
termini \cite{Klimov_PNAS99,Klimov_JPCB01}. In this case the
reweighting is carried out not only for temperature but also for
force. We collected data for six values of $T$ at $f=0$ and for
five values of $f$ at a fixed value of $T$. The duration of MD
runs for each trajectory was chosen to be long enough to get the
system fully equilibrating (9$\times 10^5 \tau_L$ from which
1.5$\times 10^5 \tau_L$ were spent on equilibration). For a given
value of $T$ and $f$ we have generated 40 independent
trajectories for thermal averaging.

\subsection{Force-Temperature diagram for single ubiquitin}

\begin{figure}
\includegraphics[width=6.3in]{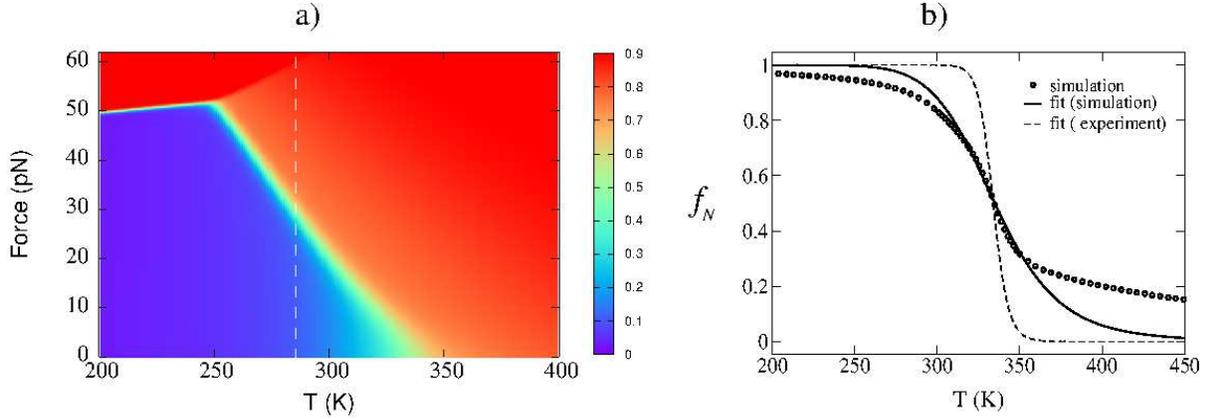}
\caption{(a) The $T-f$ phase diagram obtained by the extended
histogram method. The force is applied to termini N and C.
The color code for $1-<\chi(T,f)>$ is given on the right.
The blue color corresponds to the state in the NBA, while the red
color indicates the unfolded states. The
vertical dashed line refers to $T=0.85 T_F \approx 285$ K at which
most of simulations have been performed. (b) The temperature
dependence
 of $f_N$ (open circles) defined as the
renormalized number of native contacts.
The solid line refers to
the two-state fit to the simulation data.
The dashed line represents the experimental two-state curve with
$\Delta H_{\rm m}$ = 48.96 kcal/mol and $T_m = 332.5$K \cite{Thomas_PNAS01}.}
\label{diagram_fN_fig}
\end{figure}

The $T-f$  phase diagram, obtained by the extended
histogram method, is shown in
Fig. \ref{diagram_fN_fig}{\em a}. The folding-unfolding
transition, defined by the yellow region, is sharp in the low
temperature region but it becomes less cooperative (the fuzzy
transition region is wider) as $T$ increases.
The weak reentrancy (the critical force slightly increases with
$T$) occurs at low temperatures.
This seemingly strange phenomenon occurs as a result of
competition between the energy gain
and the entropy loss upon stretching.
The similar cold unzipping
transition was also observed in a number of models for
heteropolymers \cite{Shakhnovich_PRE02} and proteins
\cite{Klimov_PNAS99} including the C$_{\alpha}$-Go model for I27
(MS Li, unpublished results). As follows from the phase diagram,
at $T=285$ K the critical force $f_c \approx 30$ pN which is close
to $f_c \approx 25$ pN, estimated from the experimental pulling
data. To estimate $f_c$ from experimental pulling data
we use $f_{max} \approx f_c {\rm ln}(v/v_{min})$ \cite{Evans_BJ97} (see also
 Eq. \ref{f_logV_eq}),
where $f_{max}$ is the maximal force needed to unfold a protein at
the pulling speed $v$. From the raw data in Fig. 3b of
Ref. \cite{Carrion-Vazquez_NSB03} we obtain $f_c \approx$ 25 pN.
Given the simplicity
of the model this agreement can be considered satisfactory and it validates
the use of the Go model.

Figure \ref{diagram_fN_fig}{\em b} shows the temperature
dependence of population of the NS. Fitting to the
standard two-state curve $f_N = \frac{1}{1 + \exp[-\Delta
H_m(1-\frac{T}{T_m})/k_BT]}$, one can see that it works pretty well
(solid curve) around the transition temperature but it gets worse
at high $T$ due to slow decay of $f_N$. Such a behavior is
characteristic for almost all of theoretical models
\cite{Kouza_BJ05} including the all-atom ones
\cite{Phuong_Proteins05}. In fitting we have chosen the hydrogen
bond energy $\epsilon_H = 0.98$ kcal/mol in Hamiltonian
(\ref{Hamiltonian}) so that $T_F = T_m = 0.675 \epsilon_H/k_B$
coincides with the experimental value 332.5 K
\cite{Thomas_PNAS01}. From the fit we obtain $\Delta H_{\rm m} =
11.4$ kcal/mol which is smaller than the experimental value 48.96
kcal/mol indicating that the Go model is, as expected, less stable
compared to the real Ub. Taking into account non-native contacts
and more realistic interactions between side chain atoms is
expected to increase the stability of the system.

The cooperativity of the denaturation transition may be characterized by the cooperativity index, $\Omega_c$ given by Eq. (\ref{cooper_index_eq}).
 From simulation data for $f_N$ presented
\begin{wrapfigure}{r}{0.47\textwidth}
\includegraphics[width=0.46\textwidth]{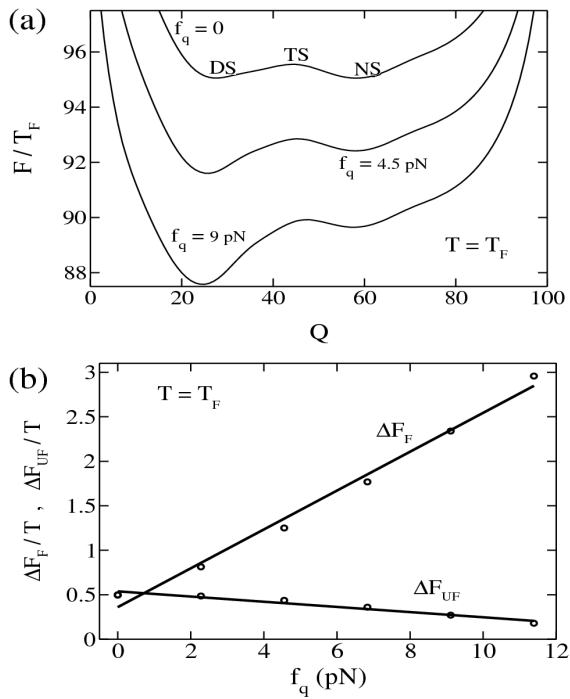}
\hfill\begin{minipage}{7.7 cm}
\linespread{0.8}
\caption{(a) The dependence of the free energy on
$Q$ for selected values of $f$ at $T=T_F$.(b) The
dependence of folding and unfolding barriers, obtained from the
free energy profiles, on $f$. The linear fits $y = 0.36 + 0.218x$
and $y=0.54 - 0.029x$ correspond to $\Delta F_{f}$ and $\Delta
F_{u}$, respectively. From these fits we obtain $x_f
\approx $ 10 nm and $x_{u} \approx$ 0.13 nm. \label{free_Q_barrier_fig}}
\end{minipage}
\end{wrapfigure}
in Fig. \ref{diagram_fN_fig}{\em b}, we have
 $\Omega_c \approx 57$ which is
considerably lower than the experimental value $\Omega_c \approx
384$ obtained with  the help of $\Delta H_{\rm m}$ = 48.96
kcal/mol and $T_m = 332.5$K \cite{Thomas_PNAS01} .
 The
underestimation of $\Omega _c$ in our simulations is not only a
shortcoming of the off-lattice Go model \cite{Kouza_JPCA06} but
also a common problem of much more sophisticated force fields in
all-atom models \cite{Phuong_Proteins05}.

Another measure of the cooperativity is the ratio between the
van't Hoff and the calorimetric enthalpy, $\kappa _2$
\cite{Kaya_PRL00}. For the Go Ub we obtained $\kappa _2 \approx
0.19$. Applying the base line subtraction \cite{Chan_ME04} gives
$\kappa _2 \approx 0.42$ which is still much below $\kappa _2
\approx 1 $ for the truly one-or-none transition. Since $\kappa
_2$ is an extensive parameter, its low value is due to the
shortcomings of the off-lattice Go models but not due to the
finite size effects. More rigid lattice models give better results
for the calorimetric cooperativity $\kappa _2$
\cite{Li_Physica05}.

Figure \ref{free_Q_barrier_fig}{\em a} shows the free energy as a
function of $Q$ for several values of force at $T=T_F$. Since
there are only two minima, our results support the two-state
picture of Ub \cite{Schlierf_PNAS04,Chung_PNAS05}. As expected,
the external
 force increases the folding barrier, $\Delta F_F$
($\Delta F_F = F_{TS} - F_{DS}$) and it lowers
the unfolding barrier, $\Delta F_{u}$ ($\Delta F_{u} = F_{TS} - F_{NS}$).
From the linear fits in
Fig. \ref{free_Q_barrier_fig}{\em b} we obtain $x_f = \Delta F_f/f \approx 1$ nm, and $x_{u} = \Delta F_{u}/f \approx 0.13$ nm.
Note that $x_f$ is very
close to $x_f \approx$ 0.96 nm obtained from refolding
times at a bit lower temperature $T=285$ K (see Fig. \ref{refold_Ub_trimer} below).
However, $x_{u}$ is lower than the experimental value 0.24 nm \cite{Carrion-Vazquez_NSB03}. 
This difference may be caused
by either sensitivity of $x_{u}$ to the temperature or the determination
of $x_{u}$ from the
approximate FEL as a function of a single
coordinate $Q$ is not sufficiently accurate.
In Chapter 8, we will show that a more accurate estimate of $x_u$
may be obtained from the dependence of unfolding times on the external force
(Eq. \ref{Bell_Ku_eq}).

We have also studied the FEL using $\Delta R$ as a reaction
coordinate.
The dependence of $F$ on $\Delta R$ was found to be smoother
(results not shown) compared
to what was obtained by Kirmizialtin {\em et al.} \cite{Kirmizialtin_JCP05}
using a more elaborated model
\cite{Sorenson_Proteins02}
which involves the non-native interactions.

\subsection{New force replica exchange method}

The equilibration of long peptides at low temperatures is a computationally
expensive job. In order to speed up computation of thermodynamic quantities
we extend the standard RE
method (with replicas at different temperatures)
developed for spin \cite{Hukushima_JPSC96} and peptide systems
\cite{Sugita_ChemPhysLett99} to the case when the RE is
performed between states with different values of
 the external force $\lbrace f_i \rbrace$.
Suppose for a given temperature
we have $M$ replicas $\lbrace x_i, f_i\rbrace$, where
$\lbrace x_i \rbrace$ denotes coordinates and velocities of
residues. Then the statistical
sum of the extended ensemble is
\begin{eqnarray}
Z \; = \; \int \ldots \int dx_1 \ldots dx_M \exp(- \sum_{i=1}^M\beta H(x_i)) = \prod_{i=1}^MZ(f_i).
\label{Z_total_eq}
\end{eqnarray}
The total distribution function has the following form
\begin{eqnarray}
P(\lbrace x,f\rbrace) &=& \prod_{i=1}^M P_{eq}(x_i,f_i), \nonumber\\
P_{eq}(x,f) &=& Z^{-1}(f)\exp(-\beta H(x,f)).
\label{P_total_eq}
\end{eqnarray}
For a Markov process the detailed balance condition reads as:
\begin{eqnarray}
P(.\,\!.\,\!.\,\!, x_m f_m, .\,\!.\,\!.\,\!, x_n f_n, .\,\!.\,\!.\,\!) W(x_m f_m \vert x_n f_n)
 \!=\! P(.\,\!.\,\!.\,\!, x_n f_m, .\,\!.\,\!.\,\!, x_m f_n, .\,\!.\,\!.\,\!) W(x_n f_m \vert x_m f_n),
\label{Markov_eq}
\end{eqnarray}
where $W(x_m f_m \vert x_n f_n)$ is the rate of transition
$\lbrace x_m, f_m \rbrace \rightarrow \lbrace x_n, f_n \rbrace$.
Using
\begin{eqnarray}
H(x,f) = H_0(x) - \vec{f}\vec{R} ,
\end{eqnarray}
and Eq. (\ref{Markov_eq})
we obtain
\begin{eqnarray}
\frac{W(x_m f_m \vert x_n f_n)}{W(x_n f_m \vert x_m f_n)} \; = \;
\frac{P(\ldots, x_m f_m, \ldots, x_n f_n, \ldots)}{P(\ldots, x_n f_m,
\ldots, x_m f_n, \ldots)} \; = \\ \nonumber \; \frac{\exp[-\beta(H_0(x_n) -
\vec{f}_m\vec{R}_n)
- \beta(H_0(x_m) - \vec{f}_n\vec{R}_m)]}{\exp[-\beta(H_0(x_m) -
\vec{f}_m\vec{R}_m)
- \beta(H_0(x_n) - \vec{f}_n\vec{R}_n)]} \; = \; \exp(-\Delta),
\end{eqnarray}
with
\begin{eqnarray}
\Delta &=& \beta (\vec{f}_m - \vec{f}_n) (\vec{R}_m - \vec{R}_n).
\label{Delta_eq}
\end{eqnarray}
This gives us the following Metropolis rule for accepting or rejecting
the exchange between replicas $f_n$ and $f_m$:
\begin{eqnarray}
W(x f_m | x' f_n) = \left\{ \begin{array}{ll}
         1&, \qquad \mbox{$\Delta < 0$}\\
        \exp(-\Delta)&, \qquad \mbox{$\Delta > 0$}\end{array} \right.
\label{Metropolis_eq}
\end{eqnarray}

\subsection{Force-Temperature diagram for three domain ubiquitin}

Since the three-domain Ub is rather
long peptide (228 residues), we apply the RE method to
obtain its $T-f$ phase diagram.
We have performed two sets of the RE simulations. In the first set we fixed
$f=0$ and the RE is carried out in the standard temperature replica space
\cite{Sugita_ChemPhysLett99}, where
12 values of $T$ were chosen in the interval $\left[0.46, 0.82\right]$
in such a way that the RE acceptance ratio was 15-33\%.
This procedure speeds up the equilibration of our system nearly
ten-fold compared to the standard computation without the use of RE.

In the second set, the RE simulation was performed in the force replica
space at $T=0.53$ using the Metropolis rule given by Eq. (\ref{Metropolis_eq}).
We have also used 12 replicas with different values of $f$
in the interval $0 \leq f \leq 0.6$ to get
the acceptance ratio about 12\%.
Even for this modest acceptance rate our new RE scheme accelerates
the equilibration
of the three-domain Ub about four-fold. One can expect better
performance by increasing the number of replicas.
However, within our computational facilities we were restricted to
parallel runs on 12 processors for 12 replicas.
The system was equilibrated
during first 10$^5 \tau_L$, after which histograms for the energy, the native
contacts and end-to-end distances were collected
for $4\times 10^5 \tau_L$ .
For each replica, we have generated 25 independent trajectories for
thermal averaging.
Using the data from two sets of the RE simulations and the
extended reweighting technique \cite{Ferrenberg_PRL89} in the
temperature and force space
\cite{Klimov_JPCB01} we obtained the $T-f$ phase diagram and the
thermodynamic quantities of the trimer.

\begin{figure}[!htbp]
\epsfxsize=6.1in
\vspace{0.2in}
\centerline{\epsffile{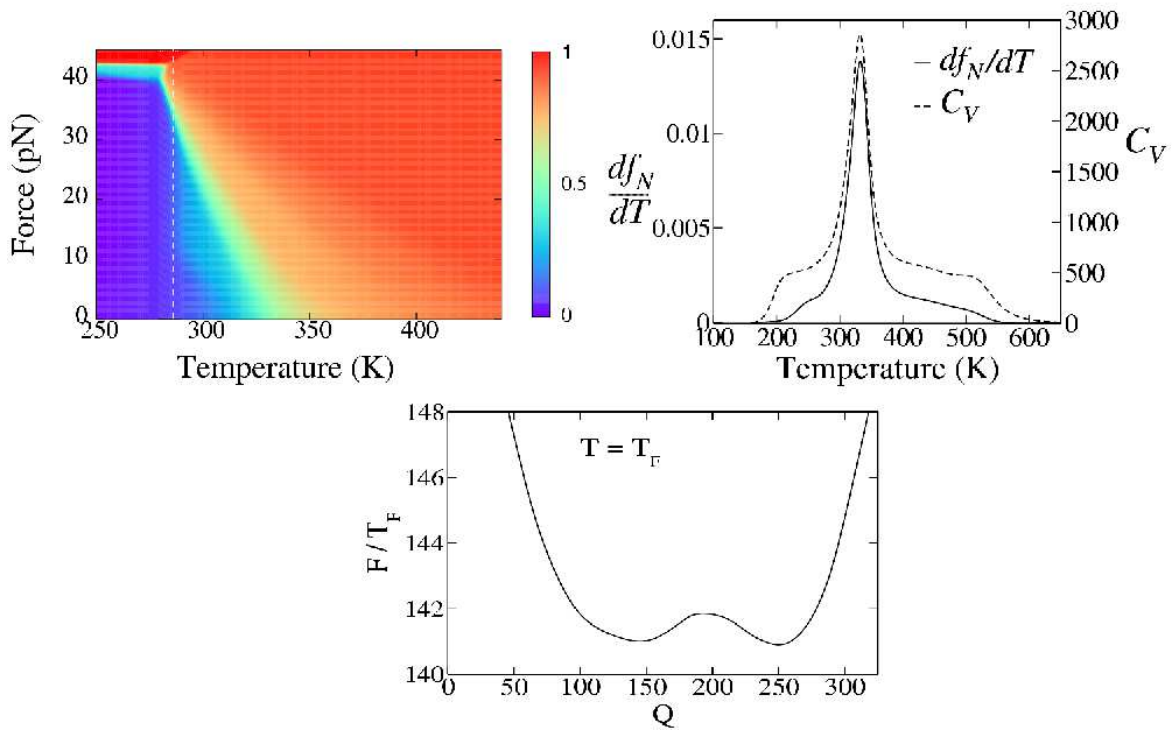}}
\linespread{0.8}
\caption{ (a) The $T-f$ phase diagram obtained by the extended
RE and
histogram method for trimer. The force is applied to termini N and C.
The color code for $1-f_N$ is given on the right.
Blue corresponds to the state in the NBA, while red
indicates the unfolded states. The
vertical dashed line denotes to $T=0.85 T_F \approx 285$ K, at which
most of simulations have been performed.
(b) Temperature dependence of the specific heat $C_V$ (right axis) and
$df_N/dT$ (left axis) at $f=0$. Their peaks coincide at $T=T_F$.
%
(c) The dependence of the free energy of the trimer on the total number
of native contacts
$Q$ at $T=T_F$.}
\label{diagram}
\end{figure}

The $T-f$ phase diagram
(Fig. \ref{diagram}a) was obtained by monitoring the probability
of being in the NS, $f_N$, as a function of $T$ and $f$.
The folding-unfolding
transition (the yellow region) is sharp in the low
temperature region, but it becomes less cooperative (the fuzzy
transition region is wider) as $T$ increases.
The folding temperature in the absence of force (peak of $C_v$ or
$df_N/dT$ in Fig. \ref{diagram}b) is equal $T_F=0.64 \epsilon_H/k_B$
which is a bit lower than $T_F=0.67 \epsilon_H/k_B$
of the single Ub \cite{MSLi_BJ07}.
This reflects the fact the folding of the trimer is less cooperative compared
to the monomer due to a small number of native contacts between domains.
One can ascertain this by calculating
the cooperativity index, $\Omega_c$ \cite{Klimov_FD98,MSLi_PRL04}
for the denaturation transition.
From simulation data for $df_N/dT$ presented in
Fig. \ref{diagram}b, we obtain
$\Omega_c \approx 40$ which is indeed lower than $\Omega_c \approx 57$ for
the single Ub \cite{MSLi_BJ07} obtained by the same Go model.
According to our previous estimate \cite{MSLi_BJ07},
the experimental value $\Omega_c \approx
384$ is considerably higher than the Go value.
Although the present Go model does not provide the realistic
estimate for cooperativity, it still mimics the experimental
fact, that
folding of a multi-domain protein remains
cooperative observed for not only Ub but also other proteins.

Fig. \ref{diagram}c shows the free energy as a function of native contacts
at $T=T_F$. The folding/unfolding barrier is rather low ($\approx$ 1 kcal/mol),
and is comparable with the case of single Ub \cite{MSLi_BJ07}.
The low barrier is probably an artifact of the simple Go modeling.
The double minimum structure suggests that the trimer is a two-state folder.

\subsection{Conclusions}

We constructed the $T$-$f$ phase diagrams of single and three-domain
Ub and showed that both are two-state folders.
The standard temperature RE method was extended to the case when the
force replicas
are considered at a fixed temperature. One can extend the RE method to
cover both temperature and force replicas, as
has been done
for all-atom simulations \cite{Paschek_PRL04} where pressure
is used  instead of force.
One caveat of the force RE method is that the acceptance depends on the
end-to-end distance (Eq. \ref{Delta_eq} and \ref{Metropolis_eq}),
and  becomes inefficient for long proteins.
We can overcome this by increasing the number of replicas,
but this will increase
CPU time substantially. Thus, the question of improving the force RE approach
for long biomolecules remains open.

\newpage

\begin{center}\section{Refolding of single and three domain ubiquitin under quenched force}\end{center}
\subsection{Introduction}

\begin{figure}
\epsfxsize=3.8in
\centerline{\epsffile{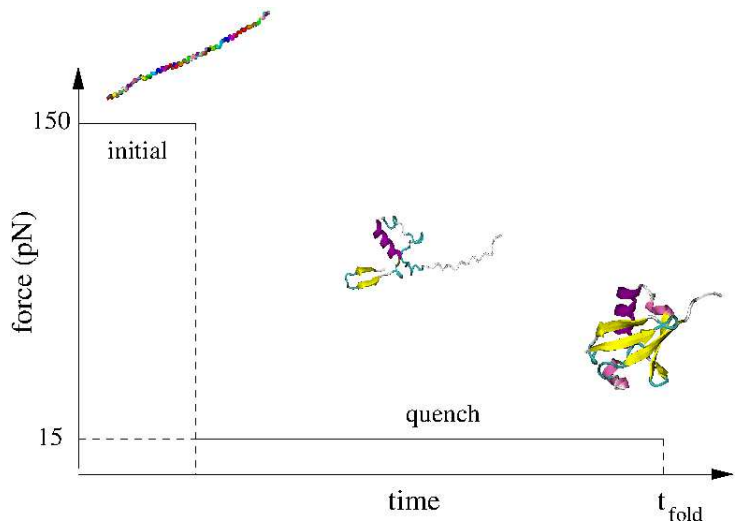}}
\linespread{0.8}
\caption{Representation of an experimental protocol of force-clamp spectroscopy.
First a protein is stretched under force of hundreds pN.
Then the external force is reduced to the quenched value
$f_q$ and this force is kept fixed during the refolding
process.}\label{force_clamp}
\end{figure}

Deciphering the folding and unfolding pathways and FEL
 of biomolecules remains a
challenge in molecular biology. Traditionally, folding and unfolding are
monitored by changing temperature or concentration of chemical denaturants.
In these experiments, due to thermal fluctuations of
initial unfolded conformations, it is difficult to describe the folding
mechanisms in an unambiguous way. 
\cite{Fisher_TBS99,Fernandez_Sci04}.
Recently, Fernandez and coworkers \cite{Fernandez_Sci04} have applied
the force-clamp technique (Fig. \ref{force_clamp}) to probe refolding of Ub under quench
force, $f_q$, which is smaller than the equilibrium critical force
separating
the folded and unfolded states.
Here, one can
control starting conformations  which are well prepared by applying
the large initial
force of several hundreds of pN. 
 Monitoring folding events as
a function of the end-to-end distance ($R$) they have made the following
important observations:
\begin{enumerate}
 \item Contrary to the standard folding from the
thermal denaturated ensemble (TDE) the refolding under the quenched
force is a multiple
stepwise process.

\item The force-quench refolding time obeys the Bell formula
\cite{Bell_Sci78}, $\tau_F \approx \tau_F^0
\exp(f_qx_f/k_BT)$, where $\tau_F^0$ is the folding time
in the absence of the quench
force and $x_f$ is the average location of the TS.
\end{enumerate}

Motivated by the experiments of Fernandez and Li  \cite{Fernandez_Sci04},
Li {\em et al} have studied  \cite{MSLi_PNAS06}
 the refolding of the domain I27 of the human muscle protein
using the C$_{\alpha}$-Go model \cite{Clementi_JMB00}
and the four-strand
 $\beta$-barrel model sequence S1 \cite{Klimov_PNAS00}
(for this sequence the nonnative interactions are also taken into account).
Basically, we have reproduced qualitatively
the major experimental findings listed above. In addition, we have
 shown that the refolding is two-state process in which the folding
 to the NBA follows the quick collapse from initial stretched
 conformations with low entropy. The corresponding kinetics can be described by the
bi-exponential time dependence, contrary to the single exponential
behavior of the folding from the TDE with high entropy.

To make the direct comparison with the experiments of
Fernandez and Li \cite{Fernandez_Sci04}, in this chapter we
performed simulations for a single domain
Ub using the C$_{\alpha}$-Go model \cite{Clementi_JMB00}.
Because the study of
refolding of 76-residue Ub (Fig. \ref{ubiquitin_struture_fig}{\em a})
by all-atom simulations is beyond
present computational facilities the Go modeling is an
appropriate choice. Most of the simulations have been carried out
at $T = 0.85T_F = 285$ K. Our present results for refolding upon
the force quench are in the qualitative agreement with the
experimental findings of Fernandez and Li, and with those obtained
for I27 and S1 theoretically \cite{MSLi_PNAS06}. A number of
quantitative differences between I27 and Ub will be also
discussed. For Ub we have found the average location of the
TS $x_f \approx 0.96$ nm which is in
reasonable agreement with the experimental value 0.8 nm
\cite{Fernandez_Sci04}.

Since the quench force slows down the folding process, it is easier to monitor refolding pathways.
However, this begs the important question as to whether the
force-clamp experiments with
one end of the protein anchored probes the same folding pathways as a free-end
protein.
Recently, using a simple Go-like model, it has been shown that
fixing the N-terminal of Ub changes its
folding pathways
\cite{Szymczak_JCP06}. If it is so, the force-clamp technique
in which the N-terminal is anchored is not useful
for prediction of folding pathways of the free-end Ub.
Using the Go model \cite{Clementi_JMB00}
we have shown that,
in agreement with an earlier study \cite{Szymczak_JCP06},
fixing
N-terminal of the single Ub changes its folding pathways. Our new finding
is that
anchoring
C-terminal leaves them unchanged. More importantly,
we have found that for the three-domain Ub
with either end fixed,
each domain follows the same folding pathways as for the free-end single
domain. Therefore,
to probe the folding pathways of Ub by the
force-clamp technique one can either use the single domain with C-terminal
fixed, or several domains with either end fixed.
In order to check if the effect of fixing one terminus is valid for other
proteins, we have studied  the titin domain I27.
It turns out
that the fixation of one end of a polypeptide chain
does not change the refolding pathways of I27.
 Therefore the
force-clamp can always predict the refolding pathways of the single as well as
multi-domain I27. Our study suggests that the effect of the end fixation
is not universal for all proteins, and the force-clamp
spectroscopy should be applied with caution.

The material of this chapter was taken from Refs. \cite{MSLi_BJ07, Kouza_JCP08}.

\subsection{Refolding of single ubiquitin under quenched force}

As in the previous chapter, we used the C$_{\alpha}$-Go model
(Eq. \ref{Hamiltonian}) to study refolding.
Folding pathways were probed by monitoring the fractions of native
contacts of secondary structures as a function of
the progressive variable $\delta$
(Eq. \ref{progress_fold_eq}).

\subsubsection{Stepwise refolding of single Ubiquitin}

Our protocol for studying the refolding of Ub is identical to what has been
done on the experiments of Fernandez and Li \cite{Fernandez_Sci04}.
We first apply the force $f_I \approx 70$ pN to prepare initial
conformations (the
protein is stretched if $R \ge 0.8 L$, where the contour length $L = 28.7 $ nm).
Starting from the force denaturated ensemble (FDE) we quenched the force
to $f_q < f_c$ and then monitored the refolding process by following
the time dependence of the number of native
contacts $Q(t)$, $R(t)$ and the radius of gyration
$R_g(t)$ for typically 50 independent trajectories.

\begin{figure}[!htbp]
\includegraphics[width=0.62\textwidth]{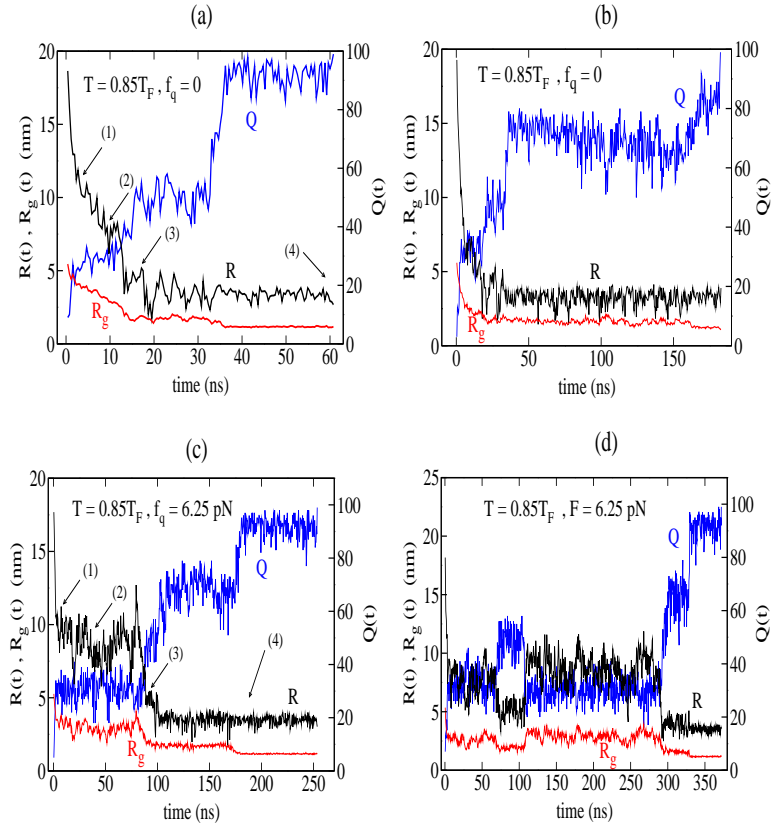}
\hfill
\linespread{0.8}
\parbox[b]{0.35\textwidth}{\caption{(a) and (b) The time dependence of $Q$, $R$ and $R_g$
for two typical trajectories starting from FDE
($f_q=0$ and $T=285$ K).
The arrows 1, 2 and 3 in (a) correspond to time 3.1 ($R=10.9$ nm),
9.3 ($R=7.9$ nm) and 17.5 ns ($R=5$ nm).
The arrow 4 marks the folding time $\tau _F$ = 62 ns ($R=2.87$ nm)
when all of 99
native contacts are formed. (c) and (d) are the same as in (a) and (b) but
for $f_q$ = 6.25 pN. The corresponding arrows refer to $t=$
7.5 ($R=11.2$ nm), 32 ($R=9.4$ nm), 95 ns
($R=4.8$ nm) and $\tau _F = 175$ ns ($R=3.65$ nm).\\\\\\\\\\\\}\label{ContRgR_time_fig}}
\vspace{5 mm}
\\
\end{figure}

Figure  \ref{ContRgR_time_fig} shows considerable diversity of
refolding pathways. In accord with experiments
\cite{Fernandez_Sci04} and simulations for I27 \cite{MSLi_PNAS06},
the reduction of $R$ occurs in a stepwise manner. In the $f_q=0$
case (Fig. \ref{ContRgR_time_fig}{\em a}) $R$ decreases
continuously from $\approx 18$ nm to 7.5 nm (stage 1) and
fluctuates around this value for about 3 ns (stage 2). The further
reduction to $R \approx 4.5$ nm (stage 3) until a transition to
the NBA. The stepwise nature of variation of $Q(t)$ is also
clearly shown up but it is more masked for $R_g(t)$. Although we
can interpret another trajectory for $f_q=0$ (Fig.
\ref{ContRgR_time_fig}b) in the same way, the time scales are
different. Thus, the refolding routes are highly heterogeneous.

The pathway diversity is also evident for $f_q >0$
(Fig. \ref{ContRgR_time_fig}{\em c}
and {\em d}). Although the picture remains qualitatively the same as in the
$f_q=0$ case, the time scales for different steps becomes much larger.
The molecule fluctuates around $R \approx 7$ nm, e.g.,
for $\approx 60$ ns (stage 2 in Fig. \ref{ContRgR_time_fig}{\em c})
which is considerably longer
than $\approx 3$ ns in Fig. \ref{ContRgR_time_fig}{\em a}.
The variation of $R_g(t)$ becomes more drastic
compared to the $f_q=0$ case.

\begin{figure}[!htbp]
\includegraphics[width=0.68\textwidth]{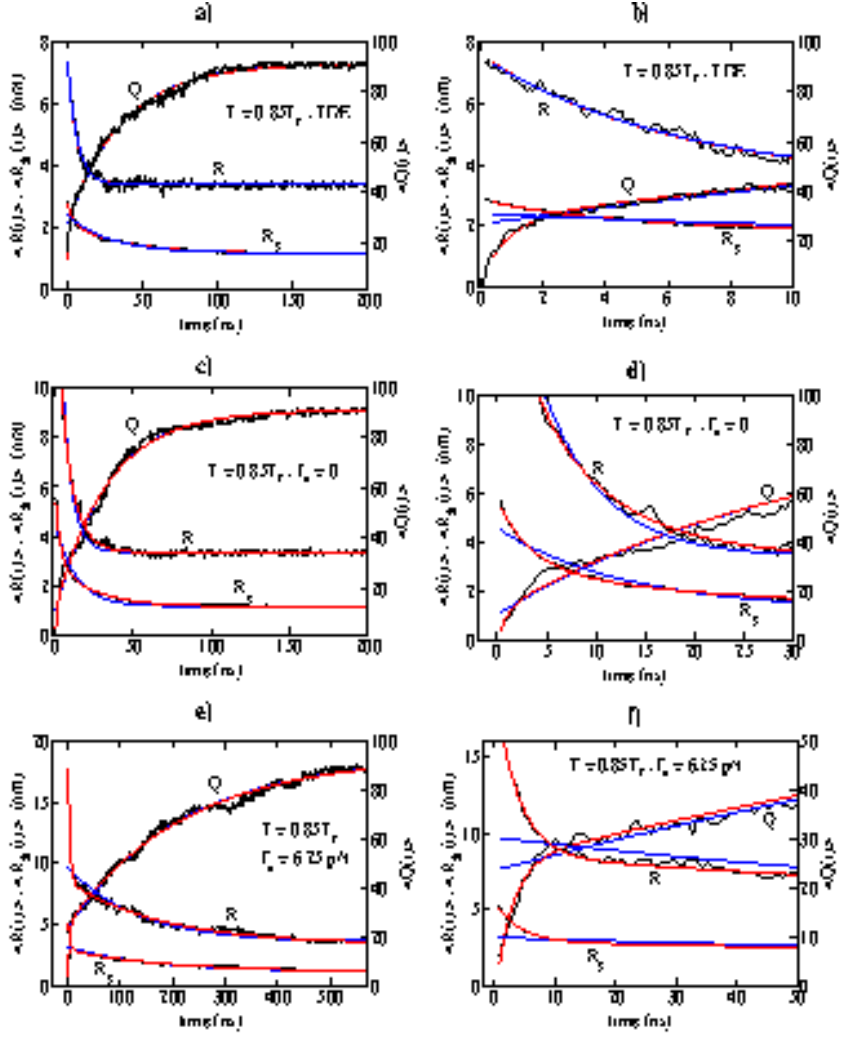}
\hfill
\linespread{0.8}
\parbox[b]{0.3\textwidth}{\caption{(a) The time dependence of $<Q(t)>$, $<R(t)>$ and $<R_g(t)>$
 when the refolding starts from TDE.
(b) The same as in (a) but for the short time scale.
(c) and (d) The same as in (a) and (b) but for FDE with $f_q=0$.
(e) and (f) The same as in (c) and (d) but for $f_q$=6.25 pN.\\\\\\\\\\\\\\\\\\\\\\\\}\label{ContRgR_time_av_fig}}
\vspace{5 mm}
\\
\end{figure}

Figure \ref{ContRgR_time_av_fig} shows the time dependence of
$<R(t)>, <Q(t)>$ and $<R_g(t)>$, where $<...>$ stands for
averaging over 50 trajectories. The left and right panels
correspond to the long and short time windows, respectively. For
the TDE case (Fig. \ref{ContRgR_time_av_fig}{\em a} and {\em b})
the single exponential fit works pretty well for $<R(t)>$ for the
whole time interval. A little departure from this behavior is seen
for $<Q(t)>$ and $<R_g(t)>$ for $t < 2$ ns (Fig.
\ref{ContRgR_time_av_fig}{\em b}). Contrary to the TDE case, even
for $f_q=0$ (Fig. \ref{ContRgR_time_av_fig}{\em c} and {\em d})
the difference between the single and bi-exponential fits is
evident not only for $<Q(t)>$ and $<R_g(t)>$  but also for
$<R(t)>$. The time scales, above which two fits become eventually
identical, are slightly different for three quantities (Fig.
\ref{ContRgR_time_av_fig}{\em d}). The failure of the single
exponential behavior becomes more and more evident with the
increase of $f_q$, as demonstrated in Figs.
\ref{ContRgR_time_av_fig}{\em e} and {\em f} for the FDE case with
$f_q = 6.25$ pN.

Thus, in agreement with our previous results, obtained for I27 and
the sequence S1 \cite{MSLi_PNAS06},
starting from FDE the refolding kinetics compiles of the fast and
slow phase. The characteristic time scales for these phases may be
obtained using a sum of two exponentials,$<A(t)> = A_0 + A_1
\exp(-t/\tau^A_1) + A_2 \exp(-t/\tau^A_2)$, where $A$ stands for
$R$, $R_g$ or $Q$. Here $\tau^A_1$ characterizes the burst-phase
(first stage) while $\tau^A_2$ may be either the collapse time
(for $R$ and $R_g$) or the folding time (for $Q$) ($\tau^A_1 <
 \tau^A_2$). As in the case of I27 and S1 \cite{MSLi_PNAS06},
$\tau^R_1$ and $\tau^{R_g}_1$ are almost independent on $f_q$
(results not shown). We attribute this to the fact that the quench
force ($f_q^{max} \approx 9$ pN) is much lower than the entropy
force ($f_e$) needed to stretch the protein. At $T=285$ K, one has
to apply $f_e \approx 140$ pN for stretching Ub to 0.8 $L$. Since
$f_q^{max} << f_e$ the initial compaction of the chain that is
driven by $f_e$ is not sensitive to the small values of $f_q$.
Contrary to $\tau^A_1$, $\tau^A_2$ was found to increase with $f_q$
exponentially. Moreover,
$\tau^R_2 < \tau^{R_g}_2 < \tau _F$ implying that the chain compaction
occurs before the acquisition of the NS.

\subsubsection{Refolding pathways of single Ubiquitin}

\begin{figure}
\vspace{5mm}
\epsfxsize=6.3in
\vspace{0.2in}
\centerline{\epsffile{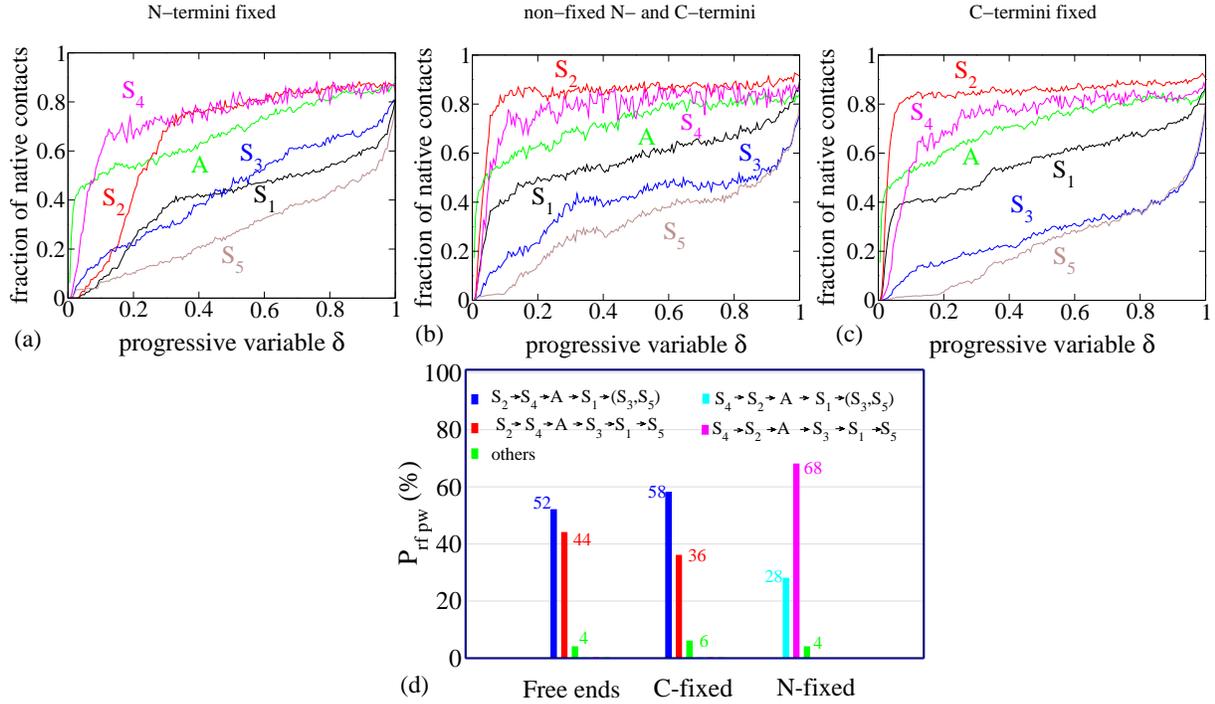}}
\linespread{0.8}
\caption{ The dependence of native contacts of $\beta$-strands
and the helix A on the progressive variable $\delta$ when the N-terminal
is fixed (a), both ends are free (b), and C-terminal is fixed
(c). The results are averaged over 200 trajectories.
(d) The probability of refolding pathways in three cases.
each value is written on top of the histograms.}
\label{single_ub_pathways_fig}
\end{figure}

In order to study refolding under small quenched force we follow the same
protocol as in the experiments \cite{Fernandez_Sci04}.
First, a large force ($\approx 130$ pN) is applied to both termini
to prepare the initial stretched conformations. This force is
then released, but a weak quench force, $f_q$, is applied to study the refolding process.
The refolding  of a single Ub was studied
\cite{MSLi_BJ07,Szymczak_JCP06} in the presence or absence of
the quench force.
Fixing the N-terminal
was found to change the refolding pathways of the free-end
Ub \cite{Szymczak_JCP06}, but the effect of anchoring the C-terminal
has not been studied yet. Here we study this problem in detail, monitoring
the time dependence of native contacts of secondary structures
(Fig. \ref{single_ub_pathways_fig}).
Since the quench force increases the folding time but leaves
the folding pathways unchanged, we present only the results for $f_q=0$
(Fig. \ref{single_ub_pathways_fig}).
Interestingly, the fixed C-terminal and free-end cases have the identical
folding sequencing
\begin{equation}
S2 \rightarrow S4 \rightarrow  A \rightarrow
S1 \rightarrow (S3,S5).
\label{free-end_pathways_eq}
\end{equation}

This is reverse of the unfolding pathway under thermal fluctuations
\cite{MSLi_BJ07}.
As discussed in detail by Li {\em et al.}
\cite{MSLi_BJ07}, Eq. (\ref{free-end_pathways_eq})
 partially agrees with the folding \cite{Went_PEDS05}
and unfolding \cite{Cordier_JMB02} experiments, and simulations
\cite{Fernandez_JCP01,Fernandez_Proteins02,Sorenson_Proteins02}.
Our new finding here is that keeping the C-terminal fixed does not
change the folding pathways.
One should keep in mind that the dominant pathway given by
 Eq. (\ref{free-end_pathways_eq})
is valid in the statistical sense.
It occurs in about 52\% and 58\% of events for the free end and C-anchored
cases (Fig. \ref{single_ub_pathways_fig}d), respectively.
The probability of observing an alternate pathway
($S2 \rightarrow S4 \rightarrow  A \rightarrow
S3 \rightarrow S1 \rightarrow S5)$ is $\approx 44$ \% and 36 \% for these
two cases
(Fig. \ref{single_ub_pathways_fig}d). The difference between these two pathways
is only in sequencing  of S1 and S3. Other pathways, denoted in green,
 are also possible
but they are rather minor.

In the case when the N-terminal is fixed (Fig. \ref{single_ub_pathways_fig})
we have the following sequencing
\begin{equation}
S4 \rightarrow S2 \rightarrow A \rightarrow S3 \rightarrow S1 \rightarrow S5
\label{fixedN_pathways_eq}
\end{equation}
which is, in agreement with Ref. \onlinecite{Szymczak_JCP06},
different from the free-end case. We present
folding pathways as the sequencing of
secondary structures,  making comparison with experiments easier
than an approach based on the time
evolution of individual contacts \cite{Szymczak_JCP06}.
The main pathway (Eq. \ref{fixedN_pathways_eq})
occurs in $\approx 68$ \% of events (Fig. \ref{single_ub_pathways_fig}d),
while the competing sequencing $S4 \rightarrow S2 \rightarrow A \rightarrow S1
 \rightarrow (S1, S5)$ (28 \%) and other minor pathways are also possible.
From Eq. (\ref{free-end_pathways_eq}) and (\ref{fixedN_pathways_eq}) it follows
that the force-clamp technique can probe the folding pathways of Ub if one anchors
the C-terminal but not the N-one.

In order to check the robustness of our prediction for refolding pathways
(Eqs. \ref{free-end_pathways_eq} and \ref{fixedN_pathways_eq}),
obtained for the friction $\zeta = 2 \frac{m}{\tau_L}$, we have performed
simulations for the water friction $\zeta = 50 \frac{m}{\tau_L}$. Our
results (not shown) demonstrate that although the folding time
is about twenty times longer compared to the
$\zeta = 2 \frac{m}{\tau_L}$ case, the pathways remain the same.
Thus, within the framework of Go modeling,
the effect of the N-terminus fixation
on refolding pathways of Ub is not an artifact of fast dynamics,
occurring for both large and small friction.
It would be very interesting to verify our prediction
using more sophisticated models. This question is left for future studies.

\subsection{Refolding pathways of three-domain Ubiquitin}

\begin{figure}
\epsfxsize=6.2in
\vspace{0.2in}
\centerline{\epsffile{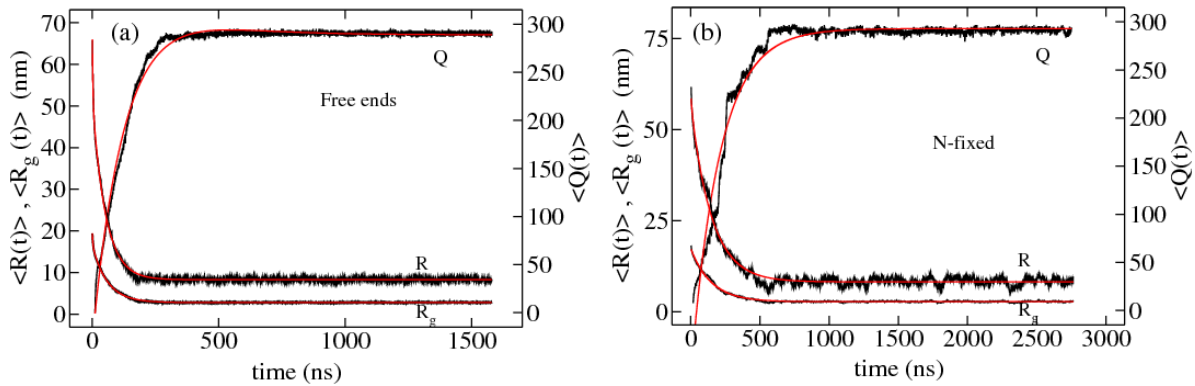}}
\linespread{0.8}
\caption{(a) The time dependence of $Q$, $R$ and $R_g$ at $T=285$ K for the free end case for trimer. (b) The same as in (a) but for the N-fixed case.
The red line is a bi-exponential fit $A(t) = A_0 + a_1\exp(-t/\tau_1)
+ a_2\exp(-t/\tau_2)$.
Results for the C-fixed case are similar to the
$N$-fixed case, and are not shown.}
\label{Q_Rnc_Rg_trimer_fig}
\end{figure}

The time dependence of the total number of native contacts, $Q$, $R$ and
the  gyration radius, $R_g$, is presented in
Fig. \ref{Q_Rnc_Rg_trimer_fig} for the trimer.
The folding time
$\tau _f \approx$ 553 ns and 936 ns for the free end and N-fixed cases,
respectively. The fact that anchoring one
end slows down refolding by a factor of nearly 2
implies that diffusion-collision processes
\cite{Karplus_Nature76} play an important role in
the Ub folding. Namely, as follows from the diffusion-collision model,
the time required for formation contacts is inversely
proportional to the diffusion coefficient, $D$, of a pair of spherical
units. If one of them is idle, $D$ is halved and
the time needed to form contacts increases accordingly.
The similar effect for unfolding was observed in our recent
work \cite{MSLi_BJ07}.

From the bi-exponential fitting, we obtain two time scales
for collapsing ($\tau_1$) and compaction ($\tau_2$) where $\tau_1 < \tau_2$.
For $R$, e.g., $\tau_1^R \approx 2.4$ ns and $\tau_2^R \approx 52.3$ ns if
two ends are free, and $\tau_1^R \approx 8.8$ ns and $\tau_2^R \approx 148$ ns
for the fixed-N case. Similar results hold for the time evolution of
$R_g$. Thus, the collapse is much faster than the barrier
limited folding process.
%
Monitoring the time evolution of $\Delta R$ and of
the number of native contacts, one can show (results not shown)
that the refolding of the trimer is staircase-like as observed in the
simulations \cite{Best_Science05,MSLi_BJ07} and the experiments
\cite{Fernandez_Sci04}.
\begin{figure}[!hbtp]
\epsfxsize=5.6in
\vspace{0.2in}
\linespread{0.8}
\centerline{\epsffile{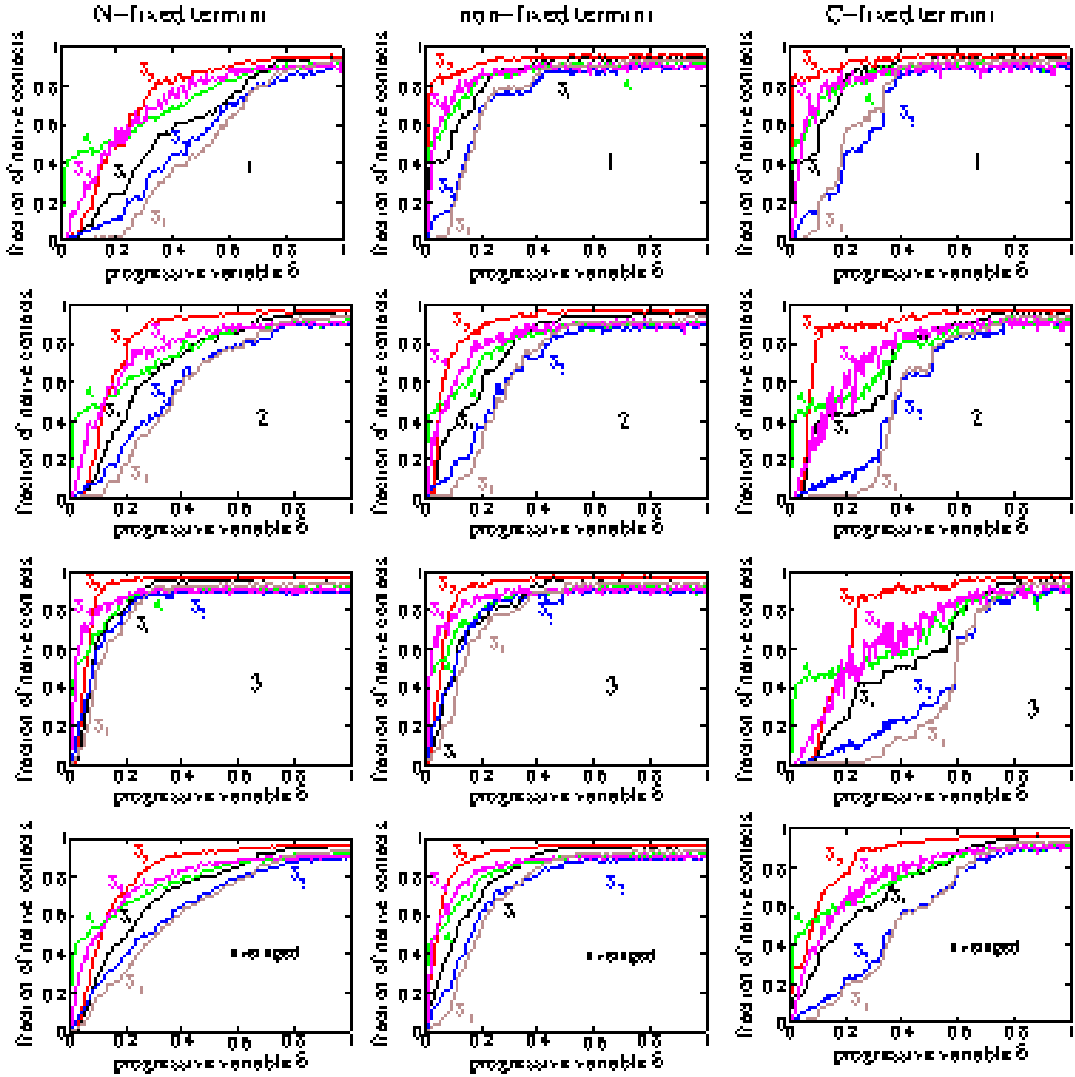}}
\caption{The same as in Fig. \ref{single_ub_pathways_fig} but for the trimer.
The numbers 1, 2 and 3 refer to the first, second and third domain.
The last row represents the results averaged over three domains.
The fractions of native contacts of each secondary
structure are averaged over 100 trajectories.}
\label{trimer_pathways_detail_fig}
\end{figure}

\begin{wrapfigure}{r}{0.46\textwidth}
\includegraphics[width=0.44\textwidth]{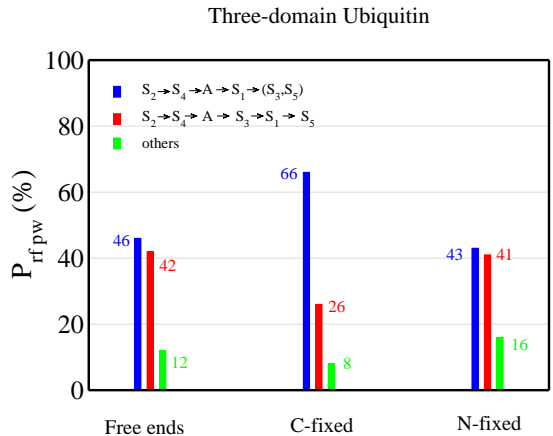}
\hfill
\begin{minipage}{6.9 cm}
\linespread{0.8}
\caption{(a) The probability of different refolding
pathways for the trimer. Each value is shown on top of
the histograms. \label{trimer_Prfpw_fig}}
\end{minipage}
\end{wrapfigure}

Fig. \ref{trimer_pathways_detail_fig} shows the dependence of the number
of native contacts of the secondary structures of each domain on $\delta$
for three situations: both termini are free and one or the other
of them is fixed.
In each of these cases the folding pathways of three domains
are identical. Interestingly, they are the same,
as given by Eq. (\ref{free-end_pathways_eq}), regardless
of we keep one end fixed or not.
As evident from Fig. \ref{trimer_Prfpw_fig},
although the dominant pathway is  the same for three cases its
probabilities are different. It is equal 68\%,
44\% and 43\% for the
C-fixed, free-end and N-fixed cases, respectively. For the last two cases,
the competing pathway
S$_2 \rightarrow$ S$_4 \rightarrow$ A $\rightarrow$ S$_3 \rightarrow$ S$_1 \rightarrow$ S$_5$
 has a reasonably  high
probability of $\approx$ 40\%.

The irrelevance of one-end fixation for refolding
pathways of a multi-domain Ub may be understood
as follows.
Recall that applying the low
quenched force to both termini does not change folding pathways of
single Ub \cite{MSLi_BJ07}. So in the three-domain case,
with the N-end of the first domain fixed,
both termini of the first and second domains are
effectively subjected to external force, and their pathways should remain the
same as in the free-end case. The N-terminal of the third domain is tethered
to the second domain but this would have much weaker effect compared to
the case when it is anchored to a surface. Thus this unit has almost free
ends and its pathways remain unchanged.
Overall, the "boundary" effect
gets weaker as the number of domains becomes
bigger. In order to check this argument, we have performed simulations for
the two-domain Ub. It turns out that the sequencing is roughly the same as
in Fig. \ref{trimer_pathways_detail_fig}, but the common tendency is less
pronounced (results not shown) compared to the trimer case.
Thus we predict that the force-clamp technique can probe
folding pathways of free Ub if one uses either the single domain with the C-terminus
anchored, or the multi-domain construction.

\begin{figure}[!htbp]
\epsfxsize=5.2in
\vspace{0.2in}
\centerline{\epsffile{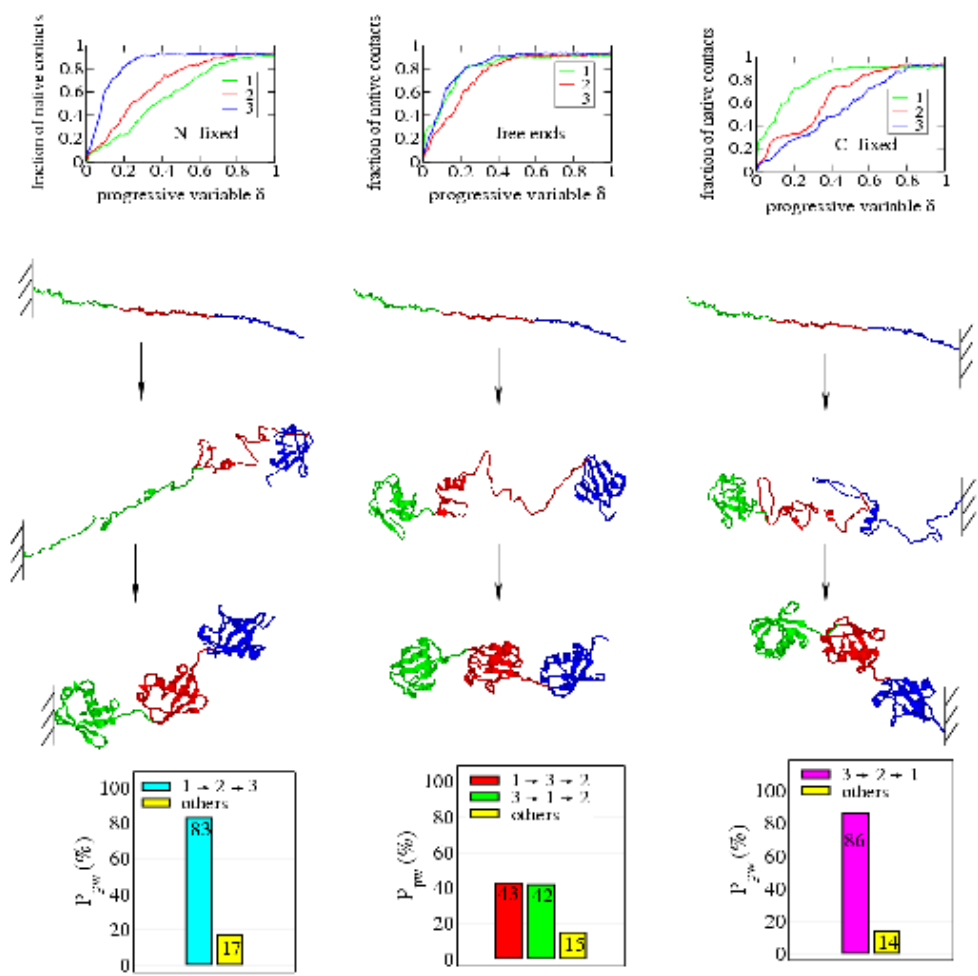}}
\linespread{0.8}
\caption{The dependence of the total number of native
contacts on $\delta$ for the first (green), second (red) and third
(blue) domains. Typical snapshots of the initial, middle and final
conformations for three cases when both two ends are free or one
of them is fixed. The effect of anchoring one terminus
on the folding sequencing of domains is clearly evident.
In the bottom we show the probability of refolding pathways for three
cases. Its value is written on the top of histograms.}
\label{trimer_pathways_overall_fig}
\vspace{5 mm}
\end{figure}

Although fixing one end of the trimer does not influence folding pathways
of individual secondary structures, it affects the folding sequencing
of individual domains (Fig. \ref{trimer_pathways_overall_fig}).
We have the following sequencing $(1,3) \rightarrow 2$,
$3 \rightarrow 2 \rightarrow 1$ and $1 \rightarrow 2 \rightarrow 3$
for the free-end, N-terminal fixed and C-terminal fixed, respectively.
These scenarios are supported by typical snapshots shown in
Fig. \ref{trimer_pathways_overall_fig}. It should be noted that the domain at
the free end folds first in all of three cases in statistical
sense (also true for the two-domain case).
As follows from the bottom of Fig. \ref{trimer_pathways_overall_fig}, if
two ends are free then each of them folds first in about 40 out of 100
observations. The middle unit may fold first, but with much lower probability
of about 15\%. This value remains almost unchanged when one of the ends
is anchored,
and the probability that
the non-fixed unit folds increases to $\ge 80$\%.


\subsection{Is the effect of fixing one terminus on refolding pathways universal?}

\begin{wrapfigure}{r}{0.49\textwidth}
\includegraphics[width=0.46\textwidth]{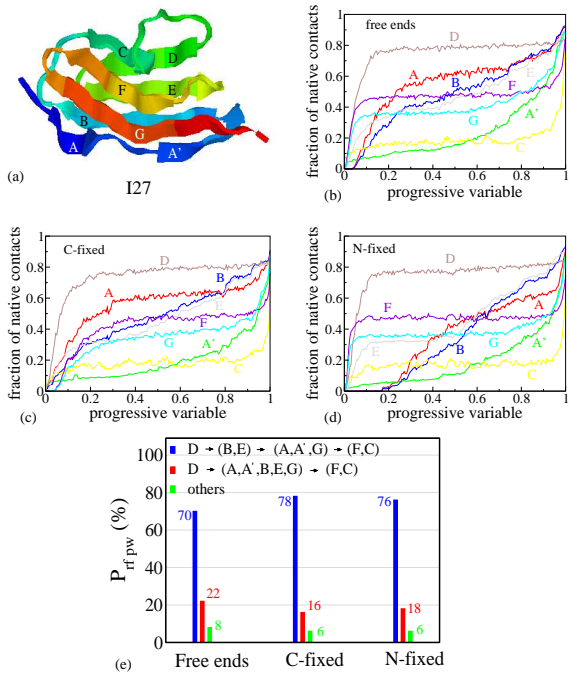}
\hfill\begin{minipage}{8.0 cm}
\linespread{0.8}
\caption{(a) NS conformation of Ig27 domain of titin(PDB ID: 1tit). There are 8 $\beta$-strands: A (4-7), A' (11-15),
B (18-25), C (32-36), D (47-52), E (55-61), F(69-75) and
G (78-88). The dependence of native contacts of different
$\beta$-strands on the progressive variable $\delta$ for the case
when two ends are free (b), the N-terminus is fixed (c) and the
C-terminal is anchored (d).
(e) The probability of observing refolding pathways for three
cases. Each value is written on top of the histograms. \label{titin_str_ref_pathways_fig}}
\end{minipage}
\end{wrapfigure}

We now ask if the effect of fixing one end
on refolding pathway, observed for Ub, is also valid for other proteins?
To answer this question, we study the single domain I27 from
the muscle protein titin.
We choose this protein as a good candidate
from the conceptual point of view
because its $\beta$-sandwich structure
(see Fig. \ref{titin_str_ref_pathways_fig}a) is very
different from $\alpha/\beta$-structure of Ub.

Moreover, because I27 is subject to
mechanical stress under physiological conditions
\cite{Erickson_Science97}, it is instructive to study
refolding from extended conformations generated by force.
There have been extensive unfolding (see recent review \cite{Sotomayor_Science07}
for
references) and refolding \cite{MSLi_PNAS06} studies
on this system, but the effect of one-end fixation on folding
sequencing of individual secondary structures have not been considered
either theoretically or experimentally.

As follows from Fig. \ref{titin_str_ref_pathways_fig}b,
if two ends are
free then strands A, B and E fold at nearly the same rate.
The pathways of the N-fixed and C-fixed cases are identical,
and they are almost the same as in the free end case
except that the strand A seems to fold after B and E.
Thus, keeping the N-terminus fixed has much weaker effect on the folding
sequencing as compared to the single Ub.
Overall the
effect of anchoring one terminus
has a little effect on the refolding pathways of I27, and
we have the following common sequencing
\begin{equation}
D \rightarrow (B,E) \rightarrow (A,G,A') \rightarrow F \rightarrow C
\end{equation}
for all three cases.
The probability of observing this main pathways varies between 70 and 78\%
(Fig. \ref{titin_str_ref_pathways_fig}e). The second pathway,
D $\rightarrow$ (A,A',B,E,G) $\rightarrow$ (F,C), has considerably lower
probability. Other minor routes to the NS are also possible.

Because the multi-domain construction weakens this effect, we expect that
the force-clamp spectroscopy can probe refolding pathways for a single and
poly-I27. More importantly, our study reveals that the influence of fixation
on refolding
pathways may depend on the native topology of proteins.

\subsection{Free energy landscape}

\begin{wrapfigure}{r}{0.47\textwidth}
\includegraphics[width=0.45\textwidth]{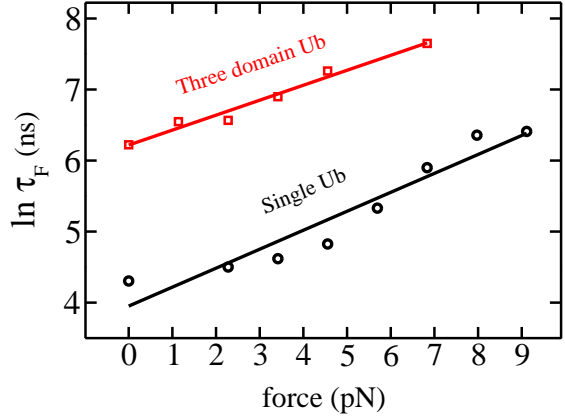}
\hfill\begin{minipage}{7.5 cm}
\linespread{0.8}
\caption{The dependence of folding times on the quench force at
$T=285$ K. $\tau_F$ was computed as the average of the first passage times
($\tau_F$ is the same as $\tau^Q_2$ extracted from the
bi-exponential fit for $<Q(t)>$). The result is averaged over 30 -
50 trajectories for each value of $f_q$. From the linear fits $y = 3.951
+ 0.267x$  and $y = 6.257 + 0.207x$ we obtain $x_f = 0.96 \pm 0.15$ nm for single Ub (black circles and curve) and $x_f = 0.74 \pm 0.07$ nm for trimer (red squares and curve), respectively.
 \label{refold_Ub_trimer}}
\end{minipage}
\end{wrapfigure}

Figure \ref{refold_Ub_trimer} shows the dependence of the folding
times on $f_q$. Using the Bell-type formula (Eq. \ref{Bell_Kf_eq}) and
the linear fit  in Fig. \ref{refold_Ub_trimer}, we obtain $x_f =  0.96 \pm 0.15$
 nm which is in acceptable agreement with the
experimental value $x_f \approx 0.8$ nm
\cite{Fernandez_Sci04}.

The linear growth of the free energy barrier to folding with $f_q$
is due to
the stabilization of the random coil states under the force.
 Our estimate for Ub is higher than
$x_f \approx 0.6$ nm obtained for I27 \cite{MSLi_PNAS06}.
One of possible reasons for such a pronounced difference is that we used
the cutoff distance $d_c=0.65$ and 0.6 nm in the Go model for Ub and I27, respectively. The larger value of $d_c$ would make a protein
more stable (more native contacts) and it may change the FEL
leading to enhancement of $x_f$. This problem requires
further investigation.

From Fig. \ref{refold_Ub_trimer} we obtain $x_f = 0.74 \pm 0.07$ nm
for trimer. Within the error bars this value coincides with
$x_f = 0.96 \pm 0.15$ nm  for Ub, and also with the experimental result $x_f \approx 0.80$ nm
\cite{Fernandez_Sci04}. Our results suggest that the multi-domain structure
leaves  $x_f$ almost unchanged.

\subsection{Conclusions}

We have shown that, in agreement with the experiments \cite{Fernandez_Sci04},
refolding of Ub under quenched force proceeds in a stepwise manner.
The effect of the one-terminal fixation on refolding pathways
depends on individual protein and it gets weaker
by a multi-domain construction. 
Our theoretical estimate of $x_f$ for single Ub
is close to the experimental one and it remains almost the same for
three-domain case.

\newpage

\begin{center}
\section{Mechanical and thermal unfolding of single and three domain Ubiquitin}
\end{center}

\subsection{Introduction}
Experimentally,
the unfolding  of the poly-Ub has been studied by applying a constant
force \cite{Schlierf_PNAS04}.
The mechanical unfolding of Ub has previously investigated using Go-like
\cite{West_BJ06} and all-atom models \cite{West_BJ06,Irback_PNAS05}.
In particular,
Irb\"ack {\em et al.} have explored mechanical
unfolding pathways 
of structures A, B, C, D and E
(see the definition of these structures and the $\beta$-strands
 in the caption to 
Fig. \ref{ubiquitin_struture_fig}) and the existence of intermediates in detail.
We present our results on mechanical unfolding of Ub
for five following reasons.

\begin{enumerate}
\item The barrier to the mechanical unfolding has not been computed.

\item Experiments of Schlierf {\em et al.} \cite{Schlierf_PNAS04} have suggested that
cluster 1 (strands S1, S2 and the helix A) unfolds after cluster 2
(strands S3, S4 and S5). However, this observation has not yet
been studied theoretically.

\item Since the structure C, which consists of the strands S1 and S5,
unzips first, Irb\"ack {\em et al.} pointed out that the strand S5 unfolds before S2 or the
terminal strands follows the unfolding pathway 
S1 $\rightarrow$ S5 $\rightarrow$ S2. This conclusion may be incorrect because
it has been obtained from the breaking of the contacts within the structure C.

\item In pulling and force-clamp experiments the external force is applied
to one end of proteins whereas the other end is kept fixed. Therefore, 
one important question emerges is how fixing one terminus affects the unfolding
sequencing of Ub. This issue has not been addressed by Irb\"ack {\em et al.}
\cite{Irback_PNAS05}. 

\item Using a simplified all-atom model it was shown \cite{Irback_PNAS05}
that mechanical intermediates occur more frequently than
in experiments \cite{Schlierf_PNAS04}. It is relevant to ask
if a C$_{\alpha}$-Go model can capture similar intermediates as this may shed
light on the role of non-native interactions.
\end{enumerate}

From the force dependence of mechanical unfolding times, we
estimated the distance between the NS and the TS to be  $x _{u} \approx 0.24$ nm which is close to the
experimental results of Carrion-Vazquez {\em et al.}
\cite{Carrion-Vazquez_NSB03} and Schlierf {\em et al.}
\cite{Schlierf_PNAS04}. In agreement with the experiments
\cite{Schlierf_PNAS04}, cluster 1 was found to unfold after cluster 2
in our simulations.
Applying the force to the both termini,
we studied the mechanical unfolding pathways of the terminal strands
in detail and obtained the sequencing S1 $\rightarrow$ S2 $\rightarrow$ S5
which is different from the result of Irb\"ack {\em et al.}.
When the N-terminus is fixed and the C-terminus is pulled by a
constant force the unfolding sequencing was found to be very different
from the previous case. The unzipping initiates, for example,
from the C-terminus
but not from the N-one. Anchoring the C-end is shown to have a little effect
 on unfolding pathways.
We have
demonstrated that the present C$_{\alpha}$-Go model does not capture rare
mechanical intermediates, presumably due to the lack of non-native interactions.
Nevertheless, it can correctly describe the two-state 
unfolding of Ub \cite{Schlierf_PNAS04}.

It is well known that thermal unfolding pathways may be very different
from the mechanical ones, as has been shown for the
domain I27 \cite{Paci_PNAS00}.
This is because the force is applied locally to the termini while 
thermal fluctuations have the
global effect on the entire protein. In the force case unzipping should
propagate from the termini whereas under thermal fluctuations the most
unstable part of a polypeptide chain unfolds first.

The unfolding of Ub under thermal fluctuations was investigated experimentally
by Cordier and Grzesiek \cite{Cordier_JMB02} and by Chung {\em et al.}
\cite{Chung_PNAS05}. If one assumes that unfolding is the reverse of the
 refolding process then one can infer information about the unfolding
pathways from the experimentally determined $\phi$-values \cite{Went_PEDS05}
and $\psi$-values \cite{Krantz_JMB04,Sosnick_ChemRev06}.
The most comprehensive $\phi$-value
analysis is that of Went and Jackson. They found  that the
C-terminal region which has very low $\phi$-values unfolds first and then the
strand S1 breaks before full unfolding of the $\alpha$ helix fragment A occurs.
However, the detailed unfolding sequencing of the other strands remains unknown.
 
Theoretically, the thermal unfolding of Ub at high temperatures has been
studied by all-atom
MD simulations by Alonso and Daggett
\cite{Alonso_ProSci98} and Larios {\em et al.} \cite{Larios_JMB04}. In
the latter work the unfolding pathways were not explored. Alonso and Daggett
have found that the $\alpha$-helix fragment A is the most resilient towards
temperature but the structure B breaks as early as the structure C.
The fact that B unfolds early contradicts not only the results for the
$\phi$-values obtained experimentally by Went and Jackson \cite{Went_PEDS05}
but also findings from a high resolution
NMR \cite{Cordier_JMB02}. Moreover, the
sequencing of unfolding events for the structures D and E was not studied.

What information about the thermal unfolding
pathways of Ub can be inferred from the folding
simulations of various coarse-grained models?  
Using a semi-empirical approach Fernandez predicted \cite{Fernandez_JCP01}
that the nucleation site involves the $\beta$-strands S1 and S5. This 
suggests that thermal fluctuations break
these strands last but
what happens to the other parts of the protein remain unknown.
Furthermore, the late breaking of S5 contradicts the unfolding
\cite{Cordier_JMB02} and folding \cite{Went_PEDS05} experiments.
From later folding simulations of Fernandez {\em et al.} 
\cite{Fernandez_Proteins02,Fernandez_PhysicaA02} one can infer
that the structures
A, B and C unzip late. Since this information is gained from $\phi$-values,
it is difficult to determine the sequencing of unfolding events even for these
fragments.
Using the results of Gilis and Rooman \cite{Gilis_Proteins01} we can
only expect that
the structures A and B unfold last. In addition,
with the help of a three-bead model it was found
\cite{Sorenson_Proteins02} that the C-terminal
loop structure is the last to fold in the folding process and most
likely plays a spectator role in the folding kinetics. This implies that
the strands S4, S5 and the second helix (residues 38-40) would unzip first
but again the full unfolding sequencing can not be inferred from this study.

Thus, neither the  direct MD \cite{Alonso_ProSci98} nor
indirect folding simulations \cite{Fernandez_JCP01,Fernandez_Proteins02,Fernandez_PhysicaA02,Gilis_Proteins01,Sorenson_Proteins02}
provide a complete picture of the thermal unfolding pathways for Ub.
One of our aims is to decipher the complete thermal unfolding sequencing
and compare
it with the mechanical one.
The mechanical and thermal routes to the DSs have been found
to be very different from each other.
Under the force the $\beta$-strand S1, e.g.,
unfolds first, while thermal fluctuations detach strand S5 first.
The later observation
is in good agreement with NMR data of Cordier
and Grzesiek \cite{Cordier_JMB02}.
A detailed comparison with available experimental and simulation
data on the unfolding sequencing will be presented.
The free energy barrier to thermal
unfolding was also calculated.

Another part of this chapter was inspired by the recent
pulling experiments of Yang {\em et al.} \cite{Yang_RSI06}.
Performing the
experiments in the temperature interval between 278 and
318 K, they found that the unfolding
force (maximum force in the force-extension profile), $f_u$,
of Ub  depends on temperature
linearly. In addition, the corresponding slopes of the linear behavior
have been found to be
independent of pulling velocities.
An interesting question that arises is if the linear dependence
of $f_u$ on $T$ is valid for this
particular region, or it holds for the whole temperature interval.
Using the same Go model \cite{Clementi_JMB00}, we can reproduce the
experimental results of Yang {\em et al.} \cite{Yang_RSI06}
 on the quasi-quantitative level.
More importantly, we have shown that for the entire temperature
interval the dependence is not linear, because a protein is not an
entropic spring in the temperature regime studied.

We have
 studied the effect of multi-domain construction
 and linkage
on the location
of the TS along the end-to-end distance reaction
coordinate, $x_u$.
It is found that the multi-domain construction has a minor effect on $x_u$
but,
in agreement with the experiments
\cite{Carrion-Vazquez_NSB03}, the Lys48-C linkage has
the strong effect on it.
Using the microscopic theory for unfolding dynamics \cite{Dudko_PRL06},
we have determined the unfolding barrier for Ub.

This chapter is based on the results presented in Refs. \cite{MSLi_BJ07, Kouza_JCP08}.

\subsection{Materials and Methods}
We use the Go-like model (Eq. \ref{Hamiltonian})
for the single as well as multi-domain Ub.
%
It should be noted that the folding thermodynamics
does not depend on the environment viscosity (or on $\zeta$)
but the folding kinetics depends
on it. Most of our simulations (if not stated otherwise)
were performed at the friction
$\zeta = 2\frac{m}{\tau_L}$, where the folding is fast.
 The
 equations of motion
were integrated using the velocity form
of the Verlet algorithm \cite{Swope_JCP82}
with the time step $\Delta t = 0.005 \tau_L$
(Chapter 3). In order to check
the robustness
of our predictions for refolding pathways, limited computations
were carried out
for the friction $\zeta = 50\frac{m}{\tau_L}$ which is believed
to correspond to the viscosity of water \cite{Veitshans_FD97}).
In this overdamped limit we use the Euler method
(Eq. \ref{Euler}) for integration
and the time step $\Delta t = 0.1 \tau_L$.

The progressive variable $\delta$
(Eq. \ref{progress_unfold_eq}) was used to probe folding pathways.
In the constant velocity force simulation, we fix the N-terminal
and follow the procedure described in Section 3.1.2.
 The pulling speeds are set equal
$\nu = 3.6\times 10^7$ nm/s and 4.55 $\times 10^8$ nm/s which are
about 5 - 6 orders of magnitude faster than those used in experiments
\cite{Yang_RSI06}.

\subsection{Mechanical unfolding pathways}
\subsubsection{Absence of mechanical unfolding intermediates in C$_{\alpha}$-Go model}

 In order to study the unfolding dynamics
of Ub,  Schlierf {\em et al.} \cite{Schlierf_PNAS04} have
performed the AFM experiments at a constant force $f = 100, 140$
and 200 pN.  The unfolding intermediates were recorded in about $5
\%$ of 800 events at different forces. The typical distance
between the initial and intermediate states is $\Delta R = 8.1 \pm
0.7$ nm \cite{Schlierf_PNAS04}. However, the intermediates do not
affect the two-state behavior of the polypeptide chain. Using the
all-atom models Irb\"ack {\em et al.} \cite{Irback_PNAS05} have also
observed the intermediates in the region 6.7 nm $< R < 18.5$ nm.
Although the percentage of intermediates is higher than in the
experiments, the two-state unfolding events remain dominating.
To check the existence of force-induced intermediates in our model, we
have performed the unfolding simulations for $f=70, 100,
140$ and 200 pN. Because the results are qualitatively similar for
all values of force, we present $f=100$ pN case only.

Figure \ref{uf100pN_long_fig}a shows the time dependence of $R(t)$
for fifteen runs starting from the native value $R_N \approx 3.9$ nm.
For all trajectories the plateau occurs at $R \approx 4.4$ nm. As
seen below, passing this plateau corresponds to breaking of intra-structure
native contacts of structure C. At this stage the chain ends get almost
stretched out, but the rest of the polypeptide chain remains
native-like. The plateau is washed out when we average over many
trajectories and $<R(t)>$ is well fitted by a single exponential
 (Fig. \ref{uf100pN_long_fig}a), in accord
with the two-state behavior of Ub \cite{Schlierf_PNAS04}.

The existence of the plateau observed for individual unfolding
events in Fig. \ref{uf100pN_long_fig}a agrees with the all-atom
simulation results of Irb\"ack {\em et al.} \cite{Irback_PNAS05} who
have also recorded the similar plateau at $R \approx 4.6$ nm at
short time scales. However unfolding intermediates at larger
extensions do not occur in our simulations. This is probably
related to neglect  of the non-native interactions in the
C$_{\alpha}$-Go model. Nevertheless, this simple model provides
the correct two-state unfolding picture of Ub in the statistical
sense.


\subsubsection{Mechanical unfolding pathways: force is applied to both termini}

\begin{figure}[!hbtp]
\epsfxsize=4.2in
\vspace{0.2in}
\centerline{\epsffile{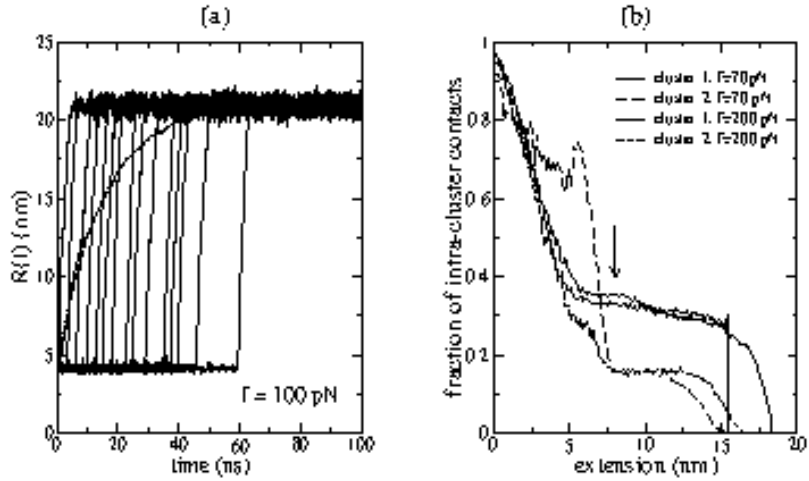}}
\linespread{0.8}
\caption{ (a) Time dependence of the end-to-end distance for $f=100$ pN.
The thin curves refer to fifteen representative trajectories. The averaged
over 200 trajectory $<R(t)>$ is represented by the thick line. The dashed curve
is the single exponential fit $<R(t)> = 21.08 -16.81\exp(-x/\tau_{u})$,
where $\tau_{u}\approx 11.8$ ns. (b) The dependence of fraction of the native contacts on $\Delta R$
for cluster 1 (solid lines) and cluster 2 (dashed lines) at
$f=70 pN$  and 200 pN.
The results are averaged over 200 independent
trajectories.
The arrow points to $\Delta R$ = 8.1 nm.}
\label{uf100pN_long_fig}
\end{figure}

Here we focus on the
mechanical unfolding pathways by monitoring the number of native
contacts as a function of the end-to-end extension $\Delta R
\equiv R-R_{\rm eq}$, where $R_{\rm eq}$ is the
equilibrium value of $R$. For $T=285$ K, $R_{\rm eq} \approx 3.4$ nm.
Following Schlierf {\em et al.}
\cite{Schlierf_PNAS04}, we first divide Ub into two clusters.
Cluster 1 consists of strands S1, S2 and the helix A (42 native
contacts) and cluster 2 - strands S3, S4 and S5 (35 native
contacts). The dependence of fraction of intra-cluster native
contacts is shown in Fig. \ref{uf100pN_long_fig}b
 for $f = 70$
and 200 pN (similar results for $f = 100$  and 140 pN are not shown). In
agreement with the experiments \cite{Schlierf_PNAS04} the cluster
2 unfolds first. The unfolding of these clusters becomes more and
more synchronous upon decreasing $f$. At $f = 70$ pN the competition
with thermal
fluctuations becomes so important that two clusters may unzip
almost simultaneously. Experiments at low forces are needed to
verify this observation.

The arrow in Fig. \ref{uf100pN_long_fig}b
 marks the position
$\Delta R = 8.1$ nm,
where some intermediates were recorded in the experiments
\cite{Schlierf_PNAS04}. At this point
there is intensive loss of native contacts of the cluster 2 suggesting that
the intermediates observed on the experiments are conformations in which
most of the contacts of this
cluster are already broken but the cluster 1 remains relatively
structured ($\approx 40\%$ contacts). One can expect that the cluster 1
is more ordered in the intermediate conformations if the side chains and
realistic interactions between amino acids are taken into account.

\begin{figure}[!htbp]
\epsfxsize=6.3in
\vspace{0.2in}
\centerline{\epsffile{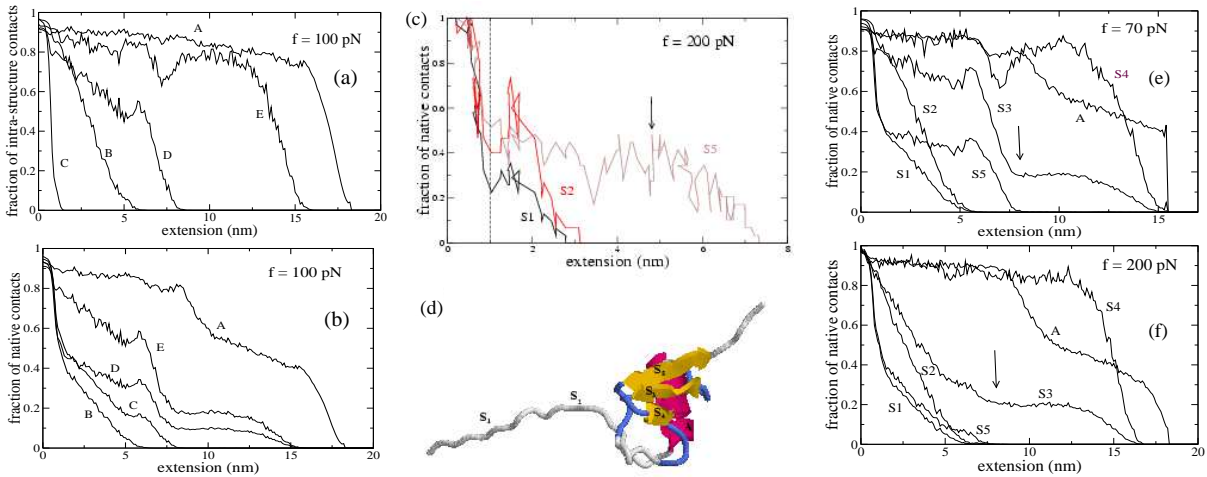}}
\linespread{0.8}
\caption{(a) The dependence of fraction of the intra-structure native contacts
on $\Delta R$ for structures A, B, C, D and E  at
$f=100$ pN. (b) The same as in a) but for all native contacts.
(c) The dependence of fraction of the native contacts on $\Delta R$
for strand S1, S2 and S5
($f=200 pN$). The vertical dashed line marks the position of the plateau
at $\Delta R \approx$ 1 nm. (d) The snapshot, chosen at the extension
marked by the arrow in c), shows that S2 unfolds before S5.
At this point all native contacts of S1 and S2 have already broken
while 50$\%$ of the native contacts of S5 are still present. (e) The dependence of fraction of the native contacts on extension
for A and all  $\beta$-strands at
$f=70 pN$. (f) The same as in e) but for $f=200$ pN. The arrow points to
$\Delta R = 8.1$ nm where the intermediates are recorded on the experiments
\cite{Schlierf_PNAS04}. The results are averaged over 200 trajectories.}
\label{dom_ext_100pN_fig}
\end{figure}

To compare the mechanical unfolding pathways of Ub with the
all-atom simulation results \cite{Irback_PNAS05} we discuss the
sequencing of helix A and structures B, C, D and E in more detail. We
monitor the intra-structure native contacts and all contacts
separately. The later include not only the contacts within a given
structure but also the contacts between it and the rest of the
protein. It should be noted that Irb\"ack {\em et al.} have studied
the unfolding pathways based on the evolution of the
intra-structure contacts. Fig. \ref{dom_ext_100pN_fig}a shows the
dependence of the fraction of intra-structure contacts on $\Delta
R$ at $f=100$ pN. At $\Delta R \approx $ 1nm, which corresponds to
the plateau in Fig. \ref{uf100pN_long_fig}a, most of the contacts
of C are broken.
In agreement with the all-atom simulations
\cite{Irback_PNAS05}, the unzipping follows C $\rightarrow$ B
$\rightarrow$ D $\rightarrow$ E $\rightarrow$ A. Since C consists
of the terminal strands S1 and S5, it was suggested that these
fragments unfold first. However, this scenario may be no
longer valid if one considers not only intra-structure
contacts but also other possible ones (Fig.
\ref{dom_ext_100pN_fig}{\em b}). In this case the statistically
preferred sequencing is B $\rightarrow$ C $\rightarrow$ D
$\rightarrow$ E $\rightarrow$ A which holds not only for $f$=100
pN but also for other values of $f$. If it is true then S2 unfold
even before S5. To make this point more transparent,
 we plot the fraction of contacts for S1, S2 and S5 as a
 function of $\Delta R$ (Fig.
\ref{dom_ext_100pN_fig}{\em c})
for a typical trajectory.
Clearly, S5 detaches from the core part of a protein
after S2 (see also the snapshot
in Fig.
 \ref{dom_ext_100pN_fig}{\em d}).
So, instead of the sequencing S1 $\rightarrow$ S5 $\rightarrow$ S2
proposed by Irb\"ack {\em et al.}, we obtain
S1 $\rightarrow$ S2 $\rightarrow$ S5.

The dependence of the fraction
of native contacts on $\Delta R$ for individual strands
is shown in Fig. \ref{dom_ext_100pN_fig}{\em e}
 ($f=70$ pN) and
Fig.  \ref{dom_ext_100pN_fig}{\em f}
($f$=200 pN). At $\Delta = 8.1$ nm contacts of S1, S2 and S5 are already broken
whereas S4 and A remain largely structured. In terms of $\beta$-strands and A
we can interpret the intermediates observed in the experiments of
Schlierf {\em et al.} \cite{Schlierf_PNAS04} as conformations with
well structured S4 and A, and low ordering of S3. This interpretation is
more precise compared to the above argument
based on unfolding of two clusters because if one considers the average
number of native contacts, then
the cluster 2 is unstructured in the IS
(Fig. \ref{uf100pN_long_fig}b,
 but its strand S4 remains highly structured
(Figs. \ref{dom_ext_100pN_fig}{\em e-f}).

From Figs. \ref{dom_ext_100pN_fig}{\em e-f}
 we obtain the following mechanical unfolding
sequencing
\begin{equation}
{\rm S1} \rightarrow {\rm S2} \rightarrow {\rm S5} 
 \rightarrow {\rm S3} \rightarrow {\rm S4} \rightarrow {\rm A}.
\label{mechanical_sequencing}
\end{equation}
It should be noted that the sequencing
(\ref{mechanical_sequencing}) is valid in the statistical  sense.
In some trajectories S5 unfolds even before S1 and S2 or the
native contacts of S1, S2 and S5 may be broken at the same time
scale (Table \ref{SimTime_trimer}).
\begin{table}
\begin{tabular}{c|c|c|c} \hline
Force (pN)~&~S1 $\rightarrow$ S2 $\rightarrow$ S5 ($\%$) &~S5
$\rightarrow$ S1 $\rightarrow$ S2 ($\%$)& (S1,S2,S5) ($\%$) \\\hline
70 & 81 & 8 & 11\\
100 & 76 & 10 & 14\\
140 & 53 & 23 & 24\\
200 & 49 & 26 & 25\\\hline
\end{tabular}
\linespread{0.8}
\caption{Dependence of unfolding pathways on the external
 force. There are three possible scenarios: S1 $\rightarrow$
 S2 $\rightarrow$ S5, S5 $\rightarrow$ S1 $\rightarrow$ S2, and three strands
unzip almost simultaneously (S1,S2,S5). The probabilities of
observing these events are given in percentage.\label{SimTime_trimer}}
\vspace{5 mm}
\end{table}
From the Table \ref{SimTime_trimer}
it follows that the probability of having
S1 unfolded first decreases with lowering $f$ but the main
trend Eq. (\ref{mechanical_sequencing}) remains unchanged.
One has to stress again that the sequencing of the terminal strands
S1, S2 and S5 given by Eq. (\ref{mechanical_sequencing}) is different from
what proposed by Irb\"ack {\em et al.} based on the breaking of 
the intra-structure contacts of C.
Unfortunately, there are no experimental data available for
comparison with our theoretical prediction.

\subsubsection{Mechanical unfolding pathways: One end is fixed}

{\em N-terminus is fixed}.
Here we adopted the same procedure as in the previous section
except the N-terminus is held fixed during simulations. As in the process
where both of the termini are subjected to force, one can show that
the cluster 1
unfolds after the cluster 2 (results not shown).

\begin{figure}
\epsfxsize=3.5in
\vspace{0.2in}
\centerline{\epsffile{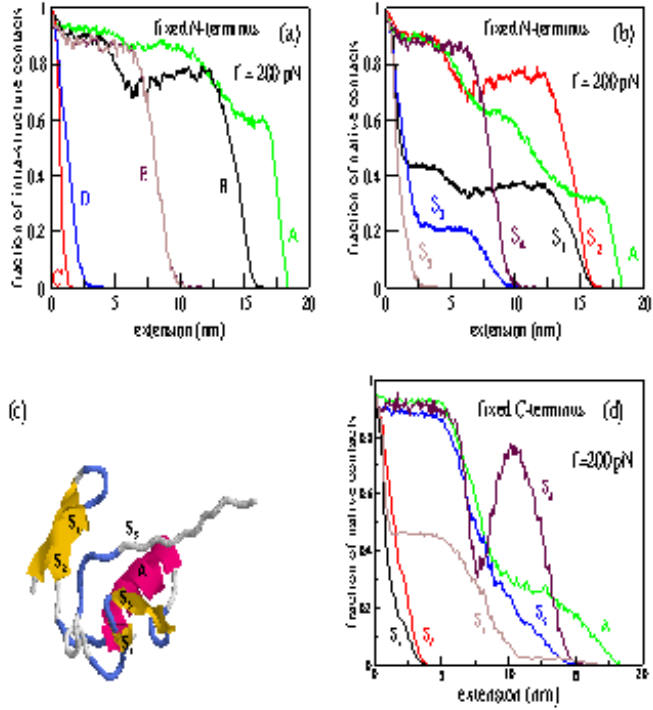}}
\linespread{0.8}
\caption{(a) The dependence of fraction of the
intra-structure native contacts on extension
for all  structures at
$f=200 pN$. The N-terminus is fixed and the external force is applied via
the C-terminus. (b) The same as in (a) but for the native contacts of all
individual $\beta$-strands and helix A .
The results are averaged over 200 trajectories.
(c) A typical snapshot which shows that S$_5$ is fully detached from
the core while S$_1$
 and S$_2$ still have $\approx 50\%$ and 100\% contacts, respectively.
(d) The same as in (b) but the C-end is anchored and N-end is pulled.
The strong drop in the fraction of native contacts of S$_4$ at $\Delta R \approx 7.5$ nm does not correspond
to the substantial change of structure as it has only 3 native
contacts in total.}
\label{cont_snap_fixN_f200pN_fig}
\end{figure}

 From Fig. \ref{cont_snap_fixN_f200pN_fig}
we obtain the following unfolding pathways
\begin{subequations}
\begin{equation}
{\rm C} \rightarrow {\rm D} \rightarrow {\rm E} \rightarrow {\rm B} \rightarrow {\rm A},
\label{mechan_fixN_sequencing_struc}
\end{equation}
\begin{equation}
{\rm S5} \rightarrow {\rm S3} \rightarrow {\rm S4} \rightarrow {\rm S1} \rightarrow {\rm S2}
 \rightarrow {\rm A},
\label{mechan_fixN_sequencing}
\end{equation}
\end{subequations}
which are also valid for the other values of force ($f$=70, 100 and 140 pN).
Similar to the case when the force is applied to both ends, the structure
C unravels first and the helix A remains the most stable. However, the
sequencing of B, D and E changes markedly compared to the result
obtained by Irb\"ack {\em et al} \cite{Irback_PNAS05} 
(Fig. \ref{dom_ext_100pN_fig}a).

As evident from Eqs. \ref{mechanical_sequencing} and \ref{mechan_fixN_sequencing}, anchoring the first terminal has a much more pronounced effect on the 
unfolding pathways of individual strands. In particular,
unzipping commences from the
C-terminus instead of from the N-one.
Fig. \ref{cont_snap_fixN_f200pN_fig}{\em c} shows a typical snapshot
where one can see clearly that S$_5$ detaches first. At the first glance,
this fact may seem trivial because S$_5$ experiences the external force
directly.
However,  our experience on unfolding pathways of the well studied domain I27
from the human cardiac titin, e.g., shows that it may be not the case.
Namely, as follows from the pulling experiments \cite{Marszalek_Nature99}
and simulations \cite{Lu_Proteins99},
the strand A from the N-terminus 
 unravels first although this terminus
is kept fixed. From this point of view, what strand of Ub detaches first
is not {\em a priory} clear. In our opinion, it depends on the
interplay between the native topology and the speed of tension propagation.
The later factor probably plays a more important role for Ub while the
opposite situation happens with I27.
One of possible reasons is related to the high stability of the helix A
which does not allow either for the N-terminal to unravel first or
for seriality in unfolding starting from the C-end. 

{\em C-terminus is fixed}.
One can show that unfolding pathways of structures A,B, C, D and E remain
exactly the same as in the case when Ub has been pulled from both termini
(see Fig. \ref{dom_ext_100pN_fig}{\em a-b}). Concerning the individual strands,
a slight difference is observed for S$_5$ 
(compare Fig. \ref{cont_snap_fixN_f200pN_fig}{\em d} and
Fig. \ref{dom_ext_100pN_fig}{\em e}). Most of the native contacts of this domain break
before S$_3$ and S$_4$, except
the long tail at extension $\Delta R \gtrsim$ 11 nm due to high
mechanical stability of
only one
contact between residues 61 and 65 (the highest resistance of
this pair is probably due to the fact that among 25 possible contacts
of S$_5$ it has the shortest distance $|61-65|=4$ in sequence).
This scenario holds in about 90\%
of trajectories whereas
S$_5$ unravels 
completely earlier than S$_3$ and S$_4$ in the remaining trajectories.
Thus, anchoring C-terminus has much less effect on unfolding pathways compared
to the case when the N-end is immobile.

It is worthwhile to note that, experimentally one has studied the effect of
 extension geometry on the 
mechanical stability of Ub fixing its C-terminus \cite{Carrion-Vazquez_NSB03}.
The greatest mechanical strength (the longest unfolding time) occurs when
the protein is extended between N- and C-termini. This result has been
supported by Monte Carlo \cite{Carrion-Vazquez_NSB03} as well as MD
\cite{West_BJ06} simulations. However the mechanical
unfolding sequencing has not
been studied yet. It would be interesting to check our results on the
effect of fixing one end on Ub mechanical unfolding pathways
by experiments.

\subsection{Free energy landscape}

In experiments one usually uses the Bell formula \cite{Bell_Sci78}
(Eq. \ref{Bell_Ku_eq})
to extract $x_u$ for two-state proteins from the force dependence
 of unfolding times. This formula is valid if the location of
the TS does not move under external force.
However, under external force the TS moves toward NS. In this case, one can
use Eq. (\ref{Dudko_eq})
to estimate not only
$x_u$ but also $G^{\ddagger}$ for $\nu = 1/2$ and 2/3.
This will be done in this section for the single Ub and the trimer.

\begin{figure}
\epsfxsize=4.1in
\centerline{\epsffile{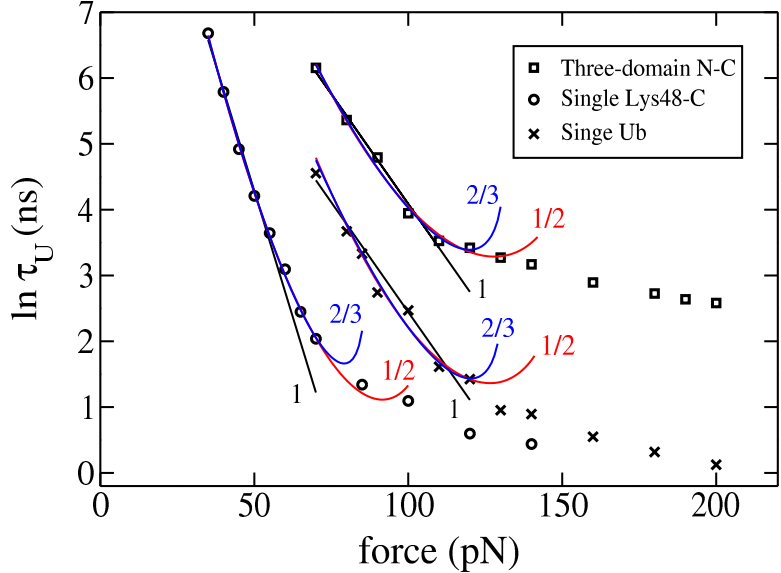}}
\linespread{0.8}
\caption{The semi-log plot for the force dependence of unfolding times at $T=285$ K.
Crosses and squares refer the the single Ub and the trimer
with the force applied to N- and C-terminal, respectively.
Circles refer to the
single Ub with the force applied to Lys48 and C-terminal.
Depending on $f$, 30-50 folding events
were using for averaging. In the Bell approximation,
if the N- and C-terminal of the trimer are pulled then
we have the  linear fit $y = 10.448 - 0.066x$ (black line) and $x_u \approx$ 0.24 nm.
The same value of $x_u$ was obtained for the single Ub \cite{MSLi_BJ07}.
In the case when we pull at Lys48 and C-terminal of single Ub the linear fit
(black line) at
low forces is $y = 11.963 - 0.168x$ and $x_u = 0.61$ nm.
The correlation level of fitting is about 0.99.
The red and blue curves correspond to the fits with $\nu =1/2$ and
$2/3$, respectively (Eq. \ref{Dudko_eq2}).}
\label{refold_unfold_vs_force_fig}
\end{figure}

\subsubsection{Single Ub}

Using the Bell approximation and Fig. \ref{refold_unfold_vs_force_fig},
 we have $x_u \approx 2.4 \AA \,$
\cite{MSLi_BJ07,Kouza_JCP08} which is consistent
with the experimental data $x_u = 1.4 - 2.5$ \AA
\cite{Carrion-Vazquez_NSB03,Schlierf_PNAS04,Chyan_BJ04}.
With the help of an all-atom simulation Li {\em et al.} \cite{Li_JCP04}
have shown that $x_u$ does depend on $f$. At low forces,
where the Bell approximation
is valid \cite{MSLi_BJ07}, they
obtained $x_u = 10$ \AA , which is noticeably higher than our and
the experimental value. Presumably,
this is due to the fact
that these authors computed $x_u$ from equilibrium
data, but their sampling was not good enough for such a long protein as Ub.

We now use Eq. (\ref{Dudko_eq2}) with
$\nu = 2/3$ and $\nu = 1/2$ to compute $x_u$ and $\Delta G^{\ddagger}$.
The regions,
where the $\nu = 2/3$ and $\nu = 1/2$ fits works well, are wider  than that for
the Bell scenario (Fig. \ref{refold_unfold_vs_force_fig}). However these fits
can not to cover the entire
force interval. The values of $\tau _u^0, x_u$ and $\Delta G^{\ddagger}$ obtained from
the fitting procedure are listed in Table \ref{Dudkotable}.
According to Ref. \onlinecite{Dudko_PRL06},
all of these quantities increase with decreasing $\nu$.
In our opinion, the microscopic theory
($\nu = 2/3$ and $\nu = 1/2$) gives too high a value for
$x_u$ compared to its typical
experimental value \cite{Carrion-Vazquez_NSB03,Schlierf_PNAS04,Chyan_BJ04}.
However, the latter was calculated from fitting experimental
data to the Bell formula,
and it is not clear how much the microscopic theory would change the result.

\begin{center}\begin{table}[!htbp] 

\begin{tabular}{|c|*{9}{c|}}
\hline
& \multicolumn{3}{c|}{Ub}& \multicolumn{3}{c|}{Lys48-C}& \multicolumn{3}{c|}{trimer}\\
\hline
$\nu $& 1/2& 2/3& 1& 1/2 &2/3 &1& 1/2& 2/3 & 1\\
$\quad\tau_U^0 (\mu s)\quad$&13200&1289&9.1&4627&2304&157&1814&756&47\\
$x_{u} (\AA)\quad$ &7.92&5.86&2.4&12.35&10.59&6.1&6.21&5.09&2.4\\
$\quad\Delta G^{\ddagger}(k_BT)\quad$&17.39&14.22&-&15.90&13.94&-&13.49&11.64&-\\
\hline
\end{tabular}
\linespread{0.8}
\caption{Dependence of $x_u$ on fitting procedures for the three-domain Ub and Lys48-C. $\nu =1$ corresponds to the phenomenological Bell approximation (Eq. \ref{Bell_Ku_eq}). $\nu = 1/2$ and 2/3 refer to the microscopic theory (Eq. \ref{Dudko_eq2}). For Ub and trimer
the force is applied to both termini.\label{Dudkotable}}
\end{table} 
\end{center}

In order to estimate the unfolding barrier of Ub from the available
experimental data
and compare it with our theoretical estimate, we use the
following formula
\begin{equation}
\Delta G^{\ddagger} = -k_BT\ln(\tau _A/\tau _u^0)
\label{UnfBarrier_eq}
\end{equation}
where $\tau _u^0$ denotes the unfolding time in the absence of force and
$\tau _A$ is a typical unfolding prefactor. Since $\tau _A$ for unfolding is
not known, we use the typical value for folding $\tau _A = 1 \mu$s
\cite{MSLi_Polymer04,Schuler_Nature02}.
Using $\tau _u^0 = 10^4/4$ s
\cite{Khorasanizadeh_Biochem93} and Eq. (\ref{UnfBarrier_eq}) we obtain
$\Delta G^{\ddagger} =  21.6 k_BT$ which is in reasonable agreement
with our result
$\Delta G^{\ddagger} \approx 17.4 k_BT$, followed from the microscopic fit
with $\nu = 1/2$.
Using the GB/SA continuum solvation model \cite{Qiu_JPCA97} and the
CHARMM27 force
field \cite{MacKerell_JPCB98}
Li and Makarov \cite{Li_JCP04,Li_JPCB04}
obtained a much
higher
value $\Delta G^{\ddagger} = 29$ kcal/mol $\approx 48.6 k_BT$.
Again,
the large
departure from the experimental result may be related to
poor sampling or to the force filed they used.

\subsubsection{The effect of linkage on $x_u$ for single Ub}

One of the most interesting experimental results of
Carrion-Vazquez {\em et al.}\cite{Carrion-Vazquez_NSB03}
is that pulling Ub at different positions changes $x_u$ drastically. Namely,
if the force is applied at the C-terminal and Lys48, then
in the Bell approximation $x_u \approx 6.3$ \AA ,
which is about two and half times larger than the case when the termini N and C
are pulled.
Using the all-atom model
Li and Makarov \cite{Li_JCP04} have shown
that $x_u$ is much larger than 10 \AA.  Thus, a
theoretical reliable estimate for $x_u$ of Lys48-C Ub is not available.
Our aim is to compute $x_u$
employing the present Go-like model \cite{Clementi_JMB00} as
it is successful
in predicting $x_u$ for the N-C Ub.
Fig. \ref{refold_unfold_vs_force_fig} shows the force dependence of
unfolding time of the fragment Lys48-C when the force is
applied to Lys48 and C-terminus. The unfolding time is defined
as the averaged time to stretch this fragment. From the linear fit
($\nu =1$ in Fig. \ref{refold_unfold_vs_force_fig}) at
low forces we obtain $x_u \approx 0.61$ nm which is in good agreement
with the experiment \cite{Carrion-Vazquez_NSB03}.
The Go model is suitable for estimating $x_u$ for not only Ub,
but also for other proteins \cite{MSLi_BJ07a} because
the unfolding is mainly governed by the native topology.
The fact that $x_u$ for the linkage Lys48-C is larger than that of the N-C
Ub may be understood using our recent observation \cite{MSLi_BJ07a}
that it anti-correlates with the contact order (CO)  \cite{Plaxco_JMB98}.
Defining contact formation between any two amino acids ($|i-j| \geq 1$)
as occurring when
the distance between the centers of mass
of side chains $d_{ij} \leq 6.0$ \AA
(see also $http://depts.washington.edu/bakerpg/contact$\_$order/$),
we obtain CO equal 0.075 and 0.15 for the Lys48-C and N-C Ub,
respectively. Thus, $x_u$ of the Lys48-C linkage is larger
than that of the
N-C case because its CO is smaller. This result suggests that
the anti-correlation between $x_u$ and CO may hold
not only when proteins are pulled at termini \cite{MSLi_BJ07a}, but also
when the force is applied to different positions.
Note that the linker (not linkage) effect on $x_u$ has been
studied for protein L \cite{West_PRE06}. It seems
that this effect is
less pronounced compared the effect caused by changing pulling direction
studied here.
We have carried out the microscopic fit for $\nu =1/2$ and $2/3$
(Fig. \ref{refold_unfold_vs_force_fig}). As in the N-C Ub case,
$x_u$ is larger than its Bell value.
However the linkage at Lys48 has a little effect on the activation energy
$\Delta G^{\ddagger}$ (Table \ref{Dudkotable}).

\subsubsection{Determination of $x_u$ for the three-domain ubiquitin}

Since the trimer is a two-state folder (Fig. \ref{diagram}c),
one can determine
 its averaged distance between the NS and TS, $x_u$,
along the end-to-end distance reaction coordinate using kinetic
theory \cite{Bell_Sci78,Dudko_PRL06}.
We now ask if the multi-domain structure of Ub changes $x_u$.
As in the
single Ub case \cite{MSLi_BJ07}, there exists a critical force
$f_c \approx 120$pN
separating the low force
and high force regimes (Fig. \ref{refold_unfold_vs_force_fig}).
In the high force region, where the
unfolding barrier disappears, the unfolding time depends on $f$ linearly
(fitting curve not shown) as predicted
theoretically by Evans and Ritchie \cite{Evans_BJ97}.
In the Bell approximation, from the linear fit
(Fig. \ref{refold_unfold_vs_force_fig}) we obtain
$x_u\approx$ 0.24 nm which is exactly
the same as for the single Ub \cite{MSLi_BJ07}.
The values of $\tau _U^0, x_u$ and $\Delta G^{\ddagger}$, extracted
from
the nonlinear fit (Fig. \ref{refold_unfold_vs_force_fig}), are presented
in Table \ref{Dudkotable}. For both $\nu = 1/2$ and $\nu = 2/3$,
$\Delta G^{\ddagger}$ is a bit lower than that
for the single Ub.
In the Bell approximation,
the value of $x_u$ is the same for the single and three-domain Ub but
it is no longer valid for the $\nu = 2/3$ and $\nu = 1/2$ cases.
It would be interesting to perform experiments to check this result and
to see the effect of  multiple domain structure on the FEL.

\subsection{Thermal unfolding of Ubiquitin}

\subsubsection{Thermal unfolding pathways}

\begin{figure}[!hbtp]
\epsfxsize=4.2in
\vspace{0.2in}
\centerline{\epsffile{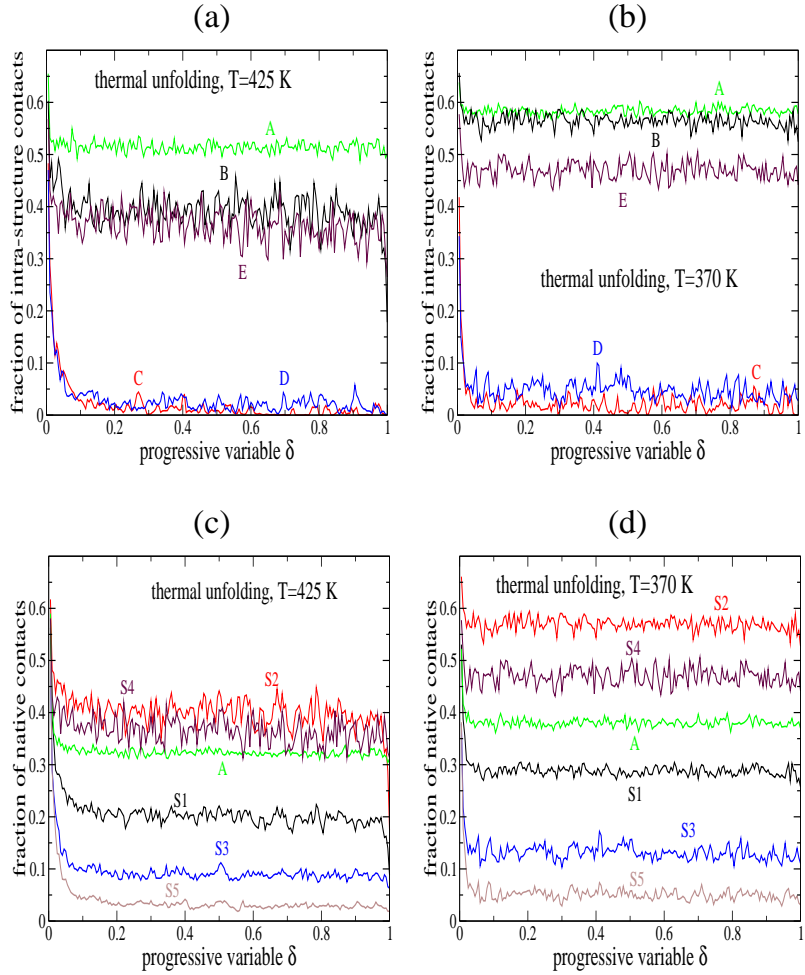}}
\linespread{0.8}
\caption{ (a) The dependence of fraction of intra-structure
native contacts on
the progressive variable $\delta$ for all structures at $T$=425 K.
(b) The same as in (a) but for $T=370$ K. (c) The dependence of
the all native contacts of the $\beta$-strands and helix A at
$T$=425 K. (d) The same as in (c) but for $T=370$ K.}
\label{cont_time_thermal_unfold_fig}
\end{figure}

To study the thermal unfolding the simulation was started from the NS
conformation and it was terminated when all of the native contacts are broken.
Two hundreds trajectories were generated with different random seed numbers. The fractions of
native contacts of helix A and five $\beta$-strands are averaged
over all trajectories for the time window $0 \le \delta \le 1$.
The unfolding routes are studied by monitoring these fractions as
a function of $\delta$. Above $T \approx 500$ K
the strong thermal fluctuations (entropy driven regime) make all
strands and helix A unfold almost simultaneously. Below this
temperature the statistical preference for the unfolding
sequencing is observed. We focus on $T=370$ and 425 K. As in the
case of the mechanical unfolding the cluster 2 unfolds before
cluster 1 (results not shown). However, the main departure from
the mechanical behavior is that the strong resistance to thermal
fluctuations of the cluster 1 is mainly due to the stability of
strand S2 but not of helix A (compare Fig.
\ref{cont_time_thermal_unfold_fig}{\em c} and {\em d} with Fig.
\ref{dom_ext_100pN_fig}{\em e-f}.
 The unfolding of cluster 2  before cluster 1
is qualitatively consistent with the experimental
observation that
the C-terminal fragment (residues 36-76)
is largely unstructured while native-like structure persists in
the N-terminal fragment (residues 1-35)
\cite{Bofill_JMB05,Cox_JMB93,Jourdan_Biochem00}.
This is also consistent with the data from the folding simulations
\cite{Sorenson_Proteins02} as well as with the experiments of Went and Jackson
\cite{Went_PEDS05} who have shown that the $\phi$-values $\approx 0$ in the
C-terminal region. However, our finding is at odds with the high $\phi$-values 
obtained for several residues in this region by all-atom simulations \cite{Marianayagam_BPC04}
and by a semi-empirical approach \cite{Fernandez_JCP01}.
One possible reason for
high $\phi$-values
in the C-terminal region is due to the force fields.
For example, Marianayagam
and Jackson have employed the GROMOS 96 force field \cite{Gunstren_96} within
the software GROMACS software package \cite{Berendsen_CPC95}.
It would be useful to check if the other force fields give the same result or
not.

The evolution of the fraction of intra-structure contacts of A, B, C, D and E
is shown in Fig. \ref{cont_time_thermal_unfold_fig}{\em a} ($T=425$ K)
and {\em b} ($T=$370 K).
Roughly we have the unfolding sequencing,
given by Eq. (\ref{thermal_sequencing_struc}),
 which strongly differs
from the mechanical one. The large stability of the $\alpha$ helix fragment
A against thermal fluctuations is consistent with the all-atom unfolding simulations
\cite{Alonso_ProSci98} and the experiments \cite{Went_PEDS05}.
The N-terminal structure B unfolds even after the core
part E and at $T=370$ K its stability is comparable with helix A.
The fact that B can withstand thermal fluctuations at high temperatures
agrees with the experimental results of Went and Jackson \cite{Went_PEDS05}
and of Cordier and Grzesiek
\cite{Cordier_JMB02} who used the notation $\beta _1/\beta _2$ instead of B.
This also agrees with the results of Gilis and Rooman \cite{Gilis_Proteins01}
who used a coarse-grained model but disagrees with results from
all-atom simulations
\cite{Alonso_ProSci98}. This disagreement is probably due to the fact that
Alonso and Daggett studied only two short trajectories and B did not
completely unfold \cite{Alonso_ProSci98}. 
The early unzipping of the structure C (Eq. \ref{thermal_sequencing_struc}) is
consistent with the MD prediction \cite{Alonso_ProSci98}.
Thus our thermal unfolding sequencing (Eq. \ref{thermal_sequencing_struc}) is more complete
compared to the all-atom simulation 
and it gives the reasonable agreement with the experiments.

We now consider the thermal unstability of individual $\beta$-strands
and helix A.
At $T$ = 370 K 
(Fig. \ref{cont_time_thermal_unfold_fig}{\em d}) the trend that S2
unfolds after S4 is more evident compared to the $T=425$ K case
(Fig. \ref{cont_time_thermal_unfold_fig}{\em c}). Overall, the
simple Go model leads to the sequencing given by Eq. (\ref{thermal_sequencing}).
\begin{subequations}
\begin{equation}
{\rm (C,D)} \rightarrow {\rm E} \rightarrow {\rm B} \rightarrow {\rm A}
\label{thermal_sequencing_struc}
\end{equation} 
\begin{equation} 
{\rm S5} \rightarrow {\rm S3} \rightarrow {\rm S1} \rightarrow {\rm A}
 \rightarrow {\rm (S4,S2)}.
\label{thermal_sequencing}
\end{equation}
\end{subequations}
From Eq. (\ref{mechanical_sequencing}), \ref{mechan_fixN_sequencing}
and \ref{thermal_sequencing} it is
obvious that the thermal unfolding pathways of individual strands
 markedly differ from
the mechanical ones. This is not surprising because the force should unfold
the termini first while under thermal fluctuations the most unstable part
is expected to detach first.
Interestingly, for the structures the thermal and mechanical
 pathways 
(compare Eq. (\ref{thermal_sequencing_struc})
and \ref{mechan_fixN_sequencing_struc}) are almost identical except that
the sequencing of C and D is less pronounced in the former case. 
This coincidence is probably accidental.

The fact that S5 unfolds first agrees with the high-resolution NMR
data of Cordier and Grzesiek \cite{Cordier_JMB02} who studied the
temperature dependence of HBs of Ub.
However, using the $\psi$-value analysis Krantz {\em et al} \cite{Krantz_JMB04}
have found that S5 (B3 in their notation) breaks even after S1 and S2.
One of possible reasons is that, as pointed out by Fersht \cite{Fersht_PNAS04},
if there is any plasticity in the TS which can accommodate the
crosslink between the metal and bi-histidines, then $\psi$-values would be
significantly greater than zero even for an unstructured region, leading to
an overestimation of structure in the TS. 
In agreement with our results,
the $\phi$-value analysis \cite{Went_PEDS05}  yields that S5 breaks before S1
and A but it fails to determine whether S5 breaks before S3.
By modeling the
amide I vibrations Chung {\em et al.} \cite{Chung_PNAS05}
 argued that S1 and S2 are
more stable than S3, S4 and S5. Eq.
\ref{thermal_sequencing} shows that the thermal stability of S1
and S2 is indeed higher than S3 and S5 but S4 may be more stable
than S1. The reason for only partial agreement between our results
and those of Chung {\em et al.} remains unclear. It may be caused
either by the simplicity of the Go model or by the model  proposed
in Ref. \cite{Chung_PNAS05}.
The relatively high stability of S4 (Eq. \ref{thermal_sequencing}) is 
supported by the $\psi$-value analysis \cite{Krantz_JMB04}.

\begin{figure}[!hbtp]
\epsfxsize=2.5in
\vspace{0.3in}
\centerline{\epsffile{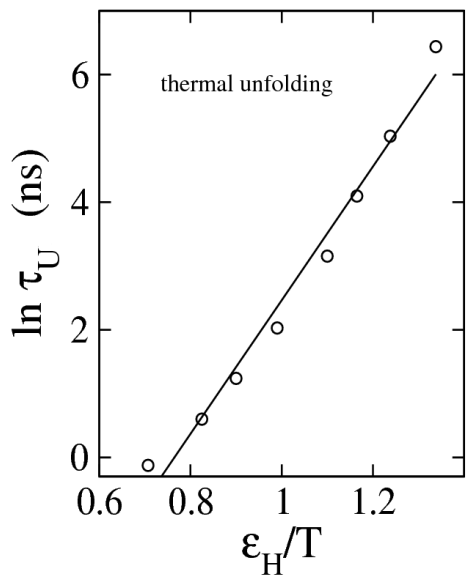}}
\linespread{0.8}
\caption{Dependence of thermal unfolding time $\tau _{u}$ on
 $\epsilon _H/T$,
where $\epsilon _H$ is the hydrogen bond energy. The straight line is a fit
 $y = -8.01 + 10.48x$. \label{uftime_T_fig}}
\end{figure}

\subsubsection{Thermal unfolding barrier}

Figure  \ref{uftime_T_fig} shows the temperature dependence of the
unfolding time $\tau_{u}$ which depends on the thermal unfolding
barrier, $\Delta F^T_{u}$, exponentially, $\tau_{u} \approx
\tau_{u}^0 \exp(\Delta F^T_{u}/k_BT)$. From the linear fit in
Fig. \ref{uftime_T_fig} we obtain $\Delta F^T_{u} \approx 10.48
\epsilon_h \approx 10.3$ kcal/mol.
It is interesting to note that $\Delta F^T_{u}$ is compatible with
$\Delta H_m \approx 11.4$ kcal/mol obtained from the equilibrium data
(Fig. \ref{diagram_fN_fig}{\em b}). However, the latter is defined
by  an equilibrium constant (the free energy difference
between NS and DS) but not by the rate constant
(see, for example, Ref. \onlinecite{Noronha_BJ04}). 

\subsection{Dependence of unfolding force of single Ubiquitin on $T$}

Recently, using the improved temperature control technique to perform
the pulling experiments
for the single Ub, Yang {\em et al.} \cite{Yang_RSI06}
have found that the unfolding force
depends on $T$ linearly
for 278 K $ \le T \le$ 318 K, and the slope of linear behavior
does not depend on pulling speeds.
 Our goal is
to see if the present Go model
can reproduce this result at least qualitatively, and more importantly,
to check whether the linear dependence holds for the whole temperature
interval where $f_{max} > 0$.

\begin{figure}
\epsfxsize=5.2in
\centerline{\epsffile{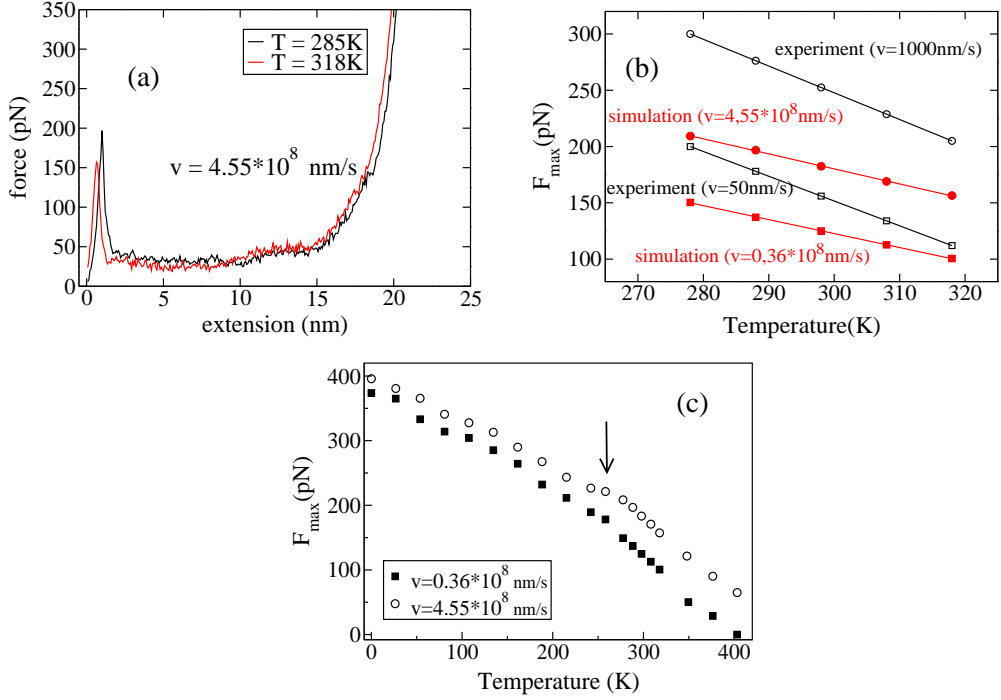}}
\linespread{0.8}
\caption{(a) The force-extension profile obtained at
$T=285$ K (black) and 318 K (red) at the pulling speed
$v= 4.55\times 10^8$ nm/s. $f_{max}$ is located at the extension
$\approx 1$ nm and 0.6 nm for $T=285$ K and 318 K, respectively.
The results
are averaged over 50 independent trajectories.
(b) The dependence of $f_{max}$ on temperature for two values
of $\nu$. The experimental data are taken from
Ref.  \onlinecite{Yang_RSI06} for comparison.
The linear fits for the simulations are
$y = 494.95 - 1.241x$ and $y = 580.69 - 1.335x$. For the experimental sets
we have $y = 811.6 - 2.2x$ and $y = 960.25 - 2.375x$.
(c) The dependence temperature of $f_{max}$ for the whole temperature
region and two values
of $\nu$. The arrow marks the crossover between two nearly linear regimes.}
\label{fmax_T_fig}
\end{figure}

The pulling simulations have been carried at two speeds following the
protocol described
in Chapter 3.
Fig. \ref{fmax_T_fig}a shows the force-extension profile of the single
Ub for $T=288$ and 318 K at the pulling speed $v= 4.55\times 10^8$ nm/s.
The peak is lowered as $T$ increases because thermal fluctuations promote
the unfolding of the system. In addition the peak moves toward a
lower extension.
This fact is also understandable, because at higher $T$ a protein can
unfold
at lower extensions due to thermal fluctuations.
For $T=318$ K, e.g., the maximum force is located at the extension
 of $\approx 0.6$ nm, which
corresponds to the plateau observed in the time dependence of
the end-to-end distance under constant force
\cite{Irback_PNAS05,MSLi_BJ07}.
One can show that, in agreement with Chyan {\em et al.}
\cite{Chyan_BJ04}, at this maximum the extension between
strands S$_1$ and S$_5$ is $\approx$ 0.25 nm. Beyond the
maximum, all of the
native contacts between strands S$_1$ and S$_5$ are broken.
At this stage, the chain ends are almost
stretched out, but the rest of the polypeptide chain remains
native-like.

The temperature dependence of the unfolding force, $f_{max}$,
is shown in Fig. \ref{fmax_T_fig}b
for 278 K $\le T \le$ 318 K, and for two pulling speeds.
The experimental results of Yang {\em et al.} are also presented
for comparison. Clearly,
in agreement with experiments \cite{Yang_RSI06}
linear behavior is observed and
the corresponding slopes do not depend on $v$.
Using the fit $f_{max} = f_{max}^0 - \gamma T$ we obtain the ratio
between the simulation and experimental slopes
$\gamma _{sim}/\gamma _{exp} \approx 0.56$.
Thus, the Go model gives
a weaker temperature dependence compared to the experiments.
Given the simplicity of this model, the agreement between theory and experiment
should be considered reasonable, but it would be interesting to check if
a fuller accounting of non-native contacts and environment can improve
our results.

As evident from Fig. \ref{fmax_T_fig}c,
the dependence of $f_{max}$ on $T$ ceases
to be linear for the whole temperature interval.
The nonlinear temperature dependence of $f_{max}$ may be understood
qualitatively using the simple theory of Evans and K. Ritchie
\cite{Evans_BJ97}. For the external force linearly ramped
with time,
the unfolding
force is given by Eq. (\ref{Bell_Ku_eq}).
(A more complicated microscopic
expression for $f_{max}$ is provided by Eq. \ref{Dudko_eq}).
Since $\tau_U^0$ is temperature dependent and $x_u$ also displays a weak
temperature dependence \cite{Imparato_PRL07}, the resulting $T$-dependence
should be nonlinear.
This result can also be understood by noting that the temperatures considered
here are low enough so that we are not in the entropic limit,
where the linear dependence would be valid for the worm-like model
\cite{Marko_Macromolecules95}.
The arrow in Fig. \ref{fmax_T_fig}c separates two regimes of the $T$-dependence
of $f_{max}$. The crossover takes place roughly in the temperature
interval where the temperature dependence of the equilibrium
critical force changes the slope (Fig. \ref{diagram_fN_fig}).
At low temperatures, thermal fluctuations are weak and
the temperature dependence of $f_{max}$
is weaker compared to the high temperature regime.
Thus the linear dependence observed in the experiments of Yang {\em et al.}
\cite{Yang_RSI06} is valid, but only in the narrow $T$-interval.

\subsection{Conclusions}

To summarize, in this chapter we have obtained the following novel results.
It was shown that the refolding of Ub is a two-stage process in which
the "burst" phase exists on very short time scales.
Using the 
 dependence of the refolding and unfolding
on $f$, $x_f$, $x_{u}$ and unfolding barriers were computed. Our results
for FEL parameters are in acceptable agreement
with the experiments. It has been demonstrated
that fixing the N-terminus of Ub has much
stronger effect on mechanical unfolding pathways compared to the case
when the C-end is anchored. In comparison with
previous studies, we provide a more
complete picture for thermal unfolding pathways which are very different
from the mechanical ones.
Mechanically strand S1 is the most unstable
whereas the thermal fluctuations break contacts of S5 first.

We have shown that, in agreement with the experiment of Carrion-Vazquez
{\em et al.}
\cite{Carrion-Vazquez_NSB03},  the Lys48-C linkage
changes $x_u$ drastically. From the point of view of
biological function,
the linkage Lys63-C is very important, but the study of its mechanical
properties is not as interesting as the Lys48-C because this fragment
is almost stretched out in the NS.
Finally, we have reproduced an experiment \cite{Yang_RSI06}
of the linear temperature dependence of unfolding force of Ub
on the quasi-quantitative level.
Moreover, we have shown that for the whole
temperature region the dependence of $f_{max}$ on $T$ is nonlinear,
and the observed linear dependence is valid only for a narrow temperature
interval. This behavior should be common for all proteins because it
reflects the fact that
the entropic limit is not applicable to all temperatures.

\newpage
\begin{center}
\section{Dependence of protein mechanical unfolding pathways on pulling speeds}
\end{center}

\subsection{Introduction}

As cytoskeletal proteins, large actin-binding proteins play a key roles
in cell organization, mechanics and signalling\cite{Stossel_NRMCB01}. 
During the process of permanent cytoskeleton reorganization, all
involved participants are subject to mechanical stress. One
of them is DDFLN4 protein,
which binds different components of
actin-binding protein. Therefore, understanding the  mechanical response of
this domain to a stretched force is of great interest.
Recently, using the AFM experiments, 
Schwaiger {\em et al.}  \cite{Schwaiger_NSMB04,Schwaiger_EMBO05} have obtained two major results for DDFLN4.
First,  this domain (Fig. \ref{native_ddfln4_strands_fig})
 unfolds via intermediates as the force-extension curve displays two peaks
centered at $\Delta R \approx 12$ nm and
$\Delta R \approx 22$ nm.
Second, with the help of  loop mutations, it was suggested
that during the first unfolding event (first peak) strands A and  B
unfold first.
Therefore, strands C - G form a stable intermediate structure, which then
unfolds in the second unfolding event (second peak).
In addition, Schwaiger {\em et al.}
\cite{Schwaiger_EMBO05} have also determined the
FEL parameters of DDFLN4. 

With the help of the C$_{\alpha}$-Go model \cite{Clementi_JMB00}, Li {\em et al.}
\cite{MSLi_JCP08}
have demonstrated that the mechanical unfolding of DDFLN4 does follow
 the three-state
scenario but the full agreement between theory and experiments was not
obtained. The simulations \cite{MSLi_JCP08} showed
that two peaks in the force-extension profile occur
at $\Delta R \approx 1.5$ nm and 11 nm, i.e.,
the Go modeling does not detect the peak
at $\Delta R \approx 22$ nm. Instead, it predicts the existence of
a peak not far from the native
conformation. More importantly, theoretical unfolding pathways
\cite{MSLi_JCP08} are very different from the
experimental ones \cite{Schwaiger_NSMB04}:
the unfolding initiates from the C-terminal,
but not from the N-terminal terminal as shown by the experiments.

It should be noted that the pulling speed used in the previous simulations
is about five orders of magnitude larger than
the experimental value \cite{Schwaiger_NSMB04}.
Therefore, a natural
question
emerges is if the discrepancy between theory and experiments is due
to huge difference in pulling speeds.
Motivated by this, we have carried low-$v$ simulations, using the Go
model \cite{Clementi_JMB00}.
Interestingly,
we uncovered that unfolding pathways of DDFLN4 depend on the pulling speed and
only at
$v \sim 10^4$ nm/s, the theoretical unfolding sequencing coincides with
the experimental one \cite{Schwaiger_NSMB04}.
However, even at low loading rates,
the existence of the peak at $\Delta R \approx 1.5$ nm
remains robust 
and the Go modeling does not capture the maximum at $\Delta R \approx 22$ nm.

\begin{figure}
\epsfxsize=4.2in
\vspace{0.2in}
\centerline{\epsffile{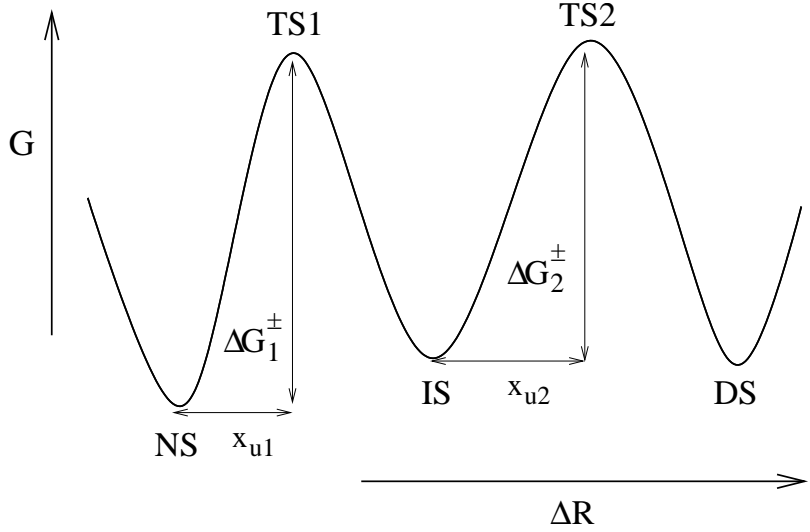}}
\linespread{0.8}
\caption{Schematic plot of
the free energy landscape
for a three-state protein as a function of the end-to-end distance.
$x_{u1}$ and $x_{u2}$ refer to the distance between the
 NS and 
TS1 
and the distance between IS and TS2.
The unfolding barrier
$\Delta G^{\ddagger}_1 = G_{TS1} - G_{NS}$ and
$\Delta G^{\ddagger}_2 = G_{TS2} - G_{IS}$.}
\label{free_3state_concept_fig}
\vspace{5 mm}
\end{figure}

In the previous work \cite{MSLi_JCP08},
using dependencies
of unfolding times on external forces,
the distance between the NS and the first transition state (TS1),
 $x_{u1}$, and the distance between IS and the second
transition state (TS2), $x_{u2}$, of DDFLN4 have been estimated
(see Fig. \ref{free_3state_concept_fig}.
In the Bell approximation, the agreement between the theory and
experiments \cite{Schwaiger_EMBO05} was reasonable.
However, in the non-Bell approximation
\cite{Dudko_PRL06}, the theoretical values of  $x_{u1}$, and $x_{u2}$
seem to be high \cite{MSLi_JCP08}.
In addition the unfolding barrier between the 
TS1 and NS, 
$\Delta G^{\ddagger}_1$, is clearly higher than its experimental
counterpart (Table \ref{DDFLN4_table}).

In this chapter \cite{MSLi_JCP09}, assuming that the microscopic kinetic theory
\cite{Dudko_PRL06} holds for a three-state protein, we calculated
$x_{ui} (i=1,2)$ and unfolding barriers
by a different method
which is based on dependencies of peaks in the force-extension curve
on $v$. Our present estimations for
the unfolding FEL parameters are more reasonable
compared to the previous ones \cite{MSLi_JCP08}.
Finally, we have also studied thermal unfolding 
pathways of DDFLN4 and shown
that the
mechanical unfolding pathways are different from the thermal ones.

This chapter is based on the results from Ref. \cite{MSLi_JCP09}.

\subsection{Method}

\begin{figure}
\includegraphics[scale=0.5]{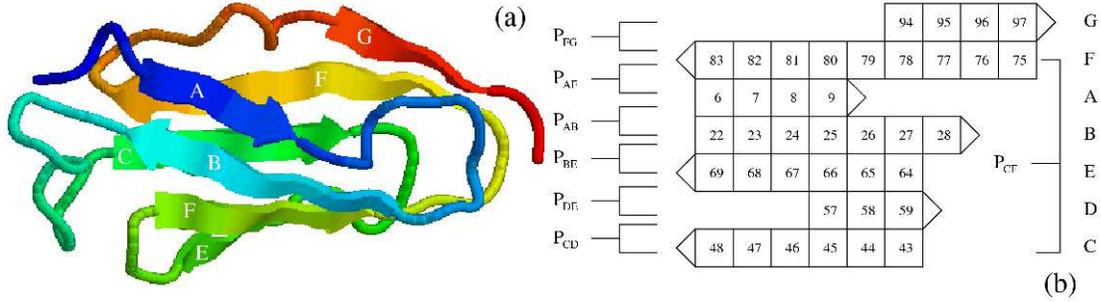}
\linespread{0.8}
\caption{ (a) NS conformation of
DDFLN4 taken from the PDB
(PDB ID: 1ksr). There are seven $\beta$-strands: A (6-9), B (22-28),
C (43-48), D (57-59), E (64-69), F (75-83), and
G (94-97).
In the NS there are 15, 39, 23, 10, 27, 49, and 20 native contacts
formed by strands A, B, C, D, E, F, and G with
the rest of the protein, respectively.
The end-to-end distance in the NS $R_{NS}=40.2$ \AA.
(b) There are 7 pairs of strands, which have the nonzero number
 of mutual native contacts
in the NS. These pairs are  P$_{\textrm{AB}}$,
P$_{\textrm{AF}}$, P$_{\textrm{BE}}$,
P$_{\textrm{CD}}$, P$_{\textrm{CF}}$, P$_{\textrm{DE}}$, and P$_{\textrm{FG}}$.
The number of native contacts between them
 are 11, 1, 13, 2, 16, 8, and 11,
respectively.}
\label{native_ddfln4_strands_fig}
\end{figure}

The native conformation
of DDFLN4, which has seven $\beta$-strands, enumerated as
A to G,
was taken from the PDB (PI: 1KSR,
Fig. \ref{native_ddfln4_strands_fig}a).
We assume that
residues $i$ and $j$ are in native contact if the distance
 between them in the native conformation,
is shorter than a cutoff distance $d_c =
6.5$ \AA.
With this choice of $d_c$, the molecule has 163 native contacts.
Native contacts exist between seven pairs
of $\beta$-strands
P$_{\textrm{AB}}$,
P$_{\textrm{AF}}$, P$_{\textrm{BE}}$,
P$_{\textrm{CD}}$, P$_{\textrm{CF}}$, P$_{\textrm{DE}}$, and P$_{\textrm{FG}}$
(Fig. \ref{native_ddfln4_strands_fig}b).

We used the C$_{\alpha}$-Go model \cite{Clementi_JMB00} for a molecule.
The corresponding parameters of this model are chosen as in Chapter 4.
The simulations were carried out in the over-damped limit
with the water viscosity $\zeta = 50\frac{m}{\tau_L}$
The Brownian dynamics equation (Eq. \ref{overdamped_eq}) was numerically solved by the simple Euler method (Eq. \ref{Euler}).
Due to the large viscosity, we can choose a large time step
$\Delta t = 0.1 \tau_L$, and this choice allows us to study unfolding at low
loading rates.
In the constant velocity force simulations, we follow the protocol described in 
section 3.1.2.
The mechanical unfolding sequencing
was studied by
monitoring the fraction of native contacts of the $\beta$-strands
and of their seven pairs as a function of $\Delta R$,
which is admitted  a good reaction coordinate.

\subsection{Results}

\subsubsection{Robustness of peak at end-to-end extension $\Delta R \approx 1.5$ nm and absence
of maximum at $\Delta R \approx 22$ nm at low pulling speeds}

\begin{figure}
\epsfxsize=5.0in
\vspace{0.2in}
\centerline{\epsffile{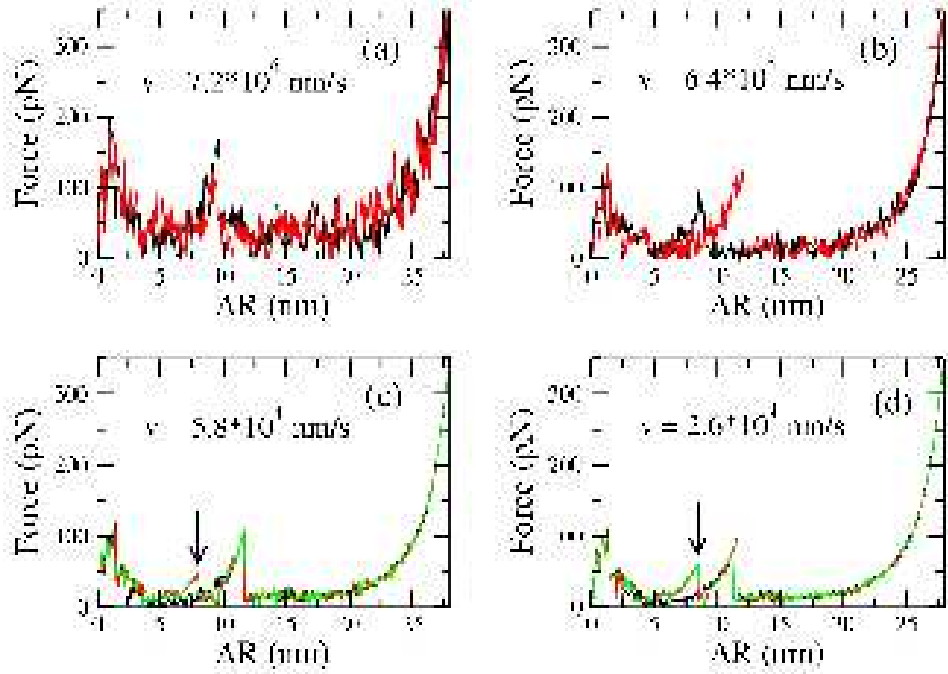}}
\linespread{0.8}
\caption{(a) Typical force-extension curves for
 $v =7.2\times 10^6$ nm/s. (b) The same as in (a) but for $v=6.4\times 10^5$ nm/s. (c) The same as in (a) but for $v=5.8\times 10^4$ nm/s. The arrow roughly refers to locations of additional peaks for two trajectories (red and green).
(d) The same as in (c) but for $v=2.6\times 10^4$  mn/s.}
\label{force_ext_traj_fig}
\vspace{2 mm}
\end{figure}

In the previous high pulling speed
($v = 3.6\times 10^7$ nm/s) Go simulations
\cite{MSLi_JCP08}, the force-extension curve shows two
peaks at $\Delta R \approx 1.5$ nm and 10 nm, while the experiments
showed that peaks appear at $\Delta R \approx 12$ nm and 22 nm.
The question we ask if one can reproduce the experimental results at
low pulling speeds. Within our computational facilities, we were
able to perform simulations at the lowest $v = 2.6\times 10^4$ nm/s
which is about three orders of magnitude lower than that used
before \cite{MSLi_JCP08}.

\begin{wrapfigure}{l}{0.42\textwidth}
\includegraphics[width=0.40\textwidth]{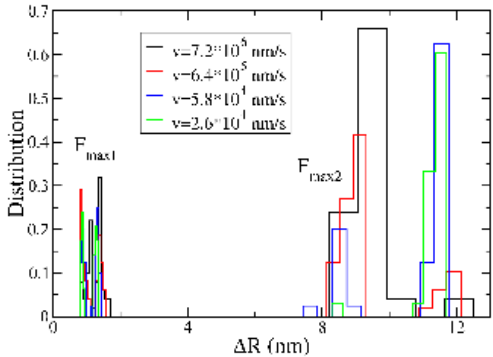}
\hfill\begin{minipage}{6.3 cm}
\linespread{0.8}
\caption{Distributions of positions of $f_{max1}$ and
$f_{max2}$ for $v =7.2\times 10^6$ (black), $ 6.4\times 10^5$ (red) , $5.8\times 10^4$ (blue) and 2.6$\times 10^4$  mn/s (green. \label{dist_fmax_pos_fig}}
\end{minipage}
\end{wrapfigure}

Fig. \ref{force_ext_traj_fig} show 
force-extension curves for four representative pulling speeds.
For the highest $v = 7.2\times 10^6$ nm/s
(Fig. \ref{force_ext_traj_fig}a), there are two peaks
located at extensions $\Delta R \approx 1.5$ nm and 9 nm.
As evident from Figs. \ref{force_ext_traj_fig}b, c and d,
the existence of the first peak remains robust against reduction of $v$.
Positions 
of $f_{max1}$ weakly fluctuate over the range
$0.9 \lesssim \Delta R \lesssim 1.8$ nm for all values of $v$
(Fig. \ref{dist_fmax_pos_fig}). As $v$ is reduced, $f_{max1}$ decreases but this peak does not
vanish if one interpolates our results to the lowest pulling speed
$v_{exp} = 200$ nm/s
used in the experiments \cite{Schwaiger_NSMB04}
(see below).
Thus, opposed to the experiments, the first peak
occurs already at small end-to-end extensions.
We do not exclude a possibility that such a peak was
overlooked  by experiments,
as it happened with the titin domain
I27. Recall that, for this domain the first
 AFM experiment \cite{Rief_Science97}
did not trace the hump which was observed in the later
simulations \cite{Lu_BJ98} and experiments \cite{Marszalek_Nature99}.

Positions of the second peak $f_{max2}$
are more scattered compared to $f_{max1}$, ranging
from about 8 nm to 12 nm (Fig. \ref{dist_fmax_pos_fig}). Overall, they
move toward higher values upon
reduction of $v$ (Fig. \ref{force_ext_traj_fig}). If at $v=6.4\times 10^5$ nm/s only about 15$\%$
trajectories display $\Delta R_{max2} > 10$ nm, then this percentage reaches
65$\%$ and 97\% for $v=5.8\times 10^4$ nm/s and $2.6\times 10^4$ nm/s,
respectively (Fig. \ref{dist_fmax_pos_fig}).

At low $v$, unfolding pathways show rich diversity. 
For $v \gtrsim 6.4\times 10^5$ nm/s, the force-extension profile shows
only two peaks in all trajectories
studied (Fig. \ref{force_ext_traj_fig}a and \ref{force_ext_traj_fig}b),while
for lower speeds $v = 5.8\times 10^4$ nm/s and $2.6\times 10^4$ nm/s,
about $4\%$ trajectories display even four peaks
(Fig. \ref{force_ext_traj_fig}c and 
\ref{force_ext_traj_fig}d), i.e. the four-state behavior.

We do not observe any peak at $\Delta R \approx 22$ nm for all 
loading rates (Fig. \ref{force_ext_traj_fig}),
and it is very unlikely that it will appear at lower
values of $v$.
Thus, the Go model, in which non-native interactions are neglected,
fails to reproduce this experimental observation.
Whether inclusion of non-native interactions would
cure this problem requires further studies.

\subsubsection{Dependence of mechanical pathways on loading rates}

\begin{figure}[!hbtp]
\epsfxsize=4.5in
\vspace{0.2in}
\centerline{\epsffile{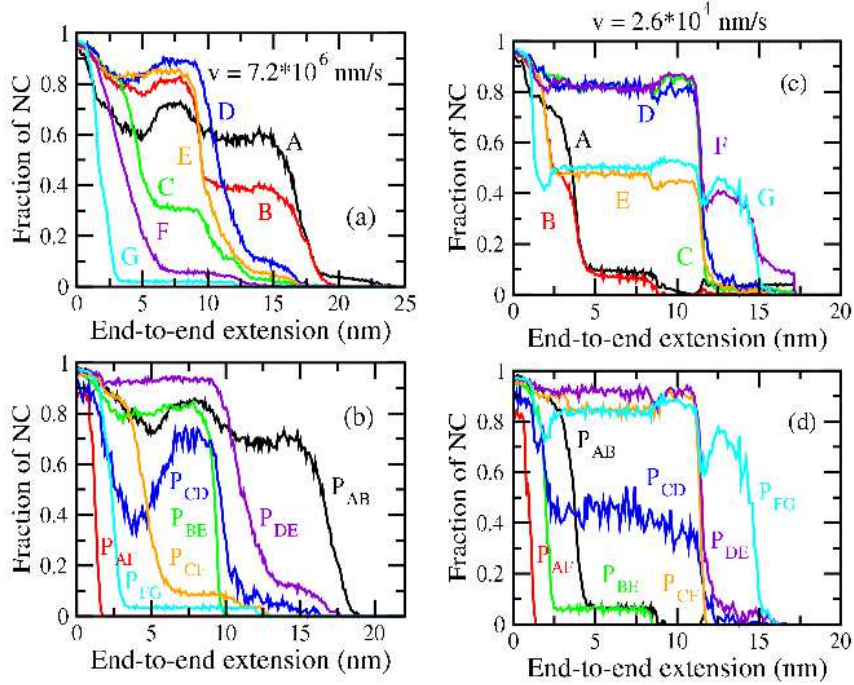}}
\linespread{0.8}
\caption{ (a) Dependences of averaged
fractions of native contacts
formed by seven strands on $\Delta R$ for $v = 7.2\times 10^6$ nm/s.
(b) The same as in (a) but for pairs of strands.
(c)-(d) The same as in a)-b) but
for $v=2.6\times 10^4$ nm/s.
Results were averaged over 50 trajectories.}
\label{cont_ext_v15_fig}
\end{figure}

The considerable fluctuations of peak positions and
occurrence of even three peaks already suggest that unfolding
pathways, which are kinetic in nature, may change if $v$ is varied.
To clarify this point in more detail, we show $\Delta R$-dependencies
of native contacts of all $\beta$-strands
and their pairs for $v=7.2\times 10^6$ nm/s
(Figs. \ref{cont_ext_v15_fig}{\em a,b}) and $v=2.6\times 10^4$ nm/s 
(Figs. \ref{cont_ext_v15_fig}{\em c,d}). For $v=7.2\times 10^6$ nm/s, one has the
following unfolding pathways:
\begin{subequations}
\begin{equation}
G \rightarrow F \rightarrow (C,E,D) \rightarrow B \rightarrow A,
\label{pathways_v6_strand_eq}
\end{equation}
\begin{equation}
P_{AF} \rightarrow P_{BE} \rightarrow  (P_{FG}, P_{CF}) \rightarrow P_{CD}
\rightarrow P_{DE}
\rightarrow P_{AB}.
\label{pathways_v6_pair_eq}
\end{equation}
\end{subequations}
According to this scenario, the unfolding initiates from the C-terminal,
while the experiments \cite{Schwaiger_NSMB04} showed that
strands A and B unfold first.
For $v=2.6\times 10^4$ nm/s, Fig. \ref{cont_ext_v15_fig}{\em c} gives
the following sequencing
\begin{subequations}
\begin{equation}
(A,B) \rightarrow (C,D,E) \rightarrow (F,G),
\label{pathways_v15_strand_eq}
\end{equation}
\begin{equation}
P_{AF} \rightarrow (P_{BE},P_{AB}) \rightarrow  P_{CF} \rightarrow 
(P_{CD},P_{DE},P_{FG}).
\label{pathways_v15_pair_eq}
\end{equation}
\end{subequations}

\begin{figure}
\epsfxsize=5.2in
\vspace{0.2in}
\centerline{\epsffile{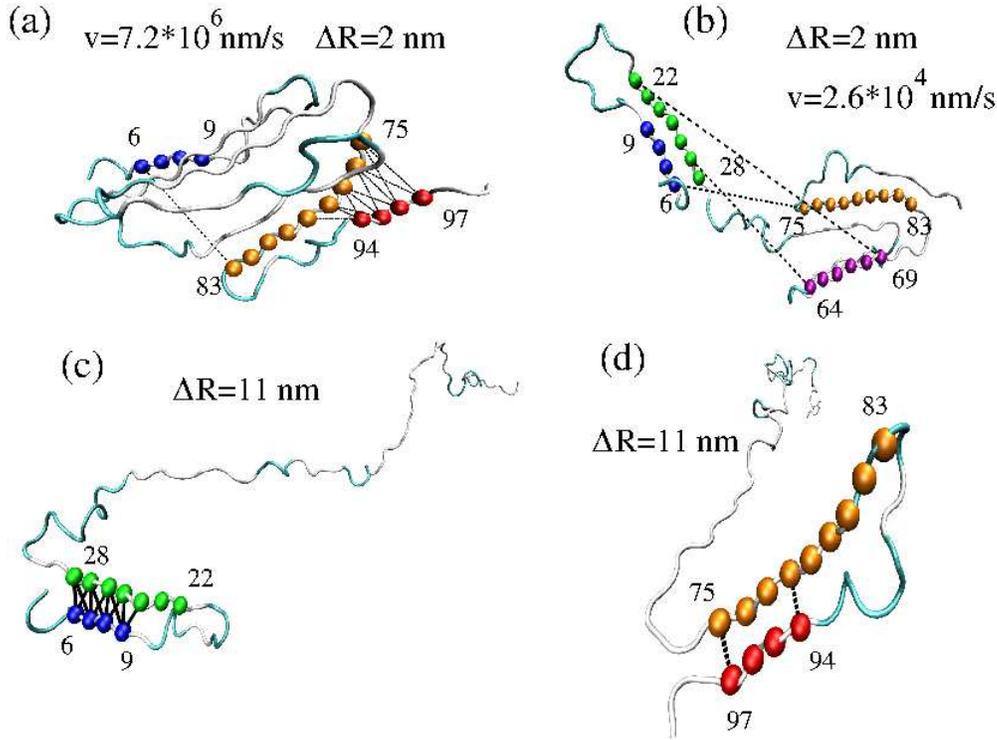}}
\linespread{0.8}
\caption{ (a) Typical snapshot obtained at $\Delta R = 2$ nm
and $v= 7.2\times 10^6$ nm/s. A single contact between strand A (blue spheres)
and strand F (orange) was broken. Native contacts between F and G (red) are also
broken and G completely unfolds. (b) The same as in (a) but
for $v=2.6\times 10^4$ nm/s. Native contacts between A and F and between
B and E are broken but all strands are remain partially structured.
(c) Typical snapshot obtained at $\Delta R = 11$ nm
and $v= 7.2\times 10^6$ nm/s. Native contacts between pairs are broken except
those between strands A and B.
All 11 unbroken contacts are marked by solid lines.  Strands A and B
do not unfold yet.
(d) The same as in (c) but for $v=2.6\times 10^4$ nm/s.
Two from 11 native contacts between F and G are broken (dashed lines).
Contacts between other pairs are already broken, but F and G remain structured.
}
\label{snapshot_v6_v15_fig}
\end{figure}

We obtain the very interesting result that at this low loading rate,
in agreement with the AFM experiments
\cite{Schwaiger_NSMB04}, the N-terminal detaches
from a protein first.
For both values of $v$, 
the first peak
corresponds to breaking of native contacts between
strands A and F (Fig. \ref{cont_ext_v15_fig}{\em d} and Fig. \ref{cont_ext_v15_fig}{\em b}).
However, the structure of unfolding intermediates, which correspond to this
peak, depends on $v$.
For $v=7.2\times 10^6$ nm/s (Fig. \ref{cont_ext_v15_fig}{\em a,b}), at
$\Delta R \approx 1.5$ nm, native contacts between F and G are
broken and strand G has already
been unstructured (Fig. \ref{cont_ext_v15_fig}a). Therefore, for this
pulling speed, the intermediate consists of
six ordered strands A-F
(see Fig. \ref{snapshot_v6_v15_fig}a
for a typical snapshot).
In the $v=2.6\times 10^4$ nm/s case, just after the first peak, 
 none of strands
unfolds completely (Fig. \ref{cont_ext_v15_fig}{\em c}),
although
(A,F) and (B,E) contacts have been already broken (Fig. \ref{cont_ext_v15_fig}{\em d}).
Thus, the intermediate looks very different from the high $v$ case, as it has
all secondary structures partially structured
(see (Fig. \ref{snapshot_v6_v15_fig}b) for a typical snapshot).
Since the experiments \cite{Schwaiger_NSMB04}
showed that intermediate structures contain five ordered strands
C-G, intermediates predicted by simulations are more ordered than the
experimental ones. Even though,
our low loading rate Go simulations  provide the same pathways
as on the experiments. The difference between theory and experiments
in intermediate structures comes from
different locations of the first peak.
It remains unclear if this is a shortcoming of Go models or of
the experiments because it is hard to imagine that a $\beta$-protein like
 DDFLN4
displays
the first peak at such  a large extension $\Delta R \approx 12$ nm
\cite{Schwaiger_NSMB04}. The force-extension curve of
the titin domain I27, which has a similar native topology, for example, displays the
first peak at $\Delta R \approx 0.8$ nm \cite{Marszalek_Nature99}.
From this prospect, the theoretical result is more favorable.

The strong dependence of unfolding pathways on loading rates is 
also clearly seen from structures around the second peak.
In the $v=7.2\times 10^6$ nm/s case,
at $\Delta R \approx 11$ nm,
strands A and B remain structured, while other strands detach
from a protein core (Fig. \ref{cont_ext_v15_fig}{\em a} and
Fig. \ref{snapshot_v6_v15_fig}c). This is entirely different from
the low loading case,
where A and B completely unfold
but F and G still survive (Fig. \ref{cont_ext_v15_fig}{\em c} and
Fig. \ref{snapshot_v6_v15_fig}d).
The result, obtained for $v=2.6\times 10^4$ nm/s,
is in full agreement with the experiments \cite{Schwaiger_NSMB04}
that at $\Delta R \approx 12$ nm, A and B detached from the core. 

\begin{figure}
\epsfxsize=4.5in
\vspace{0.2in}
\centerline{\epsffile{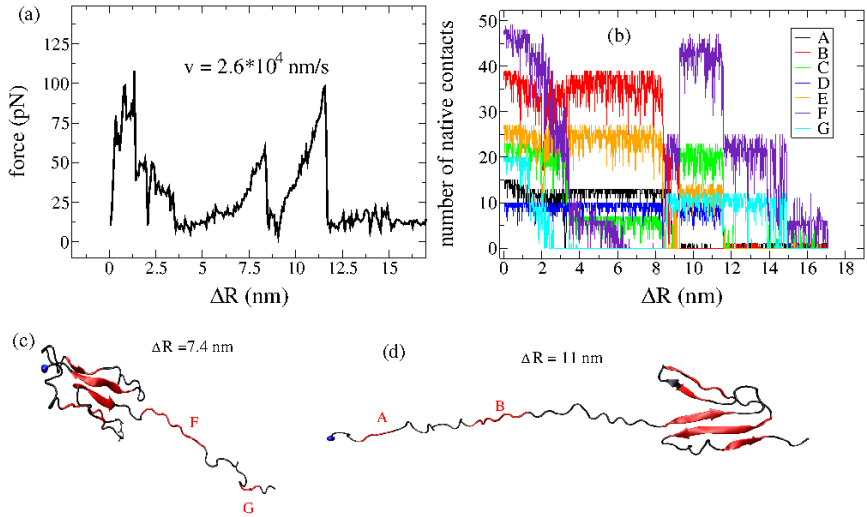}}
\linespread{0.8}
\caption{(a) Force-extension curve for
an atypical unfolding  pathway at $v = 2.6\times 10^4$ nm/s. (b)
Dependence of fractions of native contacts of seven strands on $\Delta R$.
Snapshot at $\Delta R = 7.4$ nm (c) and $\Delta R = 11$ nm (d).
\label{reentrance_FG_fig}}
\end{figure}

Note that the unfolding pathways given by Eq. (\ref{pathways_v6_strand_eq}),
\ref{pathways_v6_pair_eq},
\ref{pathways_v15_strand_eq}, and
\ref{pathways_v15_pair_eq}
are valid in the statistical sense. In all 50 trajectories studied
for $v=7.2\times 10^5$ nm/s, strands A and B always unfold last, and F and G
unfold first (Eq. \ref{pathways_v6_strand_eq}), while the sequencing of
unfolding events for C, D and E depends on individual trajectories.
At $v=2.6\times 10^4$ nm/s, most of trajectories follow
the pathway given by Eq. (\ref{pathways_v15_strand_eq}), but
we have observed  a few unusual pathways, as it is illustrated in 
Fig. \ref{reentrance_FG_fig}. Having three peaks in 
the force-extension profile, 
the evolution of native contacts of
F and G display an atypical behavior. 
At $\Delta R \approx 7$ nm, these strands  fully unfold
(Fig. \ref{reentrance_FG_fig}c),
but they refold again at $\Delta R \approx 11$ nm (Fig. \ref{reentrance_FG_fig}b
and \ref{reentrance_FG_fig}d). Their final unfolding takes place
around $\Delta R \approx 16.5$ nm. As follows from
Fig. \ref{reentrance_FG_fig}b, the first peak in
Fig. \ref{reentrance_FG_fig}a corresponds to unfolding of G. Strands
A and B unfold after passing the second peak, while the third maximum occurs
due to unfolding of C-G , i.e. of a core part
shown in Fig. \ref{reentrance_FG_fig}d.

The dependence of unfolding pathways on $v$ is understandable.
If a protein is pulled very fast, the perturbation, caused
by the external force,  does not
have enough time to propagate to the fixed N-terminal before the C-terminal
unfolds. Therefore, at very high $v$, we have the pathway given by
Eq. (\ref{pathways_v6_strand_eq}). In the opposite limit,
it does matter
what end is pulled as the external force
is uniformly felt along a chain. Then, a strand, which has
a weaker
link with the core, would unfold first.

\subsubsection{Computation of free energy landscape parameters}

As mentioned above, at low loading rates, for some trajectories,
the force-extension curve
does not show two, but three peaks. However,
the percentage of such trajectories is rather
small,
we will neglect them and consider DDFLN4 as a three-state
protein.
Recently,
using dependencies of unfolding times on the 
constant external force and
the non-linear kinetic theory \cite{Dudko_PRL06},
we obtained distances $x_{u1} \approx x_{u2} \approx 13 \AA$
\cite{MSLi_JCP08}.
These values seem to be large for $\beta$-proteins like DDFLN4,
which are supposed to have smaller $x_u$ compared to $\alpha/\beta$- and
$\alpha$-ones \cite{MSLi_BJ07a}.
A clear difference between theory and experiments was also observed
for the unfolding barrier $\Delta G^{\ddagger}_1$.
 In order to see if one can improve our previous
results,
we will extract the FEL parameters by a different approach.
Namely, assuming that all FEL parameters of the three-state DDFLN4, 
including
the barrier between the second TS and
the IS $\Delta G^{\ddagger}_2$ (see 
Ref. \onlinecite{MSLi_JCP08} for the definition),
can be determined from dependencies of $f_{max1}$ and $f_{max2}$ on $v$,
we calculate them in the the Bell-Evans-Rirchie (BER) approximation
as well as beyond this approximation. 

\paragraph{Estimation of $x_{u1}$ and $x_{u2}$ in the BER approximation}

\begin{figure}
\epsfxsize=4.2in
\vspace{0.2in}
\centerline{\epsffile{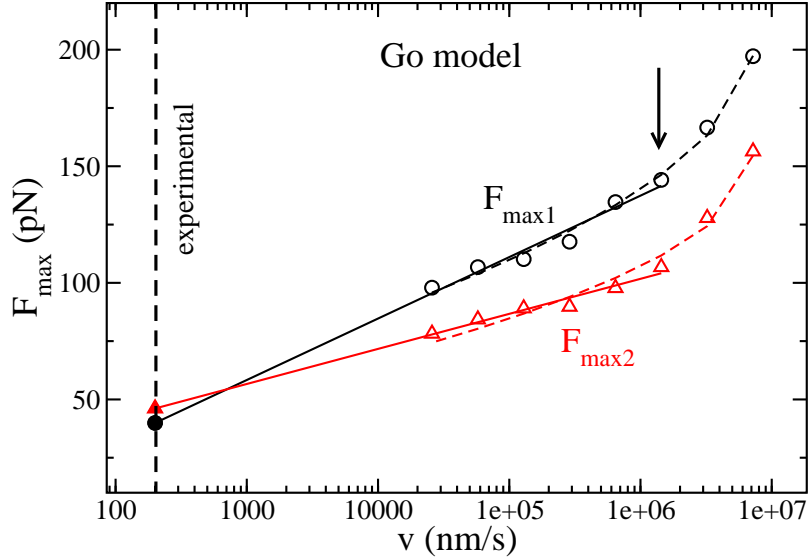}}
\linespread{0.8}
\caption{Dependences of $F_{max1}$ (open circles)
and $F_{max2}$ (open squares) on $v$.
Results were obtained by using the Go model.
Straight lines are fits to the BER equation
%
($y = -20.33 + 11.424ln(x)$ and $y= 11.54 + 6.528ln(x)$ for $F_{max1}$
and $F_{max2}$, respectively). Here $f_{max}$ and $v$ are measured in pN
and nm/s, respectively.
From these fits we obtain
$x_{u1}=3.2 \AA\,$ and $x_{u2}=5.5 \AA$.
The solid circle and triangle correspond to $f_{max1} \approx 40$ pN
and $f_{max2} \approx 46$ pN,
obtained by interpolation of linear fits to the experimental
value $v = 200$ nm/s.
Fitting to the nonlinear microscopic theory (dashed lines) gives
$x_{u1}=7.0 \AA\, \Delta G^{\ddagger}_1 = 19.9 k_BT,
x_{u2} = 9.7 \AA\,$, and $\Delta G^{\ddagger}_2 = 20.9 k_BT$.}
\label{fmax_Nfix_v_fig}
\end{figure}

In this approximation,
$x_{u1}$ and $x_{u2}$ are related to $v$, $f_{max1}$ and
$f_{max2}$ by the following equation \cite{Evans_BJ97}:
\begin{equation}
f_{maxi} \; = \; \frac{k_BT}{x_{ui} }
\ln \left[ \frac{vx_{ui}}{k_{ui}(0)k_BT}\right], i = 1,2,
\label{f_logV_eq2}
\end{equation}
where $k_{ui}(0)$ is unfolding rates at zero external force.
In the low force regime ($v \lesssim 2\times 10^6$ nm/s), the 
dependence of $f_{max}$ on $v$ is logarithmic and  
$x_{u1}$ and $x_{u2}$ are defined by
slopes of linear fits in
Fig. \ref{fmax_Nfix_v_fig}. Their values are listed in Table \ref{DDFLN4_table}.
The estimate of $x_{u2}$
agrees very well with the experimental \cite{Schwaiger_EMBO05}
as well as with the previous theoretical result \cite{MSLi_JCP08}.
The present value of
$x_{u1}$ agrees with the experiments better than the old one
\cite{MSLi_JCP08}.
Presumably, this is because it has been estimated by the same procedure
as in the experiments \cite{Schwaiger_EMBO05}.

It is important to note that
the logarithmic behavior is observed only at low enough $v$. At high
loading rates, the dependence of $f_{max}$ on $v$ becomes power-law.
This explains why all-atom simulations, performed 
at $v \sim  10^9$ nm/s for most of proteins, are not able to provide
reasonable estimations for $x_u$.

The another interesting question is if the peak at 
$\Delta R \approx 1.5$ nm disappears at loading rates used in the experiments
\cite{Schwaiger_EMBO05}. Assuming that the logarithmic dependence
in Fig. \ref{fmax_Nfix_v_fig} has the same slope at low $v$, we interpolate
our results to $v_{exp} = 200$ nm/s and obtain 
$f_{max1}(v_{exp}) \approx 40$ pN.
Thus, in the framework of the Go model, the existence of
the first peak is robust at experimental speeds.

\paragraph{Beyond the BER approximation}

In the BER approximation, one assumes that the location
of the TS does not move under the action of an
external force.
Beyond this approximation, $x_u$ and unfolding barriers can be extracted,
using the following formula 
\cite{Dudko_PRL06}:
\begin{equation}
f_{max} \,  =  \frac{\Delta G^{\ddagger}}{\nu x_u} \left\{ 1- 
\left[\frac{k_BT}{\Delta G^{\ddagger}} \textrm{ln} \frac{k_BT k_u(0) e^{\Delta G^{\ddagger}/k_BT + \gamma}}{x_u v}\right]^{\nu} \right\}
\label{Dudko_eq2}
\end{equation}
Here, $\Delta G^{\ddagger}$ is the unfolding barrier, $\nu = 1/2$ and 2/3
for the cusp \cite{Hummer_BJ03} and the
linear-cubic free energy surface \cite{Dudko_PNAS03}, respectively.
$\gamma \approx 0.577$ is the Euler-Mascheroni constant.
Note that
$\nu =1$ corresponds to the phenomenological
BER theory (Eq. \ref{f_logV_eq2}).
If $\nu \ne 1 $, then
Eq. (\ref{Dudko_eq2}) can be used to estimate not only
$x_u$, but also $\Delta G^{\ddagger}$.
Since the fitting with $\nu = 1/2$ is valid in a wider force
 interval
compared to the $\nu = 2/3$ case, we
consider the former case only.
The region,
where the $\nu = 1/2$ fit works well, is expectantly wider  than that for
the Bell scenario (Fig. \ref{fmax_Nfix_v_fig}). 
From the nonlinear fitting (Eq. \ref{Dudko_eq2}),
we obtain 
$x_{u1}=7.0 \AA\,$, and
$x_{u2} = 9.7 \AA\,$ which
are about twice as large as the Bell estimates (Table \ref{DDFLN4_table}).
Using AFM data, Schlierf and Rief \cite{Schlierf_BJ06},
have shown that beyond BER approximation
$x_u \approx 11 \AA\,$. This value is close to our estimate for $x_{u2}$.
However, a full comparison with experiments is not possible as
these authors did not consider $x_{u1}$ and $x_{u2}$ separately.
The present estimations of these quantities are
clearly lower than the previous one \cite{MSLi_JCP08} (Table \ref{DDFLN4_table}).
The lower values
of $x_{u}$ would be more favorable because they are expected to
be not high for beta-rich proteins \cite{MSLi_BJ07a} like DDFLN4. 
Thus, beyond BER approximation,
the method based on Eq. (\ref{Dudko_eq2}) provides more reasonable
estimations for $x_{ui}$ compared to the method, where these
parameters are extracted
from unfolding rates \cite{MSLi_JCP08}. However,
in order to decide what method is better,
more experimental studies are required. 

The corresponding values for $\Delta G^{\ddagger}_1$, and $\Delta G^{\ddagger}_2$
are listed in Table \ref{DDFLN4_table}.  
The experimental and previous theoretical
results \cite{MSLi_JCP08} are also shown for comparison.
The present estimates for both barriers agree with the 
experimental data, while  
the previous theoretical value of $\Delta G^{\ddagger}_1$ 
fits to experiments worse than
the current one.
\begin{table}
\begin{center}
\begin{tabular}{lll|lllr}
& \multicolumn{2}{c|}{BER approximation} &\multicolumn{4}{c}{Beyond BER approximation}\\ \cline{2-7}
& \; $x_{u1}(\AA)$ \;& \; $x_{u2}(\AA)$ \; & \; $x_{u1}(\AA)$ \;& \; $x_{u2}(\AA)$ \; & \; $\Delta G^{\ddagger}_1/k_BT \;$ &  \; $\Delta G^{\ddagger}_2/k_BT \; $ \\
\hline
Theory \cite{MSLi_JCP08}&\; 6.3 $\pm$ 0.2 \; & \; 5.1 $\pm$ 0.2 \;&\; 13.1 \; &\; 12.6 \; & \; 25.8 \; & \; 18.7 \; \\
Theory (this work)&\; 3.2 $\pm$ 0.2 \; & \; 5.5 $\pm$ 0.2 \;&\; 7.0 \;& \; 9.7 \; & \; 19.9 \; & \; \; 20.9 \; \\
Exp. \cite{Schwaiger_EMBO05,Schlierf_BJ06} & \; 4.0 $\pm 0.4$ \; & \; 5.3 $\pm$ 0.4 \; & & &\; 17.4 \; & \; 17.2 \;\\
\hline
 \end{tabular}
\end{center}
\linespread{0.8}
\caption{Parameters $x_{u1}$, and $x_{u2}$ were obtained in the
Bell and beyond-Bell approximation. Theoretical values of the unfolding
barriers were extracted from the microscopic theory of Dudko {\em et al}
(Eq. \ref{Dudko_eq})
with $\nu = 1/2$. The experimental estimates were taken from
Ref. \onlinecite{MSLi_JCP08}.\label{DDFLN4_table}}
\end{table}

\subsubsection{Thermal unfolding pathways}

\begin{figure}[!hbtp]
\epsfxsize=4.2in
\centerline{\epsffile{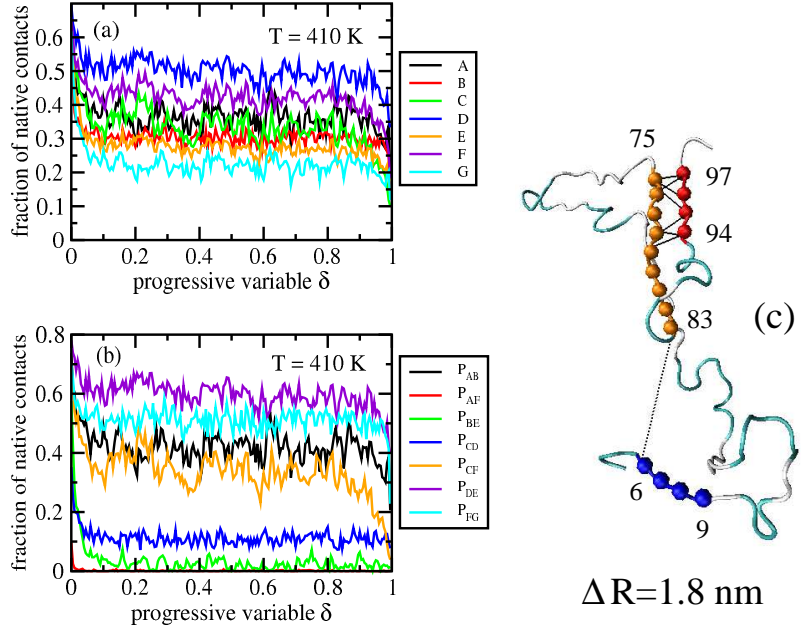}}
\linespread{0.8}
\caption{Thermal unfolding pathways. (a) Dependence of
native contact fractions of
seven strands on the progress variable $\delta$ at $T=410$ K.
(b) The same as in (a) but for seven strand pairs.
(c) A typical snapshot at
$\Delta R \approx 1.8$ nm. The contact between
strands S1 and S6 is broken but 7 contacts between strands S6 and S7
(solid lines) still
survive.}
\label{ther_unfold_pathways_snap_fig}
\end{figure}

In order to see if the thermal unfolding pathways are different
from the mechanical ones, we performed zero-force simulations
at $T=410$ K.  The progress variable $\delta$ is used
as a reaction coordinate to monitor pathways (see Chapter 3).
From Fig. \ref{ther_unfold_pathways_snap_fig}, we have  the following sequencing
for strands and their pairs:
\begin{subequations}
\begin{equation}
G \rightarrow (B, C, E)  \rightarrow (A, F, D),
\label{thermal_pathway_str_eq}
\end{equation}
\begin{equation}
P_{AF}  \rightarrow P_{BE}  \rightarrow (P_{CD}, P_{CF})  \rightarrow 
(P{AB}, P_{FG}, P_{DE}).
\label{thermal_pathway_pair_eq}
\end{equation}
\end{subequations}
It should be noted that these pathways are just 
major ones as other pathways
are also possible. The pathway given by Eq. (\ref{thermal_pathway_pair_eq}),
e.g.,  occurs in 35\% of events.
About 20\% of trajectories follow
$P_{AF}  \rightarrow P_{CF}  \rightarrow P_{BE} \rightarrow
(P_{CD},P{AB}, P_{FG}, P_{DE})$ scenario. We have also observed the sequencing
 $P_{AF}  \rightarrow P_{BE} \rightarrow
(P_{CF},P{AB}, P_{FG}, P_{DE}) \rightarrow P_{CD}$, and
$P_{BE}  \rightarrow P_{AF} \rightarrow
(P_{CD},P{CF}, P_{AB}, P_{FG},P_{DE})$ in 12\% and 10\% of runs, respectively.
Thus,
due to strong thermal fluctuations,
thermal unfolding pathways are more diverse compared to mechanical ones.
From Eqs. \ref{pathways_v6_strand_eq}, \ref{pathways_v6_pair_eq},
\ref{pathways_v15_strand_eq}, \ref{pathways_v15_pair_eq},
\ref{thermal_pathway_str_eq}, and \ref{thermal_pathway_pair_eq}, it is clear that thermal unfolding
pathways of DDFLN4 are different from the mechanical pathways.
This is also
illustrated in 
Fig. \ref{ther_unfold_pathways_snap_fig}c.
As in the mechanical case
(Fig. \ref{snapshot_v6_v15_fig}a and \ref{snapshot_v6_v15_fig}b),
 the contact between A and F is broken,
but the molecule is much
less compact at the same end-to-end distance. 
Although 7 contacts ($\approx 64$\%) between strands
F and G remain survive, all contacts of pairs
$P_{AF}, P_{BE}$ and $P_{CD}$ are already broken.

The difference between mechanical and thermal unfolding pathways
is attributed to the fact
that thermal fluctuations have a global effect on the biomolecule,
while the force acts only on its termini. Such a difference was also observed
for other proteins like I27 \cite{Paci_PNAS00} and Ub
\cite{MSLi_BJ07,Mitternacht_Proteins06}.
We have also studied folding pathways of DDFLN4 at $T=285$ K. It turns
out that they are reverse of the thermal unfolding pathways given
by Eqs. \ref{thermal_pathway_str_eq} and  \ref{thermal_pathway_pair_eq}.
It would be interesting to test our prediction on thermal folding/unfolding
of this domain experimentally.

\subsection{Conclusions}

The key result of this chapter is that
mechanical unfolding pathways of DDFLN4 depend on loading rates.
At large $v$ the C-terminal unfolds first, but the N-terminal
unfolds at low $v \sim 10^4$ nm/s. The agreement with the
experiments \cite{Schwaiger_NSMB04}
is obtained only
in low loading rate simulations.
The dependence of mechanical unfolding pathways on the loading rates
was also observed for I27 (M.S. Li, unpublished). On the other hand,
the previous studies \cite{Irback_PNAS05,MSLi_BJ07} showed that mechanical unfolding pathways
of the two-state Ub do not depend on the force strength.
Since DDFLN4 and I27 are three-state
proteins, one may think that
the unfolding pathway change with variation of the pulling
speed,
is universal 
for proteins that unfold via intermediates.
A more comprehensive study is needed to verify this 
interesting issue.

Dependencies of unfolding forces on pulling speeds have been widely used
to probe FEL of two-state proteins \cite{Best_PNAS02}.
However, to our best knowledge,
here we have made a first attempt to apply this approach
to extract not only
$x_{ui}$, but also 
$\Delta G^{\ddagger}_i$ ($i= 1,$ and 2) for a three-state protein.
This allows us to improve our previous results \cite{MSLi_JCP08}.
More importantly, a better agreement with the experimental data
\cite{Schwaiger_EMBO05,Schlierf_BJ06} suggests that this method
is also applicable to
other multi-state  biomolecules.
Our study clearly shows that the low loading
rate regime, where FEL parameters can be estimated, occurs at 
$ v \leq 10^6$ nm/s which are about two-three orders of magnitude
lower than those used in all-atom simulations.
Therefore, at present, deciphering unfolding FEL of long proteins by
all-atom simulations with explicit water is computationally prohibited.
From this prospect, coarse-grained models are of great help.

We predict the existence of a peak at $\Delta R \sim 1.5$ nm even
at pulling speeds used in now a day experimental setups.
This result would stimulate
new experiments on mechanical properties of DDFLN4.  
Capturing the experimentally observed peak at $\Delta R \sim 22$ nm
remains a challenge to theory.

\clearpage

\begin{center}
\section{Protein mechanical unfolding: importance of non-native interactions}
\end{center}

\subsection{Introduction}

In this chapter, we continue to study the mechanical unfolding of DDFLN4 using the all-atom simulations. Motivation for this is that Go model can not explain some experimental results. 
Namely, in the AFM force-extension curve (Schwaiger {\em et al.} \cite{Schwaiger_NSMB04,Schwaiger_EMBO05} observed two peaks at $\Delta R \approx 12$ and 22 nm.
However, using a Go model \cite{Clementi_JMB00}, Li {\em et al.}\cite{MSLi_JCP08} and Kouza and Li (chapter 9)   have also obtained two peaks but they are located at
$\Delta R \approx 1.5$ and 11 nm. A natural question to ask is if the disagreement
between experiments and theory is due to over-simplification of 
the Go modeling, where non-native interactions between residues are omitted.
In order to answer this question, we have performed
all-atom
MD simulations,
using the GROMOS96 force field 43a1 \cite{Gunstren_96} and the SPC explicit water solvent \cite{Berendsen81}.

We have shown that, 
two peaks do appear at almost
the same positions as in the experiments \cite{Schwaiger_NSMB04,Schwaiger_EMBO05}  and more importantly, the peak at $\Delta R \approx 22$ nm comes from the non-native interactions. It explains why it has not been seen in the previous Go simulations\cite{MSLi_JCP08}.
 In our opinion, this result is very important as it opposes to the common belief
\cite{West_BJ06,MSLi_BJ07a} that mechanical unfolding properties are governed by the native topology.
In addition to two peaks at large $\Delta R$, in agreement with the Go results \cite{MSLi_JCP08},
 we have also observed a maximum at $\Delta R \approx 2$ nm.  Because such a peak
was not detected by the AFM experiments \cite{Schwaiger_NSMB04,Schwaiger_EMBO05},
further experimental and theoretical studies are required to clarify this point.

 The results of this chapter are adapted from Ref. \cite{Kouza_JCP09}.

\subsection{Materials and Methods}

\begin{figure}[!htbp]
\epsfxsize=6.3in
\vspace{0.2in}
\centerline{\epsffile{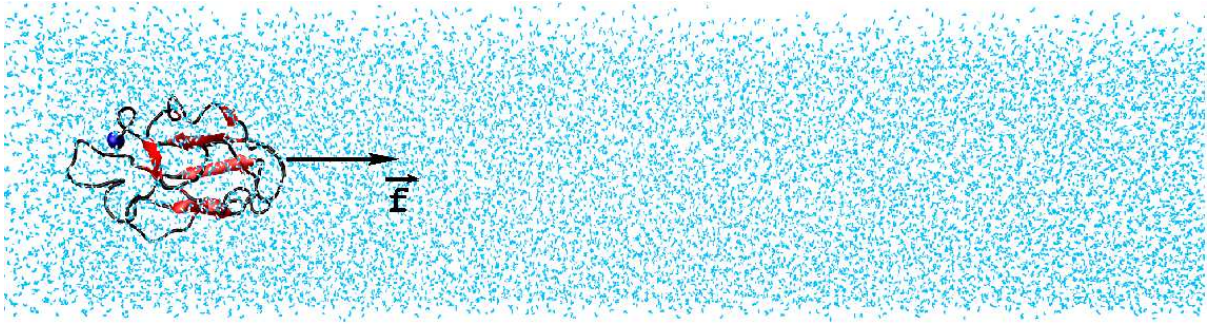}}
\linespread{0.8}
\caption{The solvated system in the orthorhombic box of water (cyan).
VMD software \cite{VMD}
 was used for a plot.
\label{native_ddfln4_strands_fig2}}
\end{figure}

 We used the GROMOS96 force field 43a1 \cite{Gunstren_96} to model
DDFLN4 which has 100 amino acids, and the SPC water model \cite{Berendsen81}
 to describe the solvent (see also chapter 4). The Gromacs version 3.3.1 has been employed.
The protein was placed in an cubic box with the 
edges of 4.0, 4.5 and 43 nm, and with 76000 - 78000 water molecules (Fig. \ref{native_ddfln4_strands_fig2}).

In all simulations, the GROMACS program suite \cite{Berendsen_CPC95,Lindahl01}
 was employed.  The equations of motion were integrated by
using a leap-frog algorithm with a time step of 2 fs.
The LINCS \cite{Hess_JCC97} was used to constrain bond lengths 
with a relative geometric tolerance of $10^{-4}$. 
We used the
particle-mesh Ewald method to treat the long-range electrostatic
interactions \cite{Darden93}.
 The nonbonded interaction pair-list were
updated every 10 fs, using a cutoff of 1.2 nm.

The protein was minimized using the steepest decent
method. Subsequently, unconstrained MD
simulation was performed to equilibrate the solvated system  for 100 ps
at constant pressure (1 atm) and temperature $T=300$ K with the help of
the Berendsen coupling procedure \cite{Berendsen84}.
The system was then equilibrated further at
constant temperature $T$ = 300 K and constant volume.
Afterward, the N-terminal was kept fixed and the
 force was applied to the C-terminal through a virtual cantilever moving
 at the constant velocity $v$ along the biggest $z$-axis of simulation box.
 During the simulations, the spring constant was chosen
 as $k=1000 kJ/(mol \times nm^2) \approx 1700 $ pN/nm which is an upper
limit for $k$ of a
 cantilever used in AFM experiments.
 Movement of the pulled termini causes an extension of
 the protein and the total force can be measured by $F=kvt$.
 The resulting force is computed for each time step to generate a
 force extension profile, which has peaks showing the most mechanically
 stable places in a protein.

Overall, the simulation procedure is similar to the experimental one,
except that pulling speeds in our simulations
 are several orders of magnitude higher than those used in experiments.
We have performed simulations for 
$v= 10^{6}, 5\times 10^{6}, 1.2\times 10^{7}$,
and $2.5\times 10^{7}$ nm/s, while in the AFM experiments one took
$v \sim 100 - 1000$ nm/s \cite{Schwaiger_NSMB04}.
For each value of $v$ we have generated 4 trajectories.

A backbone contact between amino acids $i$ and $j$ ($|i-j| > 3$)
is defined as formed if the distance between
 two corresponding C$_{\alpha}$-atoms
is smaller than a cutoff distance $d_c=6.5$ \AA . 
With this choice, the molecule has 163 native contacts.
A hydrogen bond is formed provided
 the distance between donor D (or atom N) and
acceptor A (or atom O)  $\leq 3.5 \AA \,$ and the angle D-H-A 
$\ge 145^{\circ}$.

The unfolding process was studied by monitoring the dependence
of numbers of backbone  contacts and HBs
formed by seven $\beta$-strands enumerated as
A to G (Fig. \ref{native_ddfln4_strands_fig}a)
 on the end-to-end extension.
In the NS, backbone contacts exist between seven pairs
of $\beta$-strands
P$_{\textrm{AB}}$,
P$_{\textrm{AF}}$, P$_{\textrm{BE}}$,
P$_{\textrm{CD}}$, P$_{\textrm{CF}}$, P$_{\textrm{DE}}$, and P$_{\textrm{FG}}$
as shown in Fig. \ref{native_ddfln4_strands_fig}b.
Additional information on unfolding pathways was also obtained
from the evolution of numbers of contacts
of these pairs.

\subsection{Results}

\subsubsection {Existence of three peaks in force-extension profile}

Since the results obtained for four pulling speeds ({\em Material and Methods})
are qualitatively similar, we will focus on the smallest 
$v=10^6$ nm/s case.
The force extension curve, obtained at
this speed, for the trajectory 1, can be divided into four regions (Fig. \ref{fe1}):
\begin{figure}[!htbp]
\epsfxsize=4.0in
\vspace{0.2in}
\centerline{\epsffile{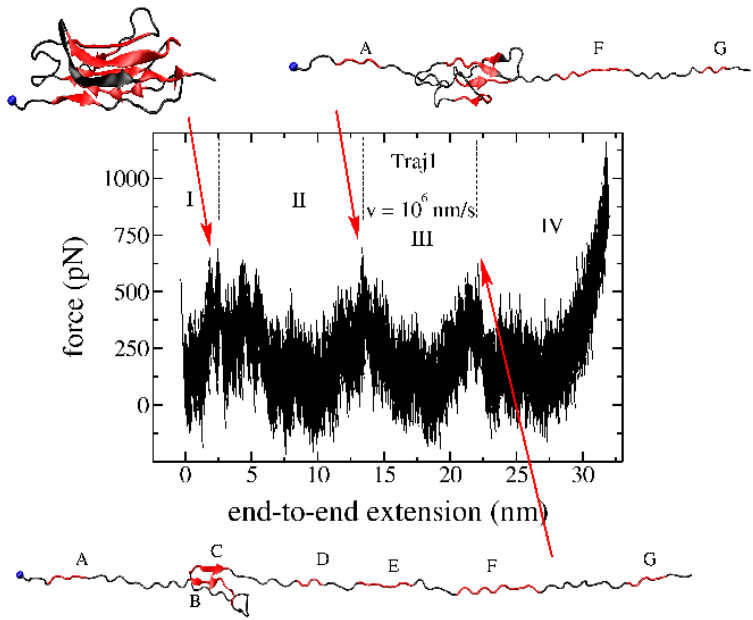}}
\linespread{0.8}
\caption{Force-extension profile for trajectory 1 for $v=10^6$ nm/s.
Vertical dashed lines separate four unfolding regimes.
Shown are typical snapshots around three peaks.
Heights of peaks (from left) are $f_{max1}=695$ pN, $f_{max2}=704$ pN,
and $f_{max3}=626$ pN.
\label{fe1}}
\end{figure}

\begin{figure}
\vspace{5.5 mm}
\includegraphics[width=0.97\textwidth]{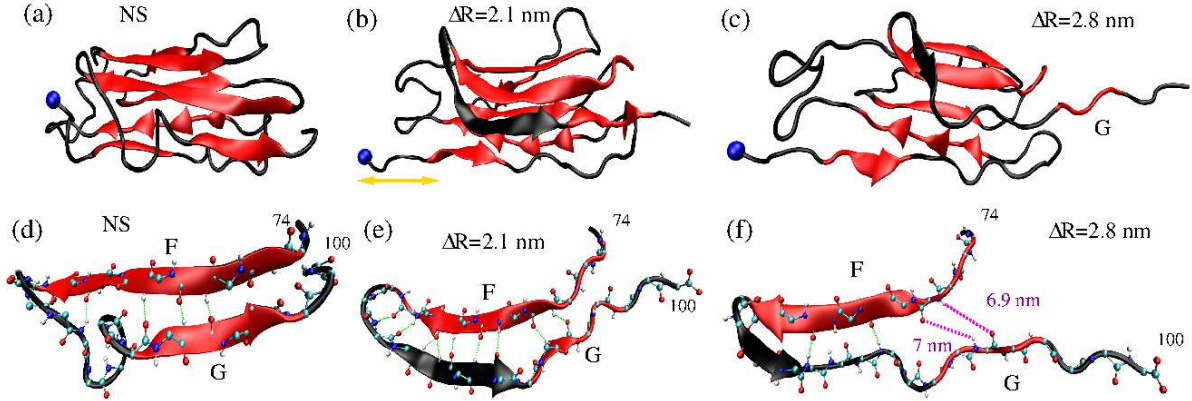}
\linespread{0.8}
\caption{(a) The NS conformation is shown for comparison with
the other ones. (b) A typical conformation before the first unfolding
event takes place ($\Delta R \approx$ 2.1 nm).
The yellow arrow shows
a part of protein which starts to unfold. An additional
non-native $\beta$-strand between amino acids 87 and 92 is marked by
black color. (c)
A conformation after the first peak, at $\Delta R \approx$ 2.8 nm,
where strand G has already
detached from the core. (d) The same as in (a) but
4 HBs
(green color) between $\beta$-strands are displayed. (e) The same as in (b)
 but all 9 HBs are shown. 
(f) The same as in (c) but 
broken HBs (purple) between F and G  are displayed.
\label{2nm}}
\end{figure}

 {\em Region I ($0 \lesssim \Delta R \lesssim 2.42$ nm)}.
Due to thermal fluctuations, the total force fluctuates a lot,
 but, in general,
it increases and reaches the first maximum $f_{max1}=695$ pN
at $\Delta R$ 2.42 nm.
A typical snapshot before the first unfolding event (Fig. \ref{fe1})
 shows that structures remain native-like.
During the first period, the N-terminal part is being extended,
 but the protein maintains all $\beta$-sheet secondary structures
(Fig. \ref{2nm}b).
 Although, the unfolding starts from the N-terminal (Fig. \ref{2nm}b),
 after the first peak, strand G from
the C-termini got unfolded first (Fig. \ref{2nm}c and \ref{2nm}f).
 In order to understand the nature of this peak
on the molecular level, we consider the evolution of HBs in detail.
As a molecule departs from the NS, 
 non-native HBs are created and at $\Delta R = 2.1$ nm, e.g.,
 a non-native $\beta$-strand between amino acids 87 and
92 (Fig. \ref{2nm}b)
 is formed. This leads to
increase of the number of HBs between F and G
from 4 (Fig. \ref{2nm}d) to 9 (Fig. \ref{2nm}e).
Structures with the enhanced number of HBs should show strong resistance to
the external perturbation and the first peak occurs due to
their unfolding (Fig. \ref{2nm}b).
It should be noted that this maximum was observed in
the Go simulations \cite{MSLi_JCP08,MSLi_JCP09},
 but not in the experiments \cite{Schwaiger_NSMB04,Schwaiger_EMBO05}.
Both all-atom and Go simulations reveal that the unfolding
of G strand is responsible for its occurrence.


{\em Region II ($2.42 nm \lesssim \Delta R \lesssim 13.36$ nm):}
After the first peak, the force drops rapidly
 from 695 to 300 pN and secondary structure elements begin to break down.
During this period, strands  A, F and G unfold completely,
whereas  B, C, D and E strands remain structured 
(see Fig. \ref{fe1}
 for a typical snapshot).

{\em Region III ($13.36 nm \lesssim \Delta R \lesssim 22.1$ nm:}
During the second  and third stages, the complete unfolding of
strands D and E takes place. Strands
B and C undergo significant conformational changes,
losing their equilibrium HBs. Even though a core
formed by these
strands remains compact (see bottom of Fig. \ref{fe1}
 for a typical snapshot).
Below we will show in detail that
the third peak is associated with breaking
of non-native HBs between strands B and C.

{\em Region IV ($\Delta R \gtrsim 22.1$ nm:} After
breaking of non-native HBs between B and C, 
the polypeptide chain gradually reaches its rod state.

\begin{figure}[!htbp]
\epsfxsize=3.5in
\vspace{0.2in}
\centerline{\epsffile{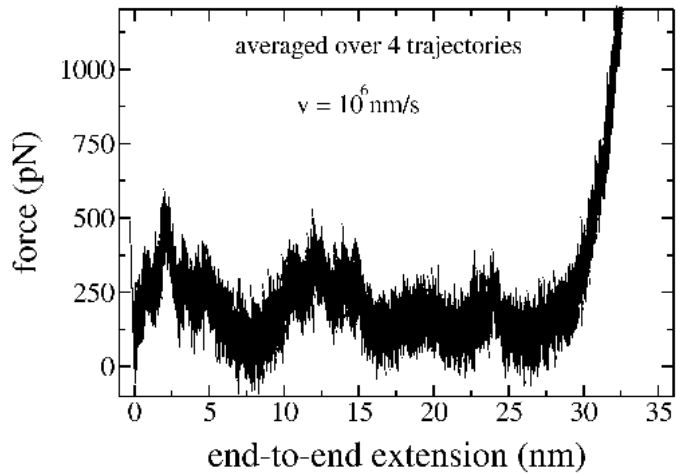}}
\linespread{0.8}
\caption{The averaged over 4 trajectories force-extension profile.
$v=10^6$ nm/s.}
\label{fea}
\end{figure}

The existence of three pronounced peaks is robust as they are observed
in all four studied trajectories
(similar results obtained in other three runs are not shown).
It is also clearly evident from Fig. \ref{fea},
 which displays the force-extension curve averaged over 4
trajectories.

\subsubsection{Importance of non-native interactions}

\begin{figure}[!htbp]
\epsfxsize=3.8in
\vspace{0.2in}
\centerline{\epsffile{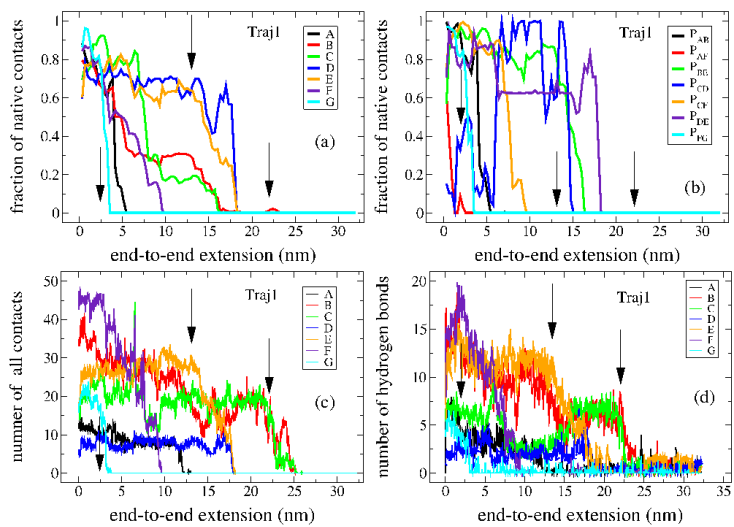}}
\linespread{0.8}
\caption{(a) Dependence of the number of native backbone contacts
formed by individual strands on $\Delta R$. Arrows refer to
positions of three peaks in the force-extension curve. (b) The same
as in (a) but for pairs of strands. (c)
The same as in (a) but for all contacts (native and non-native).
(d) The same as in (c) but for HBs.
\label{cont_ext_traj1_fig}}
\end{figure}

As mentioned above, the third peak at $\Delta R \approx 22$ nm was observed in the experiments
but not in Go models \cite{MSLi_JCP08,MSLi_JCP09}, where non-native interactions are omitted.
In this section, we show, at molecular level, that these very interactions lead to its existence. 
To this end, we plot the dependence of the number of native contacts formed by seven strands and their pairs on $\Delta R$.
The first peak corresponds to unfolding of strand G (Fig. \ref{cont_ext_traj1_fig}a)
 as all (A,F) and (F,G) contacts are broken just after passing it (Fig. \ref{cont_ext_traj1_fig}b).
Thus, the structure of the first IS1, which corresponds to this peak, consists of 6 ordered strands A-F
(see Fig. \ref{2nm}c for a typical snapshot).

The second unfolding event
is associated with full unfolding of A and F and drastic decrease of 
native contacts of B and C (Fig. \ref{cont_ext_traj1_fig}a. 
After the second peak
only (B,E), (C,D) and (D,E) native contacts survive (\ref{cont_ext_traj1_fig}b).
The structure of the second intermediate state (IS2) contains partially
structured strands B, C, D and E. A typical snapshot is displayed
in top of Fig. \ref{fe1}.

Remarkably, for $\Delta R \gtrsim 17$ nm, none of native contacts exists,
except very small fluctuation of a few contacts of strand
 B around $\Delta R \approx 22.5$ nm (Fig. \ref{cont_ext_traj1_fig}a).
Such a fluctuation is negligible as it
is not even manifested  in existence of native contacts
between corresponding pairs (A,B) and (B,E) (Fig. \ref{cont_ext_traj1_fig}b).
Therefore, we come to a very interesting conclusion that the third peak
centered at $\Delta R \approx 22.5$ nm is not related to native interactions.
This explains why it was not detected by simulations \cite{MSLi_JCP08,MSLi_JCP09}
 using the Go model \cite{Clementi_JMB00}.

\begin{figure}[!htbp]
\includegraphics[width=0.48\textwidth]{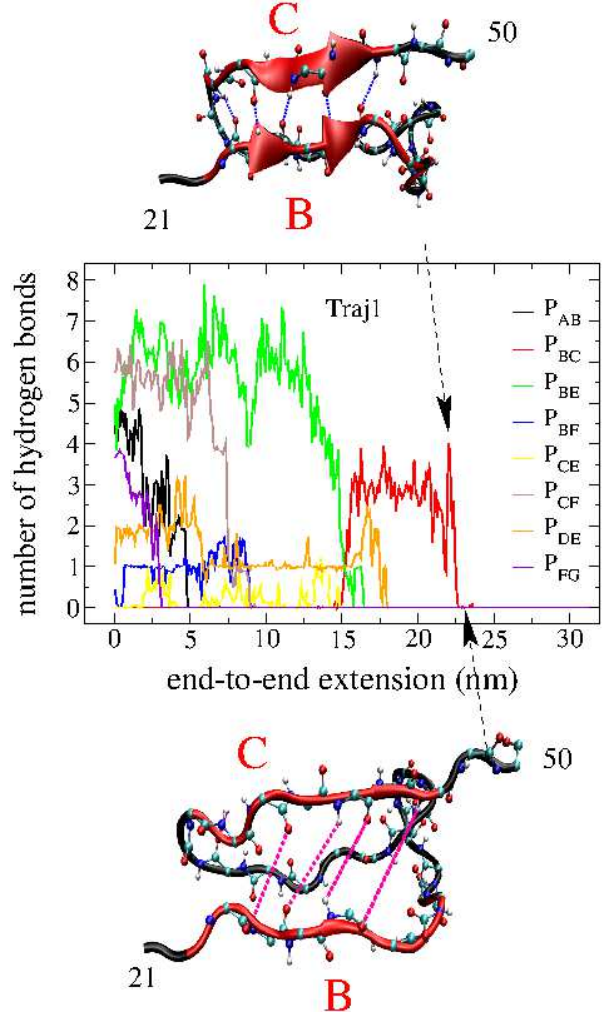}
\linespread{0.8}
{\caption {Dependence of the number of HBs between pairs of strands.
Red arrow refers to a position where non-native HBs between strands
B and C start to appear. Their
creation leads to the maximum centered at $\Delta R \approx 22.4$ nm.
Upper snapshot shows five HBs between B an C before the third
unfolding event.
Lower snapshot is a fragment after the third peak,
 where all HBs are already broken
(purple dotted lines).}
\label{HB_pair_ext_traj1_fig}}
\end{figure}

The mechanism underlying occurrence of the third peak may be revealed
using the results shown in Fig. \ref{cont_ext_traj1_fig}c, where
the number of all backbone contacts (native and non-native) 
is plotted as a function of $\Delta R$. Since,
for $\Delta R \gtrsim 17$ nm,  native contacts vanish, this peak
is associated with an abrupt decrease of non-native contacts between strands
B and C. Its nature may be also understood by monitoring
the dependence of HBs on $\Delta R$ (Fig. \ref{cont_ext_traj1_fig}d),
which shows that
the last maximum is caused by loss of
HBs of these strands.
More precisely, five HBs between B and C, which were not present in
the native conformation, are broken (Fig. \ref{HB_pair_ext_traj1_fig}). 
Interestingly, these bonds appear at $\Delta R \gtrsim$ 15 nm, i.e.
after the second unfolding event (Fig. \ref{HB_pair_ext_traj1_fig}).
Thus, our study can not only reproduce the experimentally observed
peak at $\Delta R \approx 22$ nm, but also shed light on its nature
on the molecular level.
From this perspective, all-atom simulations are superior to experiments.

One corollary from Fig. \ref{cont_ext_traj1_fig}a-d is that one can not
provide a complete description of the unfolding process based on the evolution
of only native contacts. 
It is because, as a molecule extends, its
secondary structures 
 change and new non-native secondary structures may occur.
Beyond the extension of 17-18 nm (see snapshot at bottom of Fig. \ref{fe1}),
 e.g., the protein
lost all native contacts, but it does not get a 
extended state without any structures.
Therefore, a full description of mechanical unfolding may be obtained
by monitoring either  all backbone contacts or HBs,
as these two quantities give the same unfolding picture
(Fig. \ref{cont_ext_traj1_fig}c and \ref{cont_ext_traj1_fig}d).

\subsubsection{Unfolding pathways}

\begin{figure}
\includegraphics[width=5in]{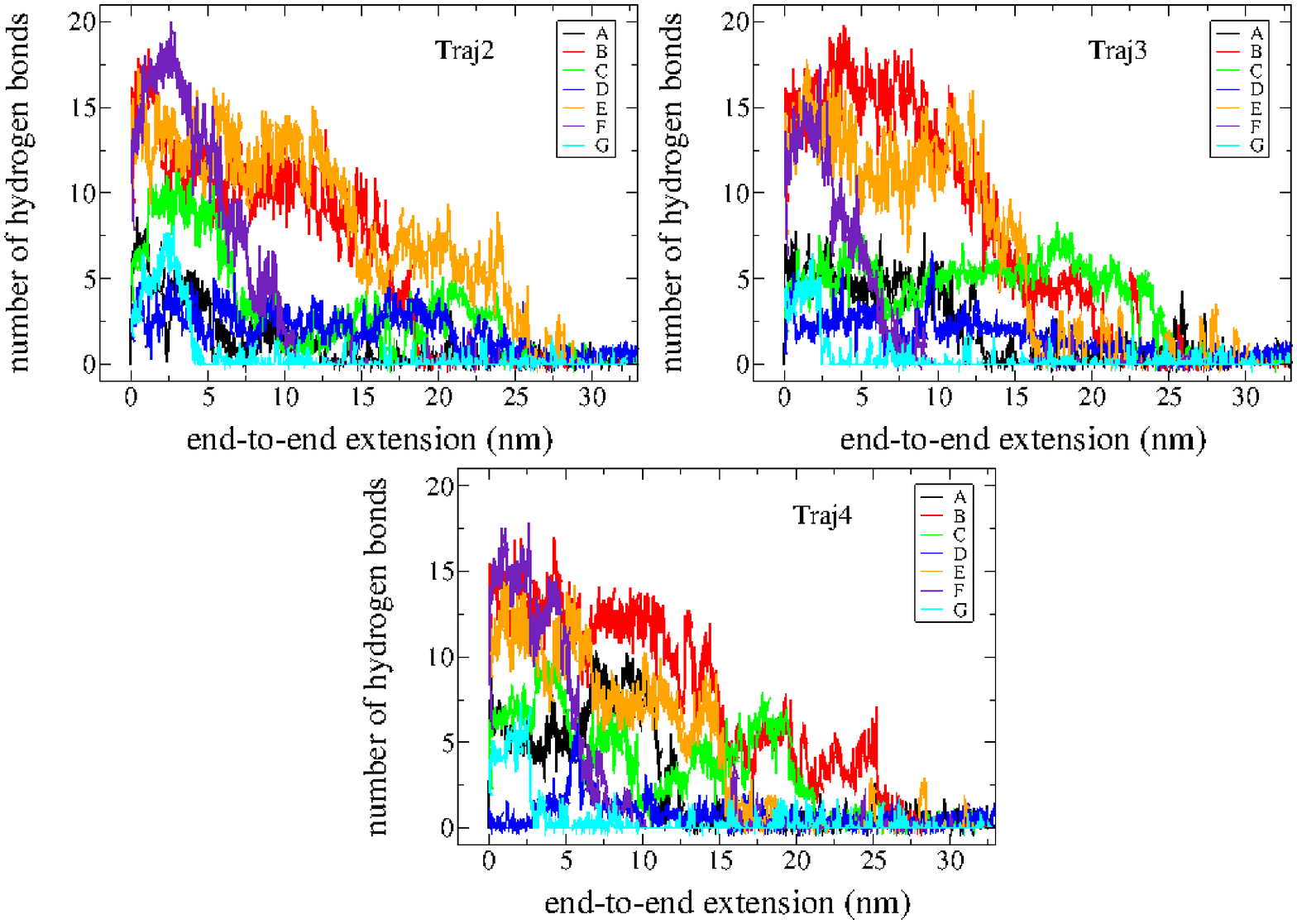}
\linespread{0.8}
\caption {The $\Delta R$ dependence of the
number of HBs, formed by
seven  strands, for trajectory 2, 3 and 4.
$v=10^6$ nm/s.
\label{HB_pair_ext_traj2_4_fig}}
\end{figure}

To obtain sequencing of unfolding events,
we use dependencies of the number of HBs on $\Delta R$.
From Fig. \ref{cont_ext_traj1_fig}d
  and  Fig. \ref{HB_pair_ext_traj2_4_fig},
 we have the following unfolding pathways for four trajectories:
\begin{eqnarray}
G \rightarrow F \rightarrow A  \rightarrow (D,E) \rightarrow (B,C), \; \;
\textrm{Trajectory 1},  \nonumber\\
G \rightarrow F \rightarrow A  \rightarrow B \rightarrow C \rightarrow (D,E),
\; \; \textrm{Trajectory 2}, \nonumber  \\
G \rightarrow F \rightarrow A  \rightarrow E \rightarrow B \rightarrow D \rightarrow
C, \; \; \textrm{Trajectory 3} \nonumber\\
G \rightarrow F \rightarrow A  \rightarrow (D,E) \rightarrow C \rightarrow B,
\; \; \textrm{Trajectory 4}.
\label{pathways_eq}
\end{eqnarray}
Although four pathways, given by Eq. (\ref{pathways_eq}) are different,
they share a common feature that the C-terminal unfolds first.
This is consistent with the results obtained by Go simulations
at high pulling speeds $v \sim 10^6$ nm/s \cite{MSLi_JCP08},
 but contradicts to the experiments \cite{Schwaiger_NSMB04,Schwaiger_EMBO05}, 
which showed that strands A and B from the N-termini unfold first.
On the other hand, our more recent Go simulations \cite{MSLi_JCP09}
have revealed that the agreement with the experimental results
is achieved if one performs simulations at relatively low
pulling speeds $v \sim 10^4$ nm/s.
Therefore, one can expect that the difference in
sequencing of unfolding events between present
all-atom results and the experimental ones is merely due to
large values of $v$ we used. In order to check this,
one has to carry out all-atom simulations, at least,
at $v \sim 10^4$ nm/s, but such a task is
far beyond present computational facilities.

\subsubsection{Dependence of unfolding forces on the pulling speed}

\begin{figure}[!hbtp]
\epsfxsize=3.4in
\centerline{\epsffile{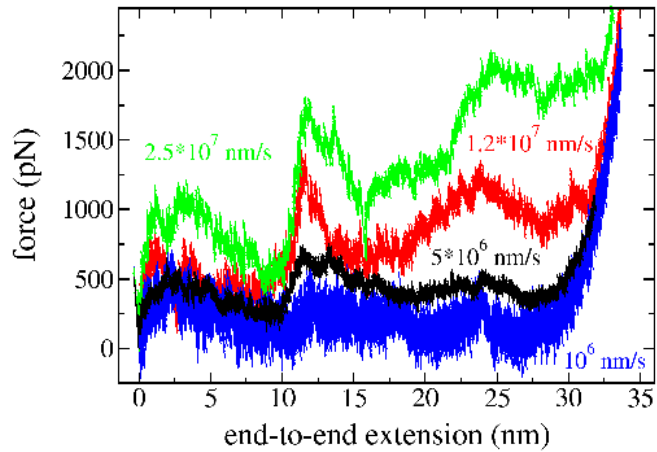}}
\linespread{0.8}
\caption{Force-extension profiles for four values of $v$ shown next to the
curves.
\label{force_extension_all_fig}}
\end{figure}

The question we now ask is whether
the unfolding FEL of DDFLN4 can be probed by
all-atom simulations with
explicit water.
To this end, we performed simulations at various loading speeds
and  monitor the dependence of
$f_{maxi} (i=1,2,$ and 3) on $v$ 
(Fig. \ref{force_extension_all_fig}).
%

\begin{figure}[!htbp]
\epsfxsize=3.5in
\centerline{\epsffile{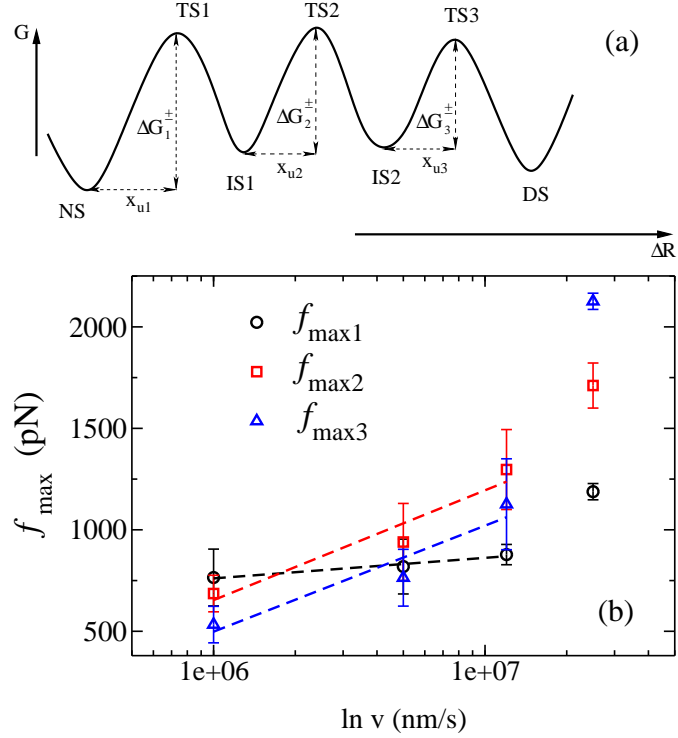}}
\linespread{0.8}
\caption{(a) Schematic plot for the free energy $G$ as a function of $\Delta R$. $\Delta G^{\ddagger}_i (i=1,2,$ and 3) refers to unfolding barriers.
The meaning of other notations is given in the text.
(b) Dependence of heights of three peaks on $v$. Results are averaged over
four trajectories for each value of $v$.
Straight lines refer to 
linear fits by Eq. (\ref{f_logV_eq2}) ($y_1 = 163 + 44x, y_2 = -2692 + 235x$ and
 $y_3 = -2630 + 227x$) through three low-$v$ data points.
These fits give $x_{u1} = 0.91 \AA, x_{u2} =
 0.17 \AA \,$,  and $x_{u3} =0.18 \AA$.
\label{fel_schem_fmax_logv_fig}}
\end{figure}

%
In accordance with theory \cite{Evans_BJ97},
 heights of three peaks decrease as $v$ is lowered (Fig. \ref{force_extension_all_fig}).
Since the force-extension curve displays three peaks, within the framework
of all-atom models, the mechanical unfolding of DDFLN4 follows 
a four-state scenario (Fig. \ref{fel_schem_fmax_logv_fig}a),
but not the three-state one as suggested by the experiments
\cite{Schwaiger_NSMB04,Schwaiger_EMBO05} and Go simulations \cite{MSLi_JCP08}.
The corresponding FEL should have three transition states
denoted by TS1, TS2 and TS3. Remember that the first
and second peaks in the force-extension profile correspond
to IS1 and IS2.

Assuming that the BER theory \cite{Bell_Sci78,Evans_BJ97}
holds for a four-state biomolecule, one can extract the distances
$x_{u1}$ (between NS and TS1), $x_{u2}$
(between IS1 and TS2),
and $x_{u3}$ (between IS2 and TS3) from Eq. (\ref{f_logV_eq2}).
%
From the linear fits (Fig. \ref{fel_schem_fmax_logv_fig}b),
 we have $x_{u1} = 0.91 \AA , x_{u2} = 0.17 \AA \, ,$ and $x_{u3} = 0.18 \AA$.
These values are far below the typical $x_u \approx 5 \AA\,$, obtained in
the experiments \cite{Schwaiger_EMBO05}
as well as in the Go simulations \cite{MSLi_JCP08,MSLi_JCP09}.
This difference comes from the fact that pulling speeds used in all-atom
simulations are to high (Fig. \ref{fel_schem_fmax_logv_fig}).
It clearly follows from Eq. (\ref{f_logV_eq2}), which shows that $x_u$ depends
on what interval of $v$ we use: the larger are values of $f_{max}$, the smaller $x_u$.
Thus, to obtain $x_{ui}$ close to its experimental counterpart, one has to reduce
$v$ by several orders of magnitude and this problem becomes unfeasible.
It is also clear why now a day all-atom simulations 
with explicit water can not be used to reproduce the FEL
parameters, obtained from experiments. From this point of view coarse-grained
models are of great help \cite{MSLi_BJ07a,MSLi_JCP08}.
The kinetic microscopic theory \cite{Dudko_PRL06}, 
which is valid beyond the BER approximation, can be applied
to extract unfolding barriers $\Delta G^{\ddagger}_i (i=1,2,$ and 3). Their values are
not presented as we are far from the interval of pulling speeds used
in experiments.

Since the first peak was not observed in the experiments
\cite{Schwaiger_NSMB04,Schwaiger_EMBO05},
 a natural question emerges is whether it is an artifact of high pulling speeds used in our simulations.
Except data at the highest value of $v$ (Fig. \ref{fel_schem_fmax_logv_fig}b),
within error bars three maxima are compatible. Therefore,
the peak centered at $\Delta R \approx 2$ nm is expected to remain
at experimental loading rates.
\cite{Schwaiger_NSMB04}.
 The force-extension curve of
the titin domain I27, which has a similar native topology, for example, displays the
first peak at $\Delta R \approx 0.8$ nm \cite{Marszalek_Nature99}.
One of possible reasons of why the experiments did not detect this
maximum is related to a strong linker effect as
a single DDFLN4 domain is sandwiched between Ig domains
I27-30 and domains I31-34 from titin \cite{Schwaiger_NSMB04}.

\subsection{Conclusions}

Using the all-atom simulations, we have reproduced the experimental result
on existence of two peaks located at $\Delta R \approx 12$ and 22 nm.
Our key result is that the later maximum occurs
due to breaking of five non-native HBs between strand B and C.
It can not be encountered by the Go models in which non-native
interactions are neglected \cite{MSLi_JCP08,MSLi_JCP09}.
Thus, our result points to the importance of these
interactions for the mechanical unfolding of DDFLN4.
The description of elastic properties of other proteins may be not
complete ignoring non-native interactions.
This conclusion is valuable as the unfolding by an external
force is widely believed
to be solely governed by native topology of proteins.

Our all-atom simulation study supports the result obtained by the Go model \cite{MSLi_JCP08,MSLi_JCP09}
 that an additional peak occurs at
$\Delta R \approx 2$ nm due to unfolding of strand G.
However, it was not observed by the AFM experiments
of Schwaiger {\em et al} \cite{Schwaiger_NSMB04,Schwaiger_EMBO05}.
In order to solve this controversy, one has to carry out not only
 simulations with
other force fields but also additional experiments.

\clearpage


\begin{center}\section*{CONCLUSIONS}\end{center}
In this thesis we have obtained the following new results.
By collecting experimental data and performing extensive on- and off-lattice
coarse-grained simulations, it was found that the scaling
exponent for the cooperativity of folding-unfolding transition
$\zeta \approx 2.2$.
This value is clearly higher than the characteristic for the first
order transition value $\zeta = 2$. Our result supports the previous
conjecture 
\cite{MSLi_PRL04} that the melting point is a tricritical point, where
the first and second order transition lines meet.
Having used CD technique and Go simulations, we studied the folding of protein
domain hbSBD in detail. Its thermodynamic parameters such as $\Delta H_G,
\Delta C_p, \Delta S_G$, and $\Delta G_S$ were determined.
Both experiments and theory support the two-state behavior of hbSBD.

With the help of the Go modeling, we have constructed
the FEL for
single and  three-domain Ub, and DDFLN4. Our estimations of $x_u$, $x_f$ and
$\Delta G^{\ddag}_u$ are in acceptable agreement with the experimental data.
The effect of pulling direction on FEL was also studied for single Ub.
Pulling at Lys48 and C-termini deforms the unfolding FEL
as it increases the distance between the NS and TS.
It has been shown that unfolding pathways of Ub depend on what terminal
is kept fixed. But it remains unclear if this is a real effect or
merely an artifact of high pulling speeds we used in simulations. This
problem requires further investigation.

It is commonly believed that protein unfolding is governed by the native 
topology and non-native interactions play a minor
role. However, having performed Gromacs all-atom simulations
for DDFLN4, for the first
time, we have demonstrated that it may depends on the non-native interactions.
Namely, they are responsible for occurrence of a peak located
at $\Delta R \approx 22$ nm in the force-extension curve. This peak was not seen in Go models
as they take into account only native interactions.
In addition, based on the Go as well as all-atom simulations, we
predict that an addition peak should appear at $\Delta R \approx 1.5$ nm.
Since such a peak was not observed in the experiments, our results
are expected to draw attention of experimentalists to this
fascinating problem.

Our new force RE method is interesting from the methodological point of view. 
Its successful application to construction of the $T-f$ phase diagram
of the three-domain Ub shows that it might be applied to other
biomolecules.

\newpage
\appendix
\section*{APPENDIX: List of abbreviations and symbols}


\begin{tabular}{ll}
AFM & Atomic Force Microscopy\\
BER & Bell-Evans-Rirchie\\
DS  &  Denaturated State\\
TS  & Transition State\\
IS  & Intermediate State\\
NS  & Native State \\
MD  & Molecular Dynamics\\
SMD  & Stereed Molecular Dynamics\\
SMFS  & Single Molecular Force Spectroscopy\\
FEL  & Free Energy Landscape\\
RE  & Replica Exchange\\
CD  & Circular Dichroism\\
 DDFLN4 & Fourth domain of {\em Dictyostelium discoideum}
filamin\\
Ub &      Ubiquitin\\
trimer & three-domain ubiquitin\\
$\Delta R$ & end-to-end extension\\
$x_f$  &  distance between TS ans DS\\
 $x_u$   &distance between NS and TS\\
  NBA     & native basin of attraction\\
  $T-f$  & Temperature-force\\
  HBs	 & Hydrogen bonds\\
TDE & Thermal denaturated ensemble\\
FDE & Force denaturated ensemble
\end{tabular}

\clearpage



\end{document}